\documentclass[11pt]{article}
\usepackage{jheppub}

\usepackage{amsmath, amssymb}
\usepackage{mathtools}
\usepackage[english]{babel}
\usepackage[utf8]{inputenc}
\usepackage{enumitem}
\usepackage{latexsym}
\usepackage{graphicx}
\usepackage{slashed}
\usepackage{xcolor}
\usepackage{color}
\usepackage{float}
\usepackage{bm}
\usepackage{braket}
\usepackage{dsfont}
\usepackage{cancel}
\usepackage{mathrsfs}
\usepackage{tikz}
\usepackage{pdfcomment}

\usepackage{hyperref}
\usepackage{tikz-feynman}
\usetikzlibrary{decorations.pathmorphing, decorations.markings, arrows.meta}
\tikzset{boson/.style={decorate, decoration={snake, amplitude=0.7mm, segment length=2mm}, draw=black, thick}}
\tikzset{scalar/.style={draw=black, thick, dashed}}
\tikzset{fermion/.style={draw=black, thick, postaction={decorate}, decoration={markings, mark=at position 0.55 with {\arrow{Latex[length=2mm]}}}}}

\DeclareMathOperator{\Tr}{Tr}
\newcommand{\Hil}{\mathcal{H}}
\newcommand{\CC}{\mathbb{C}}
\newcommand{\RR}{\mathbb{R}}

\newcommand{\ketbra}{\rangle \langle}

\usepackage{caption}

\usepackage{hhline}

\usepackage{array}

\newcolumntype{C}[1]{>{\centering\arraybackslash}p{#1}}
\graphicspath{{figs/}}

\def\be{\begin{equation}}
\def\ee{\end{equation}}
\def\ba{\begin{eqnarray}}
\def\ea{\end{eqnarray}}
\def\nn{\nonumber }

\title{\centering \boldmath
Qubit-qubit-qutrit quantum correlations in $H \to f \bar f V$
}

\author[a,b,c]{Michał Banacki,}
\author[d]{Misaki Ohta,}
\author[a]{Abhyoudai S. Shaleena,}
\author[e]{Paweł Caban,}
\author[f]{Michał Eckstein,}
\author[g]{Kazuki Sakurai,}
\author[h,i]{Michael Spannowsky,}
\author[a,j]{Paweł Horodecki}

\affiliation[a]{International Centre for Theory of Quantum Technologies (ICTQT), University of Gdańsk, Jana Bażyńskiego 8, 80-309 Gdańsk, Poland}

\affiliation[b]{Institute of Mathematics, Faculty of Mathematics, Physics and Informatics, University of Gda\'{n}sk, Wita Stwosza 57, 80-308, Gda\'{n}sk, Poland}

\affiliation[c]{Institute for Theoretical Physics, University of Cologne, Höninger Weg 100, 50969, Cologne, Germany}

\affiliation[d]{IQM Quantum Computers, Tokyo, Japan}

\affiliation[e]{Department of Theoretical Physics, University of Łódź, Pomorska 149/153, 90-236 Łódź, Poland}

\affiliation[f]{Institute of Theoretical Physics, Faculty of Physics, Astronomy and Applied Computer Science, Jagiellonian University, ul. Łojasiewicza 11, 30-348 Kraków, Poland}

\affiliation[g]{Institute of Theoretical Physics, Faculty of Physics, University of Warsaw, Pasteura 5, 02-093, Warsaw, Poland}

\affiliation[h]{Institute for Theoretical Physics, Campus Süd, Karlsruhe Institute of Technology (KIT), D-76128 Karlsruhe, Germany}

\affiliation[i]{Institute for Quantum Materials and Technologies, Karlsruhe Institute of Technology, Karlsruhe 76131, Germany}

\affiliation[j]{Faculty of Applied Physics and Mathematics, Gdańsk University of Technology, Gabriela Narutowicza 11/12, 80-233 Gdańsk, Poland}

\emailAdd{kazuki.sakurai@fuw.edu.pl}

\abstract{
We perform an extensive analysis of the quantum correlations carried by the qubit--qubit--qutrit pure state arising in the decay of a massive scalar into a fermion--antifermion pair and a massive gauge boson, $H \to f \bar f V$, specialising to the Higgs boson decay $h \to \tau^- \tau^+ Z$.
Working with the exact tree-level spin state and its systematic expansion around the massless-fermion limit, we obtain analytic control over the entire phase space: the bipartite entanglement measures, the genuine $2 \otimes 2 \otimes 3$ entanglement structure (the Miyake classification), as well as the Bell-inequality violations and the non-stabiliserness (magic) are all mapped and reproduced by compact formulas.
The bipartite measures exhibit a monogamy-like trade-off between the fermion pair and the fermion--boson pairs.
The state is genuinely $2 \otimes 2 \otimes 3$ entangled over almost the entire phase space, most strongly in the collinear regions.
We derive, for the first time, semi-analytical expressions for the tight $4 \times 4 \times 2$ Bell inequalities of the $2 \otimes 2 \otimes 3$ system, generalising the optimisation previously available only for three qubits, and find that the local-hidden-variable bound is violated over the entire phase space, reaching within a few per cent of the quantum bound at the upper endpoint of the di-tau mass spectrum.
We further extend the stabiliser R\'enyi entropy and the non-local magic to systems with unequal local dimensions, and show that the near-endpoint state carries almost exactly one bit of non-local magic, which peaks at $\log_2 \frac{27}{7} \simeq 1.95$ in the collinear regions.
The differential decay rate concentrates precisely in the most nonclassical region of the phase space.
}

\begin{document}
\compress
\renewcommand{\afterAffiliationSpace}{\vskip1pt plus1pt}
\newlength{\titlepagewiden}
\setlength{\titlepagewiden}{1.6cm}
\addtolength{\textwidth}{\titlepagewiden}
\addtolength{\oddsidemargin}{-0.5\titlepagewiden}
\addtolength{\evensidemargin}{-0.5\titlepagewiden}
\hsize=\textwidth \linewidth=\textwidth
\let\jhepbeforetochook\beforetochook
\renewcommand{\beforetochook}{%
  \addtolength{\textwidth}{-\titlepagewiden}%
  \addtolength{\oddsidemargin}{0.5\titlepagewiden}%
  \addtolength{\evensidemargin}{0.5\titlepagewiden}%
  \hsize=\textwidth \linewidth=\textwidth
  \jhepbeforetochook}
\maketitle

\section{Introduction}
\label{sec:intro}

A high-energy collider is a prolific source of quantum states: each event prepares the spins of its final-state particles afresh, and events with the same kinematics build up an ensemble of identically prepared copies, which is exactly what the measurement of quantum observables requires.
The machinery of quantum information theory, from entanglement measures to Bell inequalities and resource monotones, thus becomes directly applicable to collider data.
What began as a theoretical proposal for top-quark pairs \cite{Afik:2020onf, Fabbrichesi:2021npl, Severi:2021cnj, Aoude:2022imd, Afik:2022kwm, Aguilar-Saavedra:2022uye, Fabbrichesi:2022ovb, Afik:2022dgh, Severi:2022qjy, Dong:2023xiw, Aguilar-Saavedra:2023hss, Han:2023fci, Cheng:2023qmz, Maltoni:2024tul, Maltoni:2024csn, Dong:2024xsg, Cheng:2024btk, Han:2024ugl, Altomonte:2024upf, Cheng:2025cuv, Afik:2025grr, Nason:2025hix, Fabbrichesi:2025psr, Lin:2025eci, Low:2025aqq, Durupt:2025wuk, Gabrielli:2026tnl, Guo:2026yhz, Fang:2026ddi, Choi:2026omc, Altakach:2026fpl,
Lamba:2026jfm,Lamba:2026sni} has become an experimental reality: ATLAS and CMS have observed spin entanglement in $t \bar t$ production \cite{ATLAS:2023fsd, CMS:2024pts, CMS:2024zkc}, and entanglement between pair-produced $Z$ bosons has recently been reported \cite{ATLAS:2026hye}.
A parallel effort has mapped out the quantum correlations of tau-lepton pairs \cite{Altakach:2022ywa, Ma:2023yvd, Ehataht:2023zzt, Fabbrichesi:2024wcd, Fabbrichesi:2025ywl, Han:2025ewp, Zhang:2025mmm, Jeans:2026eys, Lee:2026wlk} and of electroweak boson pairs \cite{Barr:2021zcp, Aguilar-Saavedra:2022wam, Aguilar-Saavedra:2022mpg, Ashby-Pickering:2022umy, Fabbrichesi:2023cev, Fabbrichesi:2023jep, Morales:2023gow, Aoude:2023hxv, Fabbri:2023ncz, Bernal:2023ruk, Bi:2023uop, Bernal:2024xhm, Ruzi:2024cbt, Grossi:2024jae, Wu:2024ovc, Sullivan:2024wzl, Aguilar-Saavedra:2024jkj, Ding:2025mzj, DelGratta:2025qyp, Goncalves:2025mvl, Aguilar-Saavedra:2025byk, Goncalves:2025xer, Ruzi:2025jql, Aguilar-Saavedra:2025njw, De:2025dpo, Pelliccioli:2026ltl, Goncalves:2026njf}; comprehensive overviews can be found in Refs.~\cite{Barr:2024djo, Afik:2025ejh}.
Attention has also turned to how robust these correlations are in a realistic collider environment: the emission of QCD and electromagnetic radiation entangles the spins with unobserved degrees of freedom, and the resulting decoherence of the spin state has recently been quantified \cite{Aoude:2025ovu, Gu:2025ijz, Aoude:2026eeg, Cheng:2026zfb}.

All of these systems, however, involve two parties, and two-party quantumness is structurally simple: a pure state is either separable or entangled, a single number locates it between the two extremes, and a single inequality, Clauser--Horne--Shimony--Holt (CHSH), exhausts the tight tests of local realism.
None of this survives the addition of a third particle.
Separability branches into inequivalent classes (full, partial and genuine multipartite entanglement); entanglement becomes a \emph{structure} rather than a quantity, constrained by monogamy and organised into inequivalent stochastic local operations and classical communication (SLOCC) families of Greenberger--Horne--Zeilinger (GHZ) and W type \cite{Horodecki:2009zz, Guhne:2008qic, Coffman:1999jd, Miyake:2003tee}; and pairwise correlations no longer determine the correlations of the whole.
Adapting this richer landscape to particle physics is a young programme: the founding framework was laid out in Refs.~\cite{Sakurai:2023nsc, Horodecki:2025tpn}, first applications to Higgs and diboson systems followed \cite{Aguilar-Saavedra:2024whi, Morales:2024jhj, Subba:2024mnl, Fabbrichesi:2025zpw}, and a detailed mixed-state study of $e^+ e^- \to t \bar t Z$ production at future lepton colliders has appeared very recently \cite{Goncalves:2026nnx}.

Entanglement, even when fully resolved into its multipartite structure, does not capture all that is quantum about a state.
The Gottesman--Knill theorem \cite{Gottesman:1998hu} shows that stabiliser states, however entangled, are efficiently simulable on a classical computer; the resource that classical simulation cannot reproduce is \emph{non-stabiliserness}, or \emph{magic}, and it is largely independent of entanglement.
Thanks to a computable quantifier, the stabiliser R\'enyi entropy \cite{Leone:2021rzd, Wang:2023uog, Leone:2024lfr, Bittel:2025yhq}, magic too is entering collider physics.
It was first proposed as an observable for the spin state of top-quark pairs \cite{White:2024nuc}, and the CMS collaboration has since reported the observation of magic in $t \bar t$ events \cite{CMS-PAS-TOP-25-001}.
It has also been studied in gluon, graviton and nucleon scattering amplitudes \cite{Gargalionis:2025iqs, Gargalionis:2026onv, Robin:2025ymq,Chu:2026yxm}, and its basis-independent core, the \emph{non-local magic} that no local operations can remove, has been introduced in Ref.~\cite{Qian:2025oit}.

In this paper we study a single process in which all of these developments come together: the three-body Higgs decay $h \to \tau^- \tau^+ Z$.
This process stands out for three reasons.
Being the decay of an on-shell scalar, it delivers a \emph{pure} spin state, so the complete pure-state toolbox of concurrences, tangles and hyperdeterminants applies straightforwardly, and the stabiliser R\'enyi entropy retains its status as a faithful magic measure.
Its Hilbert space, $\mathbb{C}^2 \otimes \mathbb{C}^2 \otimes \mathbb{C}^3$, is the smallest tripartite space with unequal local dimensions, placing the state beyond the well-studied three-qubit case: several of the theoretical tools we need simply did not exist for such asymmetric systems and are constructed here.
And it is analytically tractable: organising the amplitude as an expansion around the massless-fermion limit reveals a pointer-state structure of the quantum state.
As a result, every observable we study is given by a compact closed formula in each corner of the phase space, which explains the main features of the numerical maps.
While we focus on $h \to \tau^- \tau^+ Z$ for definiteness, our analytical study assumes generic interactions, couplings and masses, and the results apply to any decay of a scalar into a fermion pair and a massive vector boson.

We study a wide range of quantum observables of this state: the bipartite entanglement, the genuine $2 \otimes 2 \otimes 3$ structure captured by the Miyake classification and the $223$-tangle, the Bell nonlocality, and the magic.
Two of our results are, to the best of our knowledge, new.
First, we derive, for the first time, semi-analytical expressions for the tight $4 \times 4 \times 2$ Bell inequalities \cite{WU2003262, laskowski} of the $2 \otimes 2 \otimes 3$ system: the optimisation over measurement settings, previously available only for three qubits \cite{Horodecki:2025tpn}, is carried out for the enlarged family of qutrit measurements with the help of the qubit--qudit CHSH result of Ref.~\cite{Bernal:2024dtg}, and the construction extends straightforwardly to any $2 \otimes 2 \otimes d$ system.
Second, we extend the stabiliser R\'enyi entropy and the non-local magic to Hilbert spaces with unequal local dimensions.
Our study is complementary to the $t \bar t Z$ analysis of Ref.~\cite{Goncalves:2026nnx}: there, the production process yields a mixed state, which is analysed with negativity-based measures; here, the decay yields a pure state, for which the full set of observables listed above becomes accessible in semi-analytic form.

The paper is organised as follows.
Section~\ref{sec:0-223} derives the spin quantum state of $H \to f \bar f V$, its expansion in the small fermion mass and its CP properties.
Section~\ref{sec:ent} analyses the entanglement structure, from one-to-other and one-to-one measures to the genuine $2 \otimes 2 \otimes 3$ classification.
Section~\ref{sec:Bell} constructs the semi-analytical optimisation of the $4 \times 4 \times 2$ Bell observables, and Sec.~\ref{sec:magic} the asymmetric stabiliser R\'enyi entropy and non-local magic, in each case followed by the phase-space maps and their analytical interpretation.
Section~\ref{sec:rate} examines the differential decay rate and the experimental prospects at the LHC and the High-Luminosity LHC (HL-LHC), and Sec.~\ref{sec:concl} concludes.
Technical material, including the sub-leading pointer states, the $I$-concurrence, the Miyake classification and the explicit Heisenberg--Weyl matrices, is collected in the appendices.

\section{\boldmath Kinematics and spin quantum state in $H \to f \bar f V$}
\label{sec:0-223}

We consider the decay of a massive CP-even scalar $H$ into a light fermion--antifermion pair, $f \bar f$, accompanied by a massive vector boson $V$.
The spin degrees of freedom of the final state span the Hilbert space $\Hil = \CC^2 \otimes \CC^2 \otimes \CC^3$, a $2 \otimes 2 \otimes 3$ system whose quantum observables we study in the following sections.
The process of primary interest is the three-body Higgs decay, $h \to \tau^- \tau^+ Z$, but in the analytic treatment we keep the masses and couplings general, specialising to $(H, f, \bar f, V) = (h, \tau^-, \tau^+, Z)$ only when the observables are evaluated numerically.
All analytical results in this paper are therefore applicable to other decay modes of the same type, such as $h \to \mu^- \mu^+ Z$ and $h \to b \bar b Z$, as well as to decays of new scalars in theories beyond the Standard Model, upon substituting the appropriate couplings and masses.

\subsection{Kinematics}

\begin{figure}[t]
\centering
\includegraphics[width=0.25\textwidth]{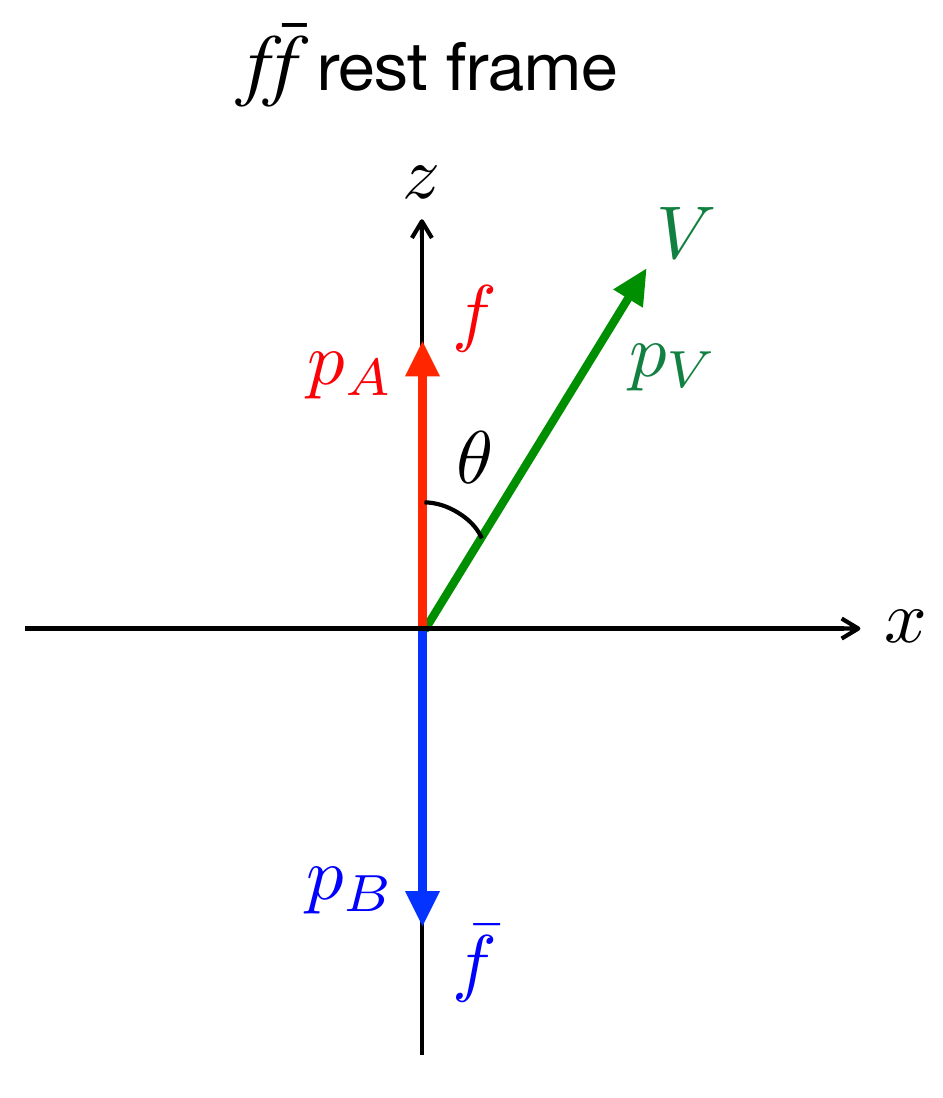}
\caption{\small Definition of the coordinate system used to parametrise the kinematics.}
\label{fig:kinematics}
\end{figure}

We work in the rest frame of the $f \bar f$ subsystem, labelling the fermion $f$ and the antifermion $\bar f$ by $A$ and $B$, respectively, so that $A$ and $B$ tag their momenta and helicities throughout.
Choosing the $z$--axis along the direction of $f$, we write the two momenta as
\be
p_A^\mu = \left( \frac{m_{AB}}{2},\, 0,\, 0,\, k_A \right),
\qquad
p_B^\mu = \left( \frac{m_{AB}}{2},\, 0,\, 0,\, -k_A \right),
\qquad
k_A = \frac{1}{2}\sqrt{ m_{AB}^2 - 4 m_f^2 }\,,
\ee
where $m^2_{AB} \equiv (p_A + p_B)^2$ is the invariant mass of the $f \bar f$ pair and $m_f$ is the common mass of $f$ and $\bar f$.
We choose the $x$--axis so that the vector boson momentum $\vec{p}_V$ lies in the $xz$--plane and $(p_V)_x \geq 0$.
The vector boson four-momentum can then be written as
\be
p_V^\mu \,=\, \left( E_V,\; k_V \sin\theta,\; 0,\; k_V \cos\theta \right),
\ee
where $\theta$ is the polar angle between $f$ and $V$ (see Fig.~\ref{fig:kinematics}), and $E_V$ and $k_V \equiv |\vec p_V|$ are the energy and momentum magnitude of $V$, respectively.

Imposing the mass-shell condition for the scalar, $p_H^2 = (p_A + p_B + p_V)^2 = m_H^2$, we obtain 
\be
E_V \,=\, \frac{ m_H^2 - m_{AB}^2 - m_V^2 }{ 2\, m_{AB} }\,,
\qquad
k_V \,=\, \sqrt{E_V^2 - m_V^2} \,=\,
\frac{ \lambda^{1/2}(m_H^2, m_{AB}^2, m_V^2) }{ 2\, m_{AB} }\,,
\ee
where $\lambda(m_H^2, m_{AB}^2, m_V^2) \equiv (m_H^2 - m_{AB}^2 - m_V^2)^2 - 4\, m_{AB}^2 m_V^2$ is the K\"all\'en function.
All momenta are thus fixed by the two independent parameters,
$m_{AB} \in \left[\, 2 m_f,\; m_H - m_V \,\right]$
and $\theta \in \left[\, 0,\, \pi \,\right]$.
At the two endpoints of the $m_{AB}$ range, the vector boson energy and momentum behave as
\ba
m_{AB}^{\rm min} = 2 m_f:&&~~
E_V, k_V \,\sim\, \frac{m_H^2 - m_V^2}{4 m_f} \,,
\nn \\
m_{AB}^{\rm max} = m_H-m_V:&&~~
E_V \,=\, m_V, ~~k_V \,=\, 0\,.
\ea
Note that for $m_f \to 0$ the vector boson energy and momentum diverge at the threshold $m_{AB} \to 2 m_f$; this observation will play an important role below.

\subsection{Spin quantum state}

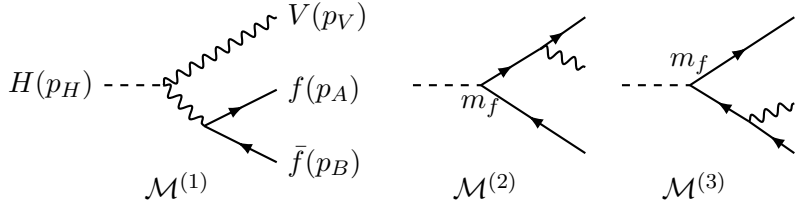
\begin{figure}[!t]
\begin{center}
\begin{tikzpicture}[scale=0.6, thick]
  \coordinate (H)    at (-1.8,  0.0);
  \coordinate (V1)   at (-0.5,  0.0);   
  \coordinate (Zext) at ( 2.0,  1.5);   
  \coordinate (V2)   at ( 0.4, -0.9);   
  \coordinate (fA)   at ( 2.0, -0.1);
  \coordinate (fbB)  at ( 2.0, -1.7);
  \draw[scalar] (H) -- (V1);
  \node[left] at (H) {$H(p_H)$};
  \draw[boson] (V1) -- (Zext);
  \node[right] at (Zext) {$V(p_V)$};
  \draw[boson] (V1) -- (V2);
  \draw[fermion] (fbB) -- (V2);
  \draw[fermion] (V2) -- (fA);
  \node[right] at (fA)  {$f(p_A)$};
  \node[right] at (fbB) {$\bar f(p_B)$};
  \node at (-0.2,-2.2) {${\cal M}^{(1)}$};
\end{tikzpicture}
\hspace{0.2cm}
\begin{tikzpicture}[scale=0.6, thick]
  \coordinate (H)   at (-1.8,  0.0);
  \coordinate (V)   at (-0.3,  0.0);   
  \coordinate (Vz)  at ( 1.0,  0.8);   
  \coordinate (fA)  at ( 2.0,  1.5);
  \coordinate (fbB) at ( 2.0, -1.5);
  \coordinate (Zout)at ( 2.0,  0.4);
  \draw[scalar] (H) -- (V);
  \node[below] at (V) {$m_f$};
  \draw[fermion] (fbB) -- (V);
  \draw[fermion] (V) -- (Vz);
  \draw[fermion] (Vz) -- (fA);
  \draw[boson] (Vz) -- (Zout);
  \node at (-0.2,-2.2) {${\cal M}^{(2)}$};
\end{tikzpicture}
\hspace{0.2cm}
\begin{tikzpicture}[scale=0.6, thick]
  \coordinate (H)   at (-1.8,  0.0);
  \coordinate (V)   at (-0.3,  0.0);   
  \coordinate (Vz)  at ( 1.0, -0.8);   
  \coordinate (fA)  at ( 2.0,  1.5);
  \coordinate (fbB) at ( 2.0, -1.5);
  \coordinate (Zout)at ( 2.0, -0.4);
  \draw[scalar] (H) -- (V);
  \node[above] at (V) {$m_f$};
  \draw[fermion] (V) -- (fA);
  \draw[fermion] (fbB) -- (Vz);
  \draw[fermion] (Vz) -- (V);
  \draw[boson] (Vz) -- (Zout);
  \node at (-0.2,-2.2) {${\cal M}^{(3)}$};
\end{tikzpicture}
\end{center}
\caption{\label{fig:diagram} Tree-level Feynman diagrams for $H \to f \bar f V$.
}
\end{figure}

The process receives contributions from three tree-level Feynman diagrams, shown in Fig.~\ref{fig:diagram}, with amplitudes ${\cal M}^{(1)}$, ${\cal M}^{(2)}$ and ${\cal M}^{(3)}$.
We compute them from the general effective interactions
\be
{\cal L} \;\supset\;
g_{Hff}\, H\, \bar\psi_f \psi_f
\,+\,
g_{HVV}\, H\, V_\mu V^\mu
\,+\,
g_{Vff}\, \bar\psi_f \gamma^\mu \left( c_L P_L + c_R P_R \right)\psi_f\, V_\mu\,,
\label{eq:Leff}
\ee
with $P_{L,R} = \tfrac{1}{2}(1\mp\gamma_5)$ the chiral projectors and $c_L,c_R \in \RR$
the left-- and right--handed couplings of the $f \bar f V$ vertex.
If $H$ is the Higgs boson, 
\be
g_{Hff} = -\,\frac{m_f}{v}\,,
\qquad
g_{HVV} = \frac{2\, m_V^2}{v}\,,
\ee
with $v$ the Higgs vacuum expectation value. 
For $(f, \bar f, V) = (\tau^-, \tau^+, Z)$, the chiral couplings are given by
\be
c_L \,=\, - \frac{1}{2} + \sin^2 \theta_W \,\simeq\, -0.269\,,
\qquad
c_R \,=\, \sin^2 \theta_W \,\simeq\, 0.231\,.
\label{eq:cLcR_SM}
\ee
It is also convenient to introduce the vector and axial-vector combinations
\be
c_V \,\equiv\, \frac{1}{2}(c_R + c_L) \,\simeq\, -0.019\,,
\qquad
c_A \,\equiv\, \frac{1}{2}(c_R - c_L) \,\simeq\, 0.250\,,
\label{eq:cVcA_SM}
\ee
whose strong hierarchy, $c_V^2 \ll c_A^2$, is an accidental consequence of the proximity of $\sin^2\theta_W$ to $1/4$; it will leave visible imprints on several observables below.
Much of this coupling dependence will enter through the combinations
\be
\hat c \,\equiv\, \frac{ 2\, c_A c_V }{ c_A^2 + c_V^2 }
\,\simeq\, -0.151\,,
\qquad
\hat s \,\equiv\, \sqrt{ 1 - \hat c^2 } \,=\, \frac{ c_A^2 - c_V^2 }{ c_A^2 + c_V^2 }
\,=\, \frac{ 2\, |c_L c_R| }{ c_L^2 + c_R^2 }\,\simeq\,0.988\,.
\label{eq:chat}
\ee

With these general interaction terms, the three amplitudes read
\ba
{\cal M}^{(1)}
&=&
g_{HVV}\, g_{Vff}\,
\left[\, \bar u_A \gamma^\mu \left( c_L P_L + c_R P_R \right) v_B \,\right]
\Pi_{\mu\nu}(p_A + p_B)\,
\epsilon^{\ast\,\nu}(p_V)\,,
\nn \\[4pt]
{\cal M}^{(2)}
&=&
g_{Hff}\, g_{Vff}\,
\bar u_A\, \gamma^\mu \left( c_L P_L + c_R P_R \right)\,
S_F(p_A + p_V)\, v_B\;
\epsilon^{\ast}_{\mu}(p_V)\,,
\nn \\[4pt]
{\cal M}^{(3)}
&=&
g_{Hff}\, g_{Vff}\,
\bar u_A\, S_F(p_B + p_V)\,
\gamma^\mu \left( c_L P_L + c_R P_R \right) v_B\;
\epsilon^{\ast}_{\mu}(p_V)\,,
\ea
where $u_A \equiv u(p_A,h_A)$ and $v_B \equiv v(p_B,h_B)$ are the external fermion spinors and $\epsilon^\ast(p_V) \equiv \epsilon^\ast(p_V,h_V)$ is the vector boson polarisation vector.
The helicity labels run over $h_A, h_B \in \{+,-\}$ and $h_V \in \{+,0,-\}$.
The massive vector and internal fermion propagators are given by
\be
\Pi_{\mu\nu}(k)
\,=\,
\frac{1}{k^2 - m_V^2 + i\, m_V \Gamma_V}
\left( -g_{\mu\nu} + \frac{k_\mu k_\nu}{m_V^2} \right),
\qquad
S_F(k)
\,=\,
\frac{\slashed{k} + m_f}{k^2 - m_f^2 + i\, m_f \Gamma_f}\,,
\ee
with $\Gamma_V$ ($\Gamma_f$) the width of the vector boson (fermion).
The total amplitude is
\be
{\cal M}_{h_A h_B h_V} \,=\,
{\cal M}^{(1)} + {\cal M}^{(2)} + {\cal M}^{(3)}\,.
\label{eq:total_amp}
\ee
The spin quantum state of the $f \bar f V$ system is given by
\be
\ket{\psi}
\,\propto\,
\sum_{h_A, h_B, h_V}
{\cal M}_{h_A h_B h_V}\, \ket{h_A, h_B, h_V}\,.
\ee

Because ${\cal M}^{(2)}$ and ${\cal M}^{(3)}$ are proportional to the Yukawa coupling $g_{Hff} \propto m_f$, the first diagram dominates in the limit $m_f \ll m_V, m_H$.
Exploiting this hierarchy, we expand the amplitudes in $m_f$ and retain terms up to linear order.
At $m_f = 0$, the amplitude ${\cal M}^{(1)}$ populates only the opposite-helicity subspace $\{\ket{-+},\ket{+-}\}$, while ${\cal M}^{(2)}$ and ${\cal M}^{(3)}$ vanish.
The spin state takes the form 
\be
\ket{ \psi } \,=\, \frac{1}{\mathcal{N}} \, \Big(\,
\ket{ \psi^{(0)} } \,+\, \varepsilon \ket{ \psi^{(1)} } + \cdots
\, \Big),
\label{eq:state}
\ee
where $\mathcal{N}$ is the normalisation factor and the expansion parameter is 
\be
\varepsilon \,\equiv\, \frac{ \sqrt{2} m_f\, (m_V^2 - m_{AB}^2) }{ m_V ( m_H^2 - m^2_{AB} - m_V^2) }
\,.
\ee
For $h \to \tau^- \tau^+ Z$ in the Standard Model, $\varepsilon$ is small and remains nearly constant across the accessible phase space, $\varepsilon \in [3.14,\,3.21] \times 10^{-2}$.

In the above expression, the \emph{unnormalised} zeroth-order state is given by
\be
\ket{ \psi^{(0)}} \,=\,
c_L\, \ket{-+} \ket{\Psi_+}
\,+\,
c_R\, \ket{+-} \ket{\Psi_-}
\label{eq:psi0}
\ee
with
\ba
\ket{\Psi_\pm} &=&
\pm\,\frac{m_V}{E_V} (1\pm\cos\theta)\,\ket{+}
\,-\, \sqrt2\, \sin\theta\,\ket{0}
\,\pm\, \frac{m_V}{E_V} (1\mp\cos\theta)\,\ket{-}\,.
\label{eq:pointer1}
\ea
Here and throughout, the phases of the helicity eigenstates $\ket{h_A, h_B, h_V}$ are fixed such that the CP transformation takes a particularly simple, phase-free form (see Sec.~\ref{sec:cp}).
The vector-boson states $\ket{\Psi_\pm}$ act as pointer states: $\ket{\Psi_+}$ and $\ket{\Psi_-}$ are each correlated with a single fermion helicity configuration, $\ket{-+}$ and $\ket{+-}$, respectively.
These pointer states are neither normalised nor mutually orthogonal, and their norms and overlap read
\ba
\braket{ \Psi_+ | \Psi_+ } \,=\,
\braket{ \Psi_- | \Psi_- } \,=\,
2 \left[ 1 + \frac{m_V^2 }{E_V^2} - \frac{k_V^2 \cos^2 \theta}{E_V^2} \right],
\quad
\braket{ \Psi_+ | \Psi_- } \,=\,
\frac{2 k_V^2}{E_V^2}\,\sin^2\theta .
\label{eq:psinorm}
\ea
The overlap vanishes only at the upper endpoint of $m_{AB}$ ($k_V = 0$) or for collinear kinematics ($\sin\theta = 0$).
The leading state $\ket{\psi^{(0)}}$ in Eq.~\eqref{eq:psi0} is thus effectively a GHZ-like $2 \otimes 2 \otimes 2$ state: only the two-dimensional subspace of the qutrit spanned by $\ket{\Psi_\pm}$ is activated.

The unnormalised state corresponding to the leading correction is given by
\be
\ket{\psi^{(1)}} \,=\, \ket{++} \ket{\Phi_+} \,+\, \ket{--} \ket{\Phi_-}\,,
\label{eq:phi1}
\ee
where the pointer states $\ket{ \Phi_\pm}$ are written as 
\ba
\ket{\Phi_\pm} &=&
a_{\pm}^{(+)}\,\ket{+}
\,+\,a_{\pm}^{(0)}\,\ket{0}
\,+\,a_{\pm}^{(-)}\,\ket{-}\,.
\label{eq:pointer}
\ea
with the coefficients $a_\pm^{(h_V)}$ given by
\ba
a^{(h_V)}_{\pm}
&=& \mp\, h_V \,\sqrt{2}\,\sin\theta\Big[
\frac{2 c_V m_V^2}{m_V^2- m_{AB}^2}
\nn \\
&& \qquad
+~\frac{ m_{AB}\, (E_V+ m_{AB}\pm h_V k_V) \big( c_V D \pm \, c_A m_{AB} k_V \cos\theta\big) }{D_+ D_-}
\Big] ,
\ea
for the transverse helicities $h_V = \pm 1$, and
\ba
a^{(0)}_{\pm}
&=& 2 \left[
 c_A \frac{k_V}{ m_V}\!\left(2+\frac{ m_{AB}^2 D\sin^2\theta}{ D_+ D_- }\right)
\mp \, c_V m_V\cos\theta\!\left(\frac{2 E_V}{m_V^2- m_{AB}^2}+\frac{ m_{AB} m_H^2}{ D_+ D_- }\right)
\right] ,
\ea
for the longitudinal helicity.
Here we used the vector and axial-vector couplings $c_{V,A}$ of Eq.~\eqref{eq:cVcA_SM}, together with the kinematic combinations $D \equiv E_V m_{AB} + m_V^2$ and $D_\pm \equiv D \pm m_{AB} k_V \cos\theta$.

As in $\ket{\psi^{(0)}}$, the sub-leading state $\ket{\psi^{(1)}}$ is effectively a GHZ-like $2 \otimes 2 \otimes 2$ state, with only a two-dimensional subspace of the qutrit activated.
The activated fermionic subspace, however, is the same-helicity one, $\{ \ket{++}, \ket{--} \}$, orthogonal to the opposite-helicity subspace supporting the leading state $\ket{\psi^{(0)}}$.
An immediate consequence is $\braket{ \psi^{(0)} | \psi^{(1)} } = 0$: the two states do not interfere.

At the threshold $m_{AB} \to 2 m_f$, the relative momentum of the $f \bar f$ pair vanishes ($k_A \to 0$) while the vector boson recoils ultra-relativistically, $E_V \simeq k_V \to (m_H^2 - m_V^2)/(4 m_f)$.
In this limit the transverse coefficients $a^{(\pm 1)}_{\pm}$ remain finite, whereas the longitudinal ones diverge as $1/m_f$,
\be
a^{(0)}_{\pm} \,\to\, \frac{4 E_V}{m_V} \left( c_A \mp c_V \cos\theta \right),
\ee
the familiar growth of longitudinally polarised gauge-boson amplitudes.
No analogous enhancement occurs in the leading state: the longitudinal coefficient of $\ket{\Psi_\pm}$ in Eq.~\eqref{eq:pointer1}, $- \sqrt{2} \sin\theta$, is energy independent.
The $1/m_f$ divergence exactly compensates the Yukawa suppression $\varepsilon \to \sqrt{2}\, m_f m_V/(m_H^2 - m_V^2)$, and $\varepsilon \ket{ \psi^{(1)} }$ approaches the finite, purely longitudinal limit
\be
\varepsilon \ket{ \psi^{(1)} } \,=\,
\sqrt{2} \Big[ \left( c_A - c_V \cos\theta \right) \ket{++}
\,+\, \left( c_A + c_V \cos\theta \right) \ket{--} \Big] \ket{0}
\,+\, \mathcal{O}(m_f)\,.
\label{eq:psi1}
\ee
In the same limit, $m_V/E_V \to 0$ and the leading pointer states \eqref{eq:pointer1} also become purely longitudinal, $\ket{\Psi_\pm} \to - \sqrt{2}\, \sin\theta \ket{0}$, so that the full state near the threshold takes the form
\ba
\ket{ \psi_{\rm thr} } &=&
\frac{\sqrt{2}}{\mathcal{N}} \Big[\,
- \left( c_V - \beta\, c_A \right) \sin\theta\, \ket{-+}
\,-\, \left( c_V + \beta\, c_A \right) \sin\theta\, \ket{+-}
\nn \\
&&\qquad
\,+\, \left( c_A - c_V \cos\theta \right) \ket{++}
\,+\, \left( c_A + c_V \cos\theta \right) \ket{--}
\,\Big] \ket{0}\,,
\label{eq:threshold-state}
\ea
where $\beta \equiv 2k_A/m_{AB} = \sqrt{ 1 - 4 m_f^2 / m_{AB}^2 }$ is the fermion velocity.
In the first two terms we have restored the exact $m_f$ dependence of the opposite-helicity couplings, $c_L \to c_V - \beta c_A$ and $c_R \to c_V + \beta c_A$, to which they reduce at $\beta = 1$.
This follows from the exact identity for the opposite-helicity currents of massive fermions,
\be
\bar u_\mp\, \gamma^\mu \left( c_L P_L + c_R P_R \right) v_\pm
\,=\,
\left( c_V \mp \beta\, c_A \right)\, \bar u_\mp\, \gamma^\mu\, v_\pm\,,
\label{eq:beta-dressing}
\ee
i.e.\ the chiral current is proportional to the pure vector current, with the axial coupling weighted by the fermion velocity --- the familiar suppression of the axial charge for non-relativistic fermions.
This dressing is invisible at any fixed order of the $m_f$ expansion --- $\beta$ is not analytic in $m_f$ at $m_{AB} = 2 m_f$ --- and $\beta \simeq 1$ everywhere except in the immediate vicinity of the threshold; it is, however, essential there, since $\beta \to 0$ at the endpoint.
Equation~\eqref{eq:threshold-state} exhibits two important features.
First, the vector boson factors out in the longitudinal state $\ket{0}$: the state is a product across the $V | AB$ partition, and all quantum correlations reside in the fermion sector.
Second, at the endpoint $\beta \to 0$ the opposite-helicity components collapse to the small vector-coupling remnant $\propto c_V$, and the fermion pair is left in the nearly maximally entangled Bell-like state
$\left( c_A - c_V \cos\theta \right) \ket{++} + \left( c_A + c_V \cos\theta \right) \ket{--}$.

Near the threshold, the norms of the leading and subleading components behave as
\be
\braket{\psi^{(0)}|\psi^{(0)}} \,\to\, 2\,\langle c^2\rangle\, \sin^2\theta\,,
\qquad
\varepsilon^2 \braket{\psi^{(1)}|\psi^{(1)}} \,\to\, 16\,\frac{m_f^2}{m_{AB}^2}\,\big( c_A^2 + c_V^2 \cos^2\theta \big)\,,
\label{eq:threshold-norms}
\ee
with $\langle c^2\rangle \equiv c_L^2 + c_R^2$. 
\begin{figure}[t!]
\centering
\includegraphics[width=0.32\textwidth]{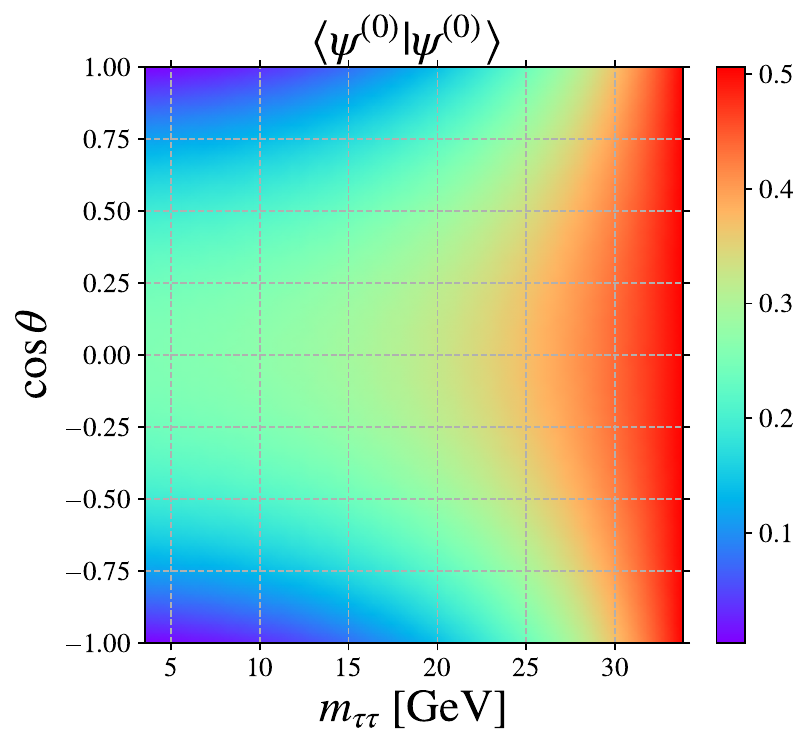}
\includegraphics[width=0.32\textwidth]{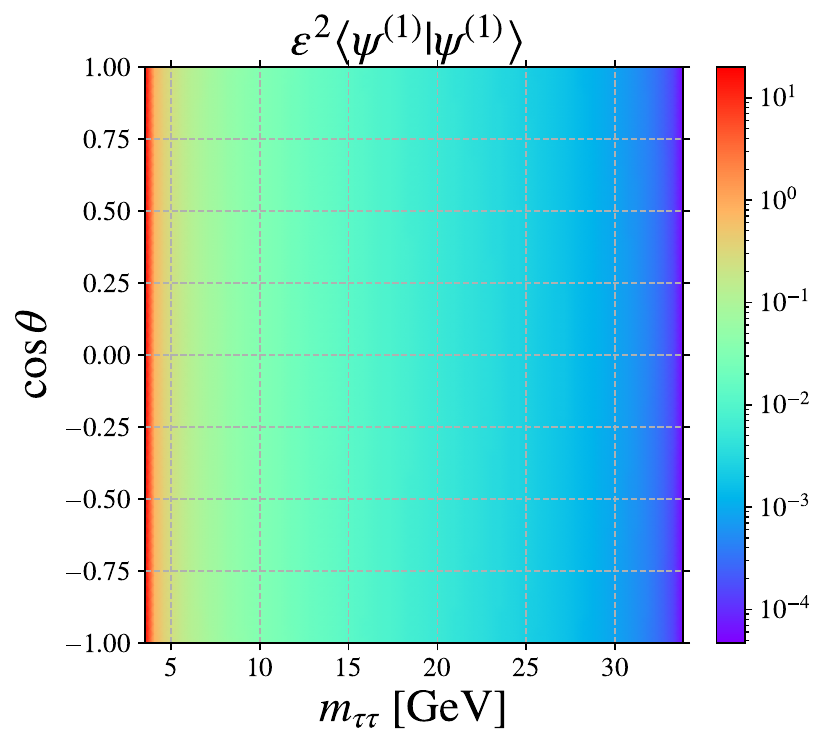}
\includegraphics[width=0.32\textwidth]{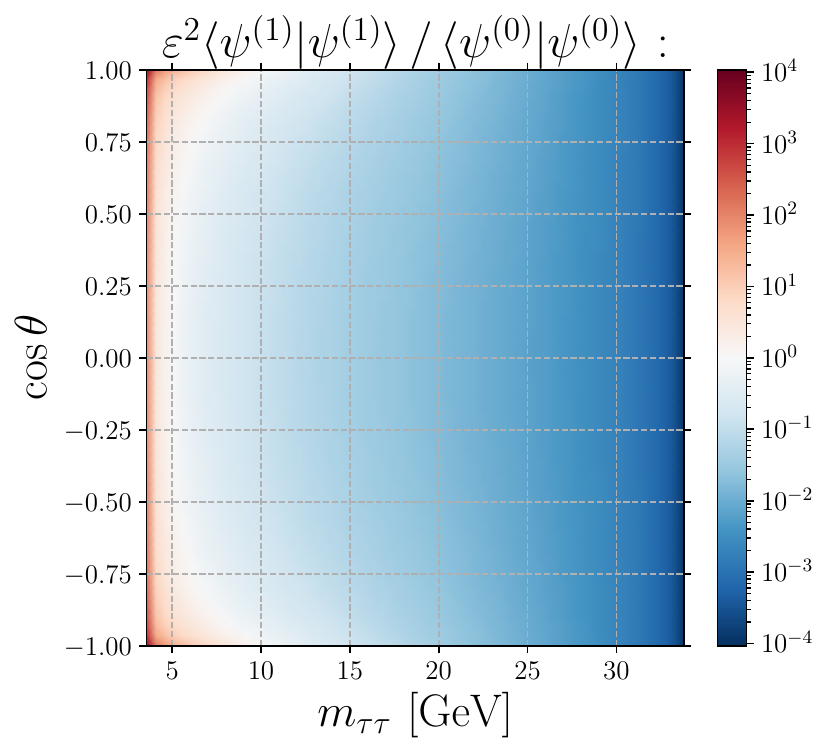}
\caption{\small 
$\braket{ \psi^{(0)} | \psi^{(0)} }$ (left), 
$\varepsilon^2\braket{  \psi^{(1)} | \psi^{(1)} }$ (centre)
and their ratio 
$\varepsilon^2\braket{  \psi^{(1)} | \psi^{(1)} } / \braket{ \psi^{(0)} | \psi^{(0)} }$ (right)
on the $(m_{\tau \tau},\, \cos \theta)$ plane, 
computed for $h \to \tau^- \tau^+ Z$ in the Standard Model.
 }
\label{fig:norms}
\end{figure}
Figure~\ref{fig:norms} shows $\braket{ \psi^{(0)} | \psi^{(0)} }$, $\varepsilon^2\braket{ \psi^{(1)} | \psi^{(1)} }$ and their ratio
$
\tfrac{\varepsilon^2 \braket{ \psi^{(1)} | \psi^{(1)} } }{ \braket{ \psi^{(0)} | \psi^{(0)} } }$
on the $(m_{\tau \tau},\, \cos \theta)$ plane, computed for $h \to \tau^- \tau^+ Z$ in the Standard Model.
Although $\varepsilon$ itself is small and nearly constant, the weight of the correction is not uniformly suppressed: it grows towards the low end of the spectrum and, in accordance with Eq.~\eqref{eq:threshold-norms}, becomes comparable to $\braket{\psi^{(0)}|\psi^{(0)}}$ near the di-tau threshold $m_{\tau\tau} \to 2 m_\tau$.
Since $c_V^2 \ll c_A^2$ in the Standard Model, the threshold value $4( c_A^2 + c_V^2 \cos^2\theta )$ is nearly independent of $\cos\theta$, as the central panel displays.
The ratio (right panel) grows further in the collinear limit, where $\braket{\psi^{(0)}|\psi^{(0)}}$ vanishes as $\sin^2\theta$, and is therefore largest in the threshold--collinear regions.

\begin{figure}[t!]
\centering
\includegraphics[width=0.32\textwidth]{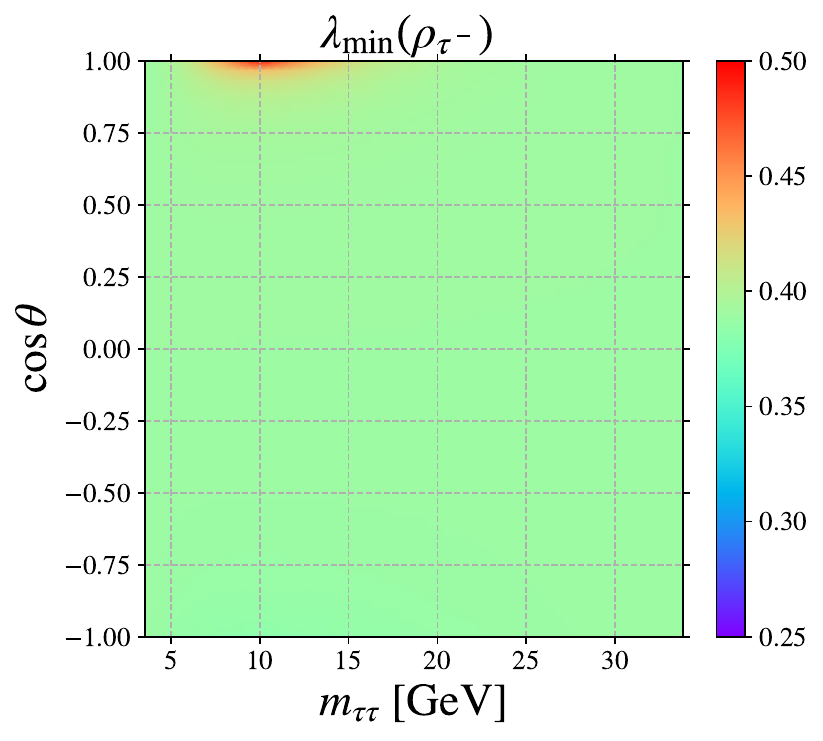}
\includegraphics[width=0.32\textwidth]{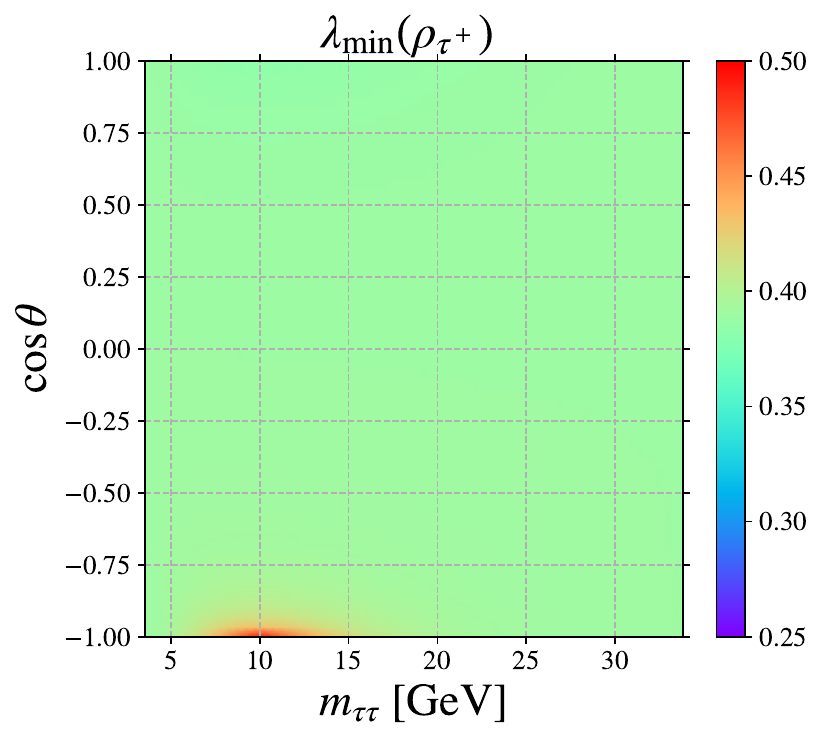}
\includegraphics[width=0.32\textwidth]{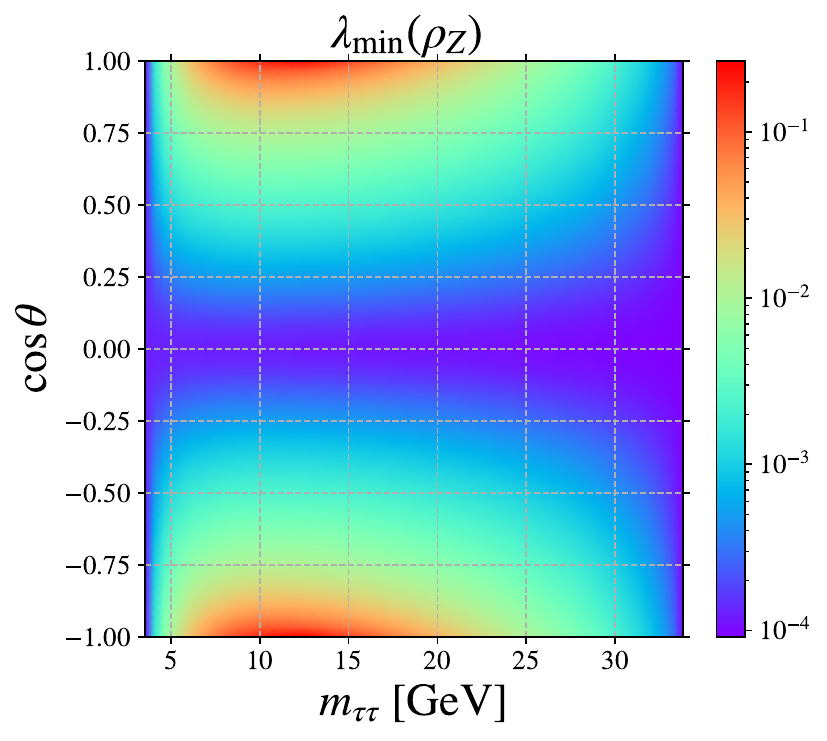}
\caption{\small The smallest eigenvalues of the single-particle reduced states,
$\lambda_{\min}(\rho_{\tau^-})$ (left),
$\lambda_{\min}(\rho_{\tau^+})$ (centre) and
$\lambda_{\min}(\rho_{Z})$ (right),
on the $(m_{\tau \tau},\, \cos \theta)$ plane,
computed for $h \to \tau^- \tau^+ Z$ in the Standard Model.
Note the logarithmic colour scale of the right panel.
 }
\label{fig:lmin}
\end{figure}

The preceding discussion suggests that the effective dimensionality of the state is often smaller than that of the full $2 \otimes 2 \otimes 3$ Hilbert space: $\ket{\psi^{(0)}}$ and $\ket{\psi^{(1)}}$ are each effectively $2 \otimes 2 \otimes 2$, and at the threshold the full state \eqref{eq:threshold-state} even reduces to a two-qubit ($2 \otimes 2$) state, the qutrit being frozen in $\ket{0}$.
It is then natural to ask whether there exists a kinematical region where the state is \emph{genuinely} $2 \otimes 2 \otimes 3$.
For a pure state, the number of local dimensions explored by each subsystem equals the rank of its reduced density matrix, so a basis-independent diagnostic is provided by the smallest eigenvalues $\lambda_{\min}(\rho_X)$, $X = A, B, V$: the state is genuinely $2 \otimes 2 \otimes 3$ wherever all three are non-zero.

Figure~\ref{fig:lmin} shows these quantities for $h \to \tau^- \tau^+ Z$ in the Standard Model, evaluated at tree level to all orders in $m_f$.
For the fermions, $\lambda_{\min}(\rho_{\tau^\mp})$ stays close to its zeroth-order value, $\min( c_L^2, c_R^2 )/\langle c^2 \rangle \simeq 0.4$, across the entire plane, rising towards the maximal value $1/2$ in the collinear hot spots; both qubits are thus fully activated everywhere.
The qutrit behaves differently: at zeroth order $\rho_V$ is supported on the two states $\ket{\Psi_\pm}$, so $\lambda_{\min}(\rho_V)$ vanishes at $\mathcal{O}(\varepsilon^0)$ and remains small over most of the phase space --- of $O( \varepsilon^2 )$, between $10^{-4}$ and a few $\times 10^{-3}$, as the logarithmic colour scale of the right panel makes visible.
It becomes sizeable only in the collinear regions $\cos\theta \to \pm 1$ at moderate invariant mass, $m_{\tau\tau} \sim 10\!-\!15$~GeV, where it reaches $\lambda_{\min}(\rho_V) \simeq 0.25$, to be compared with the maximal value $1/3$; the additional suppression along $\cos\theta \simeq 0$, also visible in the right panel, will find a natural explanation in Sec.~\ref{sec:genuine223}.

The enhancement of $\lambda_{\min}(\rho_V)$ in the collinear regions can be understood as follows.
In the collinear limit the longitudinal component of $\ket{\Psi_\pm}$ is switched off by $\sin\theta \to 0$ while its transverse components survive with weight $m_V/E_V$; the mass-induced amplitudes, on the other hand, supply the longitudinal mode with weight $\varepsilon\, a^{(0)}_\pm \propto m_f E_V$.
Indeed, in the collinear regions ($\sin\theta \ll m_V/E_V$) the state simplifies considerably.
The $\cos\theta \to \pm 1$ limit reads
\ba
\ket{ \psi } \big|_{\cos \theta = +1} &\propto&
c_L\, \ket{-+} \ket{+}
\,-\, c_R\, \ket{+-} \ket{-}
\,+\, \kappa \left(\, c_R\, \ket{--} \,-\, c_L\, \ket{++}  \,\right) \ket{0}\,,
\nn \\
\ket{ \psi } \big|_{\cos \theta = -1} &\propto&
c_L\, \ket{-+} \ket{-}
\,-\, c_R\, \ket{+-} \ket{+}
\,+\, \kappa \left(\, c_R\, \ket{++} \,-\, c_L\, \ket{--} \,\right) \ket{0}\,,
\label{eq:collinear-state}
\ea
with
\be
\kappa \,\equiv\, \frac{ m_f \left( m_H^2 - m_V^2 \right) }{ \sqrt{2}\, m_{AB}^2\, m_V }\,.
\label{eq:kappa}
\ee
Both states are genuinely qubit--qubit--qutrit: all three qutrit levels are occupied, each tagged by a distinct fermion configuration, so that $\rho_V$ is diagonal with spectrum $\big( c_L^2,\, c_R^2,\, \kappa^2 \langle c^2 \rangle \big) / \big[ (1 + \kappa^2) \langle c^2 \rangle \big]$.
The balance is controlled by $\kappa$, which decreases with $m_{AB}$ and passes through $\kappa = 1$ at $m_{AB} = \big[ m_f ( m_H^2 - m_V^2 )/(\sqrt{2}\, m_V) \big]^{1/2}$.
\begin{figure}[t!]
\centering
\includegraphics[width=0.4\textwidth]{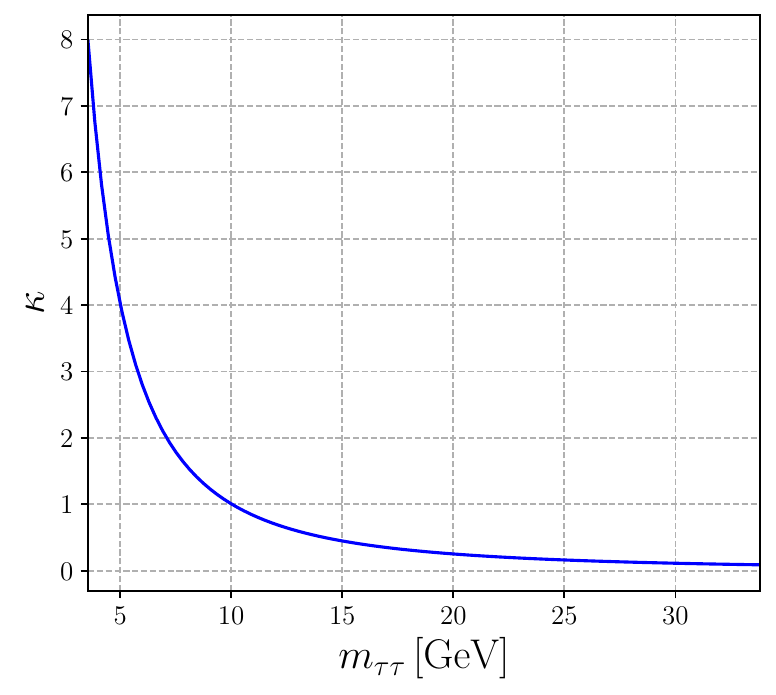}
\caption{\small The parameter $\kappa$ of Eq.~\eqref{eq:kappa} as a function of $m_{\tau\tau}$ for $h \to \tau^- \tau^+ Z$ in the Standard Model.
It decreases monotonically from the di-tau threshold towards the upper endpoint, crossing $\kappa = 1$ --- where the three qutrit populations of the collinear states \eqref{eq:collinear-state} are balanced --- at $m_{\tau\tau} \simeq 10$~GeV.
 }
\label{fig:kappa}
\end{figure}
For $h \to \tau^- \tau^+ Z$ in the Standard Model, the relation between $\kappa$ and $m_{\tau \tau}$ is displayed in Fig.~\ref{fig:kappa}; $\kappa = 1$ is crossed at $m_{\tau\tau} \simeq 10$~GeV.
Around this point the three qutrit populations are comparable, and at $\kappa = 1$ exactly $\lambda_{\min}(\rho_V) = \min( c_L^2, c_R^2 )/( 2 \langle c^2 \rangle ) \simeq 0.21$.
The maximum is attained slightly earlier, at
\be
\kappa_\ast \,=\, \sqrt{ \frac{ 1 - | \hat c | }{ 2 } } \,\simeq\, 0.65
\qquad ( m_{\tau\tau} \simeq 12~{\rm GeV} )\,,
\ee
where the mass-induced population $\kappa^2 \langle c^2 \rangle$ balances the smaller of $c_L^2$ and $c_R^2$, giving
\be
\lambda_{\min}^{\max}( \rho_V ) \,=\, \frac{ 1 - | \hat c | }{ 3 - | \hat c | } \,\simeq\, 0.30\,,
\label{eq:lminVmax}
\ee
which approaches the qutrit maximum $1/3$ in the limit $\hat c \to 0$.
This is the origin of the hot spots in the right panel of Fig.~\ref{fig:lmin}, whose values approach Eq.~\eqref{eq:lminVmax} from below as $\cos\theta \to \pm 1$; it is in these regions that the genuine $2 \otimes 2 \otimes 3$ character of the state is most pronounced.
At $\kappa = 1$ one of the two fermion reduced states --- $\rho_A$ at $\cos\theta \to +1$ and $\rho_B$ at $\cos\theta \to -1$ --- becomes maximally mixed, which explains the rise of $\lambda_{\min}( \rho_{\tau^-} )$ and $\lambda_{\min}( \rho_{\tau^+} )$ towards $1/2$ in the corresponding hot spots of Fig.~\ref{fig:lmin}.
These features will be quantified further in Sec.~\ref{sec:ent}.

Before proceeding, we comment on the robustness of this tree-level picture against radiative corrections.
For $h \to \tau^- \tau^+ Z$ the expansion parameter is almost constant across the phase space, $\varepsilon \simeq 0.03$, while one-loop corrections enter at the relative order $\alpha/\pi \sim g^2/(16\pi^2) \simeq 2\text{--}3 \times 10^{-3}$, which lies between $\varepsilon$ and $\varepsilon^2$.
Nevertheless, the hierarchy of the expansion is preserved at loop level.
The amplitudes of the same-helicity configurations, which support $\ket{\psi^{(1)}}$, flip the fermion chirality and are therefore proportional to $m_f$ at any loop order; radiative corrections cannot generate them at $O( \alpha/\pi )$, but only renormalise them at $O( \varepsilon\, \alpha/\pi )$.
In particular, the enhancements of the $\varepsilon$ terms at the threshold and in the collinear regions are tree-level chirality effects that loops cannot mimic, so radiative corrections remain a small multiplicative correction to each order of the expansion everywhere in the phase space.
The largest higher-order effect is in fact real photon emission, which makes the observed spin state slightly mixed; the soft part of the radiation is spin-diagonal and decoheres very little, and we refer to Refs.~\cite{Aoude:2025ovu, Gu:2025ijz, Aoude:2026eeg, Cheng:2026zfb} for quantitative studies of such decoherence effects.

\subsection{CP properties}
\label{sec:cp}

The effective interactions in Eq.~\eqref{eq:Leff} are CP invariant.
On the final state, CP combines charge conjugation, $f \leftrightarrow \bar f$ (i.e.,\ exchange of qubits), with parity, which reverses all momenta and helicities.
Writing 
\be
\ket{\psi} ~=\, \sum_{h_A, h_B, h_V} \psi_{h_A, h_B, h_V}(\theta) \, \ket{  h_A, h_B, h_V },
\ee
the CP transformation maps the coefficients to\footnote{In general this relation involves helicity-dependent phase factors, generated by the parity transformation and by the subsequent $\pi$ rotation about the $z$--axis needed to restore the convention $(p_V)_x \geq 0$. 
With our choice of the helicity-state phases, these factors cancel for every helicity configuration.
}
\be
\psi_{h_A, h_B, h_V}(\theta) ~\xrightarrow{\rm CP}~ \psi^{\rm CP}_{h_A, h_B, h_V}(\theta) \,=\,
\psi_{-h_B, -h_A, -h_V}(\pi - \theta )\,.
\ee
The CP invariance thus implies
\be
\psi_{h_A, h_B, h_V}(\theta) \,=\, \psi_{-h_B, -h_A, -h_V}(\pi - \theta ) \,.
\qquad (\rm CP~invariance)
\label{eq:cprel2}
\ee
Noting $\ket{-+}$ and $\ket{+-}$ of the fermion part are CP invariant
and $\ket{\Psi_\pm}|_{\cos\theta \to -\cos\theta,\, h_V \to -h_V} = \ket{\Psi_\pm}$, as can be read off Eq.~\eqref{eq:pointer1}, one can easily check the CP invariance of the leading state $\ket{ \psi^{(0)} }$ in Eq.~\eqref{eq:psi0}.
The CP invariance of the subleading state $\ket{ \psi^{(1)} }$ in Eq.\ \eqref{eq:phi1} can also be confirmed by noting $\ket{++} \xleftrightarrow{\rm CP} \ket{--}$ and
$a^{(h_V)}_- = a^{(-h_V)}_+\big|_{\cos\theta\to-\cos\theta}$.

The CP relation \eqref{eq:cprel2} has a direct consequence for the quantum observables studied in this work.
It states that the state at $\pi - \theta$ is obtained from the state at $\theta$ by exchanging the two fermions and reversing all helicity labels --- i.e.\ by the $A \leftrightarrow B$ exchange combined with a local unitary.
Since the quantum-information quantities considered below --- purities, eigenvalues of reduced states, entanglement measures and optimised Bell-inequality violations --- are invariant under local unitaries, any such observable $O$ obeys
\be
O( A, B, V; \cos\theta ) \,=\, O( B, A, V; -\cos\theta )\,,
\label{eq:cp-obs}
\ee
where the arguments indicate the roles assigned to the three parties.
Observables symmetric under $A \leftrightarrow B$, such as $C_{V|AB}$ or any quantity involving only $\rho_V$ or $\rho_{AB}$, are therefore even functions of $\cos\theta$, while observables that treat $A$ and $B$ asymmetrically come in mirror pairs, e.g.\
$\lambda_{\min}( \rho_A )( \cos\theta ) = \lambda_{\min}( \rho_B )( -\cos\theta )$ and
$C_{A|BV}( \cos\theta ) = C_{B|AV}( -\cos\theta )$.
This behaviour is already visible in Fig.~\ref{fig:lmin}: the $\tau^-$ and $\tau^+$ panels are mirror images of each other under $\cos\theta \to -\cos\theta$, while the $Z$ panel is symmetric.
The same pattern recurs throughout the observables presented in the following sections.

\section{Entanglement structure}
\label{sec:ent}

The final spin state of $H \to f \bar f V$ carries quantum correlations at several structural levels, which we now dissect.
We first quantify the bipartite entanglement between a single particle and the remaining pair (\emph{one-to-other}), then the entanglement between each pair of particles once the third is traced out (\emph{one-to-one}), and finally the correlations that are irreducibly $2 \otimes 2 \otimes 3$.

\subsection{Bipartite entanglement}

We quantify the bipartite entanglement of our $2 \otimes 2 \otimes 3$ pure state $\ket{\psi}$ with measures adapted to the two types of partition.
For the one-to-other partitions we use the $I$-concurrence of Ref.~\cite{Rungta:2001zcj}, reviewed in Appendix~\ref{app:I-conc}, which for a pure state is fixed by the single-party reduced states alone.
The one-to-one partitions involve mixed two-body states and thus a convex-roof construction: there we use Wootters' concurrence for the qubit--qubit pair and a tangle-based measure for the two qubit--qutrit pairs.
We treat the two cases in turn.

\subsubsection{One-to-other entanglement}

The normalised $I$-concurrences quantifying one-to-other entanglement of the three possible partitions of $A$-$B$-$V$ are given by 
\be
C_{A|BV} = \sqrt{2(1 - \Tr \rho_A^2)}\,,
\quad
C_{B|AV} = \sqrt{2(1 - \Tr \rho_B^2)}\,,
\quad
C_{V|AB} = \sqrt{\tfrac{3}{2}(1 - \Tr \rho_V^2)}\,,
\ee
where $\rho_X \equiv \Tr_{\overline X} | \psi \ketbra \psi |$ ($X = A, B, V$) is the reduced state of the single party $X$, and
$\Tr_{\overline X}$ denotes the partial trace over the two parties other than $X$.

At zeroth order, the pointer structure of Eq.~\eqref{eq:psi0} makes the reduced states immediate.
The two fermions are left in the diagonal states
\be
\rho_A^{(0)} = \frac{1}{ \langle c^2 \rangle }
\begin{pmatrix} c_R^2 & 0 \\ 0 & c_L^2 \end{pmatrix},
\qquad
\rho_B^{(0)} = \frac{1}{ \langle c^2 \rangle }
\begin{pmatrix} c_L^2 & 0 \\ 0 & c_R^2 \end{pmatrix}
\label{eq:rhoAB-LO}
\ee
in the $h_{A,B} = +, -$ basis; all kinematic dependence cancels because the pointer states share a common norm, $\braket{ \Psi_+ | \Psi_+ } = \braket{ \Psi_- | \Psi_- } \equiv P$.
The vector boson is left in the rank-two mixture
\be
\rho_V^{(0)} = \frac{ c_L^2\, | \Psi_+ \ketbra \Psi_+ | + c_R^2\, | \Psi_- \ketbra \Psi_- | }{ \langle c^2 \rangle\, P }\,.
\label{eq:rhoV-LO}
\ee
Evaluating the purities at this order yields the compact closed forms for the $I$-concurrences
\be
C^{(0)}_{A|BV} \,=\, C^{(0)}_{B|AV} \,=\, \frac{2\,|c_L c_R|}{\langle c^2 \rangle}
\,=\, \hat s\,,
\qquad
C^{(0)}_{V|AB} \,=\, \frac{\sqrt{3}}{2}\, \hat s\,
\sqrt{ 1 - \frac{Q^2}{P^2} }\,,
\label{eq:conc-LO}
\ee
with $Q \equiv \braket{ \Psi_+ | \Psi_- }$; inserting Eq.~\eqref{eq:psinorm}, the kinematic factor reads
\be
\sqrt{ 1 - \frac{Q^2}{P^2} } \,=\,
\frac{ 2\, m_V \sqrt{ E_V^2 - k_V^2 \cos^2\theta } }{ 2 E_V^2 - k_V^2 ( 1 + \cos^2\theta ) }\,.
\label{eq:kinfac}
\ee
At this order $C^{(0)}_{A|BV}$ and $C^{(0)}_{B|AV}$ are fixed entirely by the chiral couplings, independently of the kinematics --- a direct consequence of the GHZ-like structure of $\ket{\psi^{(0)}}$ --- and the smallness of $\hat c^2$ [Eq.~\eqref{eq:chat}] renders them nearly maximal in the Standard Model, $\simeq 0.988$.
$C^{(0)}_{V|AB}$ carries the same coupling factor but is bounded from above by $\sqrt{3}/2$: a rank-two $\rho_V$ can never reach the qutrit maximum $C_{V|AB} = 1$, which requires all three polarisations to be populated.
Its kinematic dependence resides entirely in the pointer-state overlap: the factor \eqref{eq:kinfac} equals unity where the pointer states are orthogonal ($Q = 0$, at the upper endpoint of $m_{AB}$ and for collinear kinematics), and vanishes where they become parallel ($Q \to P$, towards the threshold at fixed $\sin\theta \neq 0$, where $\rho_V^{(0)}$ degenerates to rank one and the vector-boson decouples).

At the threshold region ($m_{AB} \to 2 m_f$) and the collinear regions ($\sin\theta \ll m_V/E_V$), the $\varepsilon$ expansion breaks down and the state is approximately given by $\ket{\psi_{\rm thr}}$ of Eq.~\eqref{eq:threshold-state}
and $\ket{\psi}\big|_{\cos \theta = \pm 1}$ of Eq.~\eqref{eq:collinear-state}, respectively.

At threshold the state is a product across the $V|AB$ partition, so
\be
C_{V|AB} \,\to\, 0\,,
\qquad
C_{A|BV} \,=\, C_{B|AV} \,\to\,
\frac{ c_A^2 \left( 1 + \beta^2 \sin^2\theta \right) - c_V^2 }{ c_A^2 \left( 1 + \beta^2 \sin^2\theta \right) + c_V^2 }\,,
\qquad ( m_{AB} \to 2 m_f )
\label{eq:conc-thr}
\ee
the latter being the concurrence of the two-qubit state carried by the fermion pair in Eq.~\eqref{eq:threshold-state}.
Remarkably, at the endpoint $\beta \to 0$ and in the collinear limit $\sin\theta \to 0$ this reduces exactly to the zeroth-order value \eqref{eq:conc-LO}.

In the collinear regions, the three single party reduced states of Eq.~\eqref{eq:collinear-state} are all diagonal, and one finds (for $\cos\theta \to +1$)
\ba
C_{A|BV} &=& \frac{ 2 \sqrt{ ( c_R^2 + \kappa^2 c_L^2 )( c_L^2 + \kappa^2 c_R^2 ) } }{ ( 1 + \kappa^2 )\, \langle c^2 \rangle }\,,
\qquad
C_{B|AV} \,=\, \frac{ 2\, |c_L c_R| }{ \langle c^2 \rangle }\,,
\nn \\[4pt]
C_{V|AB} &=& \sqrt{ \frac{3}{2} \left[ 1 - \frac{ c_L^4 + c_R^4 + \kappa^4 \langle c^2 \rangle^2 }{ ( 1 + \kappa^2 )^2 \langle c^2 \rangle^2 } \right] }\,,
\label{eq:conc-collinear}
\ea
with $\kappa = m_f ( m_H^2 - m_V^2 )/( \sqrt{2}\, m_{AB}^2\, m_V )$.
For $\cos\theta \to -1$, the roles of $A$ and $B$ are interchanged, in accordance with the CP relation \eqref{eq:cp-obs}.
In terms of the coupling parameters $\hat c$ and $\hat s$ of Eq.~\eqref{eq:chat}, these expressions collapse to the compact forms
\be
C_{A|BV} \,=\, \sqrt{ \hat s^2 + \hat c^2 \left( \frac{ 2 \kappa }{ 1 + \kappa^2 } \right)^{\!2} }\,,
\qquad
C_{B|AV} \,=\, \hat s\,,
\qquad
C_{V|AB} \,=\, \frac{ \sqrt{3} }{ 2 }\, \frac{ \sqrt{ \hat s^2 + 4 \kappa^2 } }{ 1 + \kappa^2 }\,.
\label{eq:conc-collinear-simple}
\ee
The concurrence $C_{A|BV}$ thus interpolates between the zeroth-order value $\hat s$ (at $\kappa \to 0, \infty$) and exact maximality at $\kappa = 1$, where the two terms complete the Pythagorean sum $\hat s^2 + \hat c^2 = 1$, while $C_{B|AV}$ retains its zeroth-order value for any $\kappa$.
Most strikingly, $C_{V|AB}$ now exceeds the rank-two bound $\sqrt{3}/2$: it is maximised at $\kappa^2 = ( 1 + \hat c^2 )/2$, where the spectrum of $\rho_V$ is closest to uniform, reaching
\be
C_{V|AB}^{\max} \,=\, \sqrt{ \frac{ 3 }{ 3 + \hat c^2 } } \,.
\label{eq:CVmax}
\ee
For the $h \to \tau^- \tau^+ Z$ process in the Standard Model,
$C_{V|AB}^{\max} \simeq\, 0.996$.
The qutrit thus becomes almost maximally entangled with the fermion pair precisely in the genuinely $2 \otimes 2 \otimes 3$ regions identified in Fig.~\ref{fig:lmin}.

\begin{figure}[t!]
\centering
\includegraphics[width=0.32\textwidth]{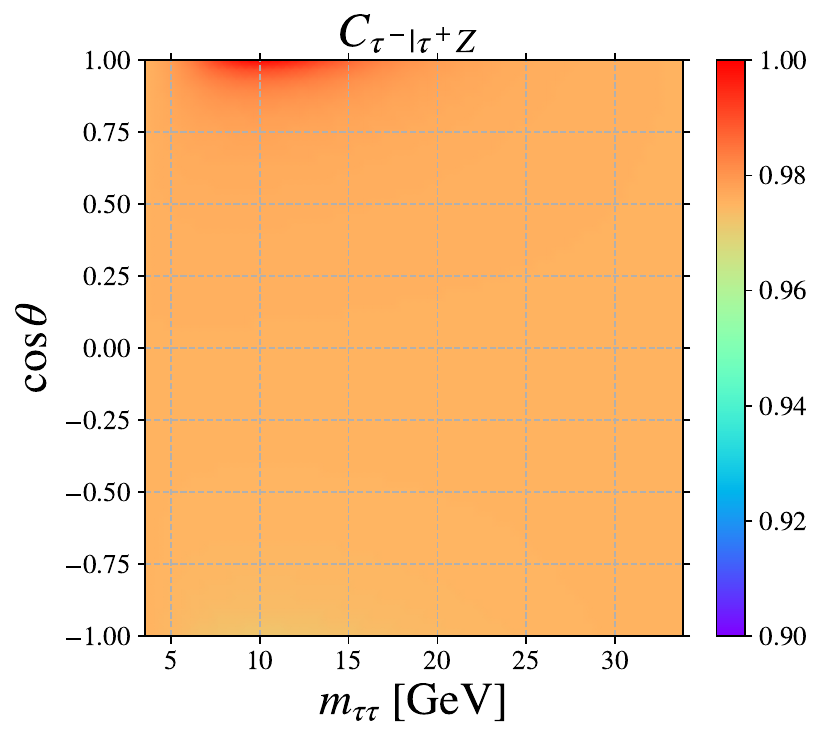}
\includegraphics[width=0.32\textwidth]{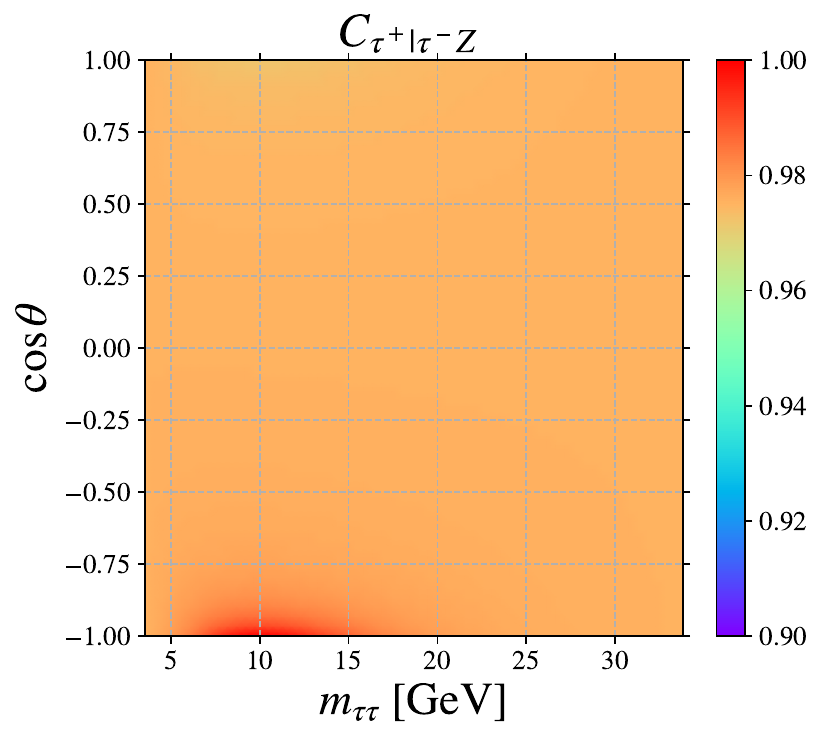}
\includegraphics[width=0.32\textwidth]{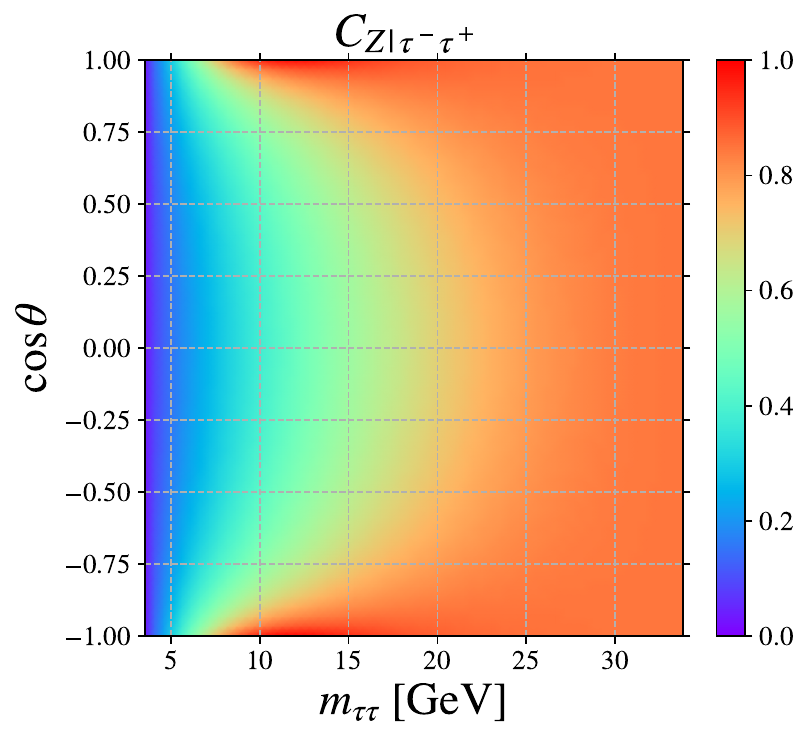}
\caption{\small One-to-other $I$-concurrences,
$C_{\tau^-|\tau^+ Z}$ (left),
$C_{\tau^+|\tau^- Z}$ (centre) and
$C_{Z|\tau^- \tau^+}$ (right),
computed for $h \to \tau^- \tau^+ Z$ in the Standard Model.
 }
\label{fig:one-to-other}
\end{figure}

In Fig.~\ref{fig:one-to-other} we show the three one-to-other concurrences for the Standard Model $h \to \tau^- \tau^+ Z$ process, evaluated at tree level to all orders in $m_f$.
The $C_{\tau^-|\tau^+ Z}$ and $C_{\tau^+|\tau^- Z}$ are mostly flat: with the Standard Model couplings \eqref{eq:cLcR_SM} the zeroth-order value is $\hat s \simeq 0.988$, and, as anticipated by Eqs.~\eqref{eq:conc-thr} and \eqref{eq:conc-collinear}, rising to $C_{\tau^-|\tau^+ Z} = 1$ in the $\cos\theta \to +1$ collinear region around $\kappa = 1$ ($m_{\tau \tau} \simeq 10$~GeV).
The $C_{Z\,|\,\tau^-\tau^+}$ map instead shows the pronounced kinematic structure of Eqs.~\eqref{eq:conc-LO} and \eqref{eq:kinfac}: a plateau at $C_{Z\,|\,\tau^-\tau^+} \simeq \tfrac{\sqrt{3}}{2}\,\hat s \simeq 0.85$ towards the upper endpoint of $m_{\tau\tau}$ (where $Q \to 0$), and a suppression towards the threshold at central angles (where $|Q| \to P$ and the $Z$ decouples), vanishing at the endpoint in accordance with Eq.~\eqref{eq:conc-thr}.
The bright spots with $C_{Z\,|\,\tau^-\tau^+} \simeq 0.99$ in the collinear regions, exceeding the rank-two bound $\sqrt{3}/2$, are quantitatively reproduced by Eq.~\eqref{eq:conc-collinear}.
Finally, all three maps obey the CP relation \eqref{eq:cp-obs}: the $C_{\tau^-|\tau^+ Z}$ and $C_{\tau^+|\tau^- Z}$ panels are mirror images of each other under $\cos\theta \to -\cos\theta$, and the $C_{Z\,|\,\tau^-\tau^+}$ panel is symmetric.

\subsubsection{One-to-one entanglement}
\label{sec:one-to-one}

The one-to-one entanglement is stored in the two-body reduced states obtained by tracing out the third party,
\be
\rho_{AB} = \Tr_V | \psi \ketbra \psi |\,, \qquad
\rho_{AV} = \Tr_B | \psi \ketbra \psi |\,, \qquad
\rho_{BV} = \Tr_A | \psi \ketbra \psi |\,,
\label{eq:bipartite_rs}
\ee
which describe a qubit--qubit pair ($AB$) and two qubit--qutrit pairs ($AV$, $BV$).
For a mixed bipartite state $\rho_{XY}$, we define the (normalised) $I$-concurrence by the convex roof as
\be
C_{XY} \,=\, C( \rho_{XY} ) \,\equiv\, \min_{ \{ p_i, \ket{\psi_i} \} }
\sum_i\,
p_i \, C( \ket{\psi_i} )\,,
\label{eq:C_XY}
\ee
where the minimisation is taken over all possible pure state decompositions of $\rho$; $\rho = \sum_i p_i | \psi_i \ketbra \psi_i |$ with $p_i \geq 0$, $\sum_i p_i = 1$.
For the reduced states \eqref{eq:bipartite_rs}, $\min( d_A, d_B) = \min( d_A, d_V) = \min( d_B, d_V) = 2$, and the $I$-concurrence is already normalised (see Appendix \ref{app:I-conc}).  

For the $2 \otimes 2$ state $\rho_{AB}$, the convex roof can be evaluated exactly with Wootters' formula~\cite{Wootters:1997id},
\be
C_{AB} \,=\, \max\big( 0,\; \lambda_1 - \lambda_2 - \lambda_3 - \lambda_4 \big)\,,
\label{eq:wootters}
\ee
where $\lambda_1 \geq \lambda_2 \geq \lambda_3 \geq \lambda_4$ are the square roots of the eigenvalues of $\rho_{AB}\, \tilde\rho_{AB}$ with  $\tilde\rho = (\sigma_y \otimes \sigma_y)\, \rho^\ast (\sigma_y \otimes \sigma_y)$.

For the $2 \otimes 3$ states $\rho_{AV}$ and $\rho_{BV}$, no closed-form solution of the convex roof is known for general mixed states.
One can, however, resort to the $I$-tangle, defined as the convex roof of the \emph{squared} $I$-concurrence,
\be
\tau(\rho_{XY}) \,=\, \min_{ \{ p_i, \ket{\psi_i} \} } \sum_i p_i\, C_I( \ket{\psi_i} )^2\,,
\qquad \rho_{XY} \,=\, \sum_i p_i |\psi_i \ketbra \psi_i |
\label{eq:Itangle}
\ee
Osborne~\cite{Osborne:2002vcf} showed that for states of rank at most two the $I$-tangle admits the closed form
\be
\tau( \rho_{XY} ) \,=\, \Tr( \rho\, \tilde\rho ) \,+\, 2\lambda_{\min}\big( 1 - \Tr \rho^2 \big)\,,
\label{eq:osborne}
\ee
where $\tilde\rho = (S_X \otimes S_Y)(\rho)$ is the universal state inverter of Eq.~\eqref{eq:spin-flip}, and $\lambda_{\min}$ is the smallest eigenvalue of a real symmetric $3 \times 3$ matrix $M$ built as follows.
Although $\rho$ acts on the full $d_X d_Y$--dimensional space, at most two of its eigenvalues are non-zero, so its support, the span of the corresponding eigenvectors, is a two-dimensional subspace.
Let $\{ \ket{v_1}, \ket{v_2} \}$ be any orthonormal basis of that subspace and set $\gamma_{ij} \equiv | v_i \ketbra v_j |$. One then forms the tensor
\be
T_{ijkl} \,=\, \Tr\big( \gamma_{ij}\, \widetilde{\gamma_{kl}} \big)\,,
\ee
where the tilde is understood in the general sense of Eq.~\eqref{eq:inverter-general}: the $\gamma_{ij}$ with $i \neq j$ are traceless and non-hermitian, so the state form \eqref{eq:spin-flip} does not apply to them.
In terms of $T_{ijkl}$, the independent entries of $M$ read
\ba
M_{11} &=& \tfrac14 T_{1221} + \tfrac12 T_{1122} + \tfrac14 T_{2112}\,,
\qquad
M_{12} \,=\, \tfrac{i}{4}\big( T_{1221} - T_{2112} \big)\,,
\nn \\[3pt]
M_{22} &=& -\tfrac14 T_{1221} + \tfrac12 T_{1122} - \tfrac14 T_{2112}\,,
\qquad
M_{13} \,=\, \tfrac14\big( T_{1121} - T_{2122} + T_{1112} - T_{1222} \big)\,,
\nn \\[3pt]
M_{33} &=& \tfrac14 T_{1111} - \tfrac12 T_{1122} + \tfrac14 T_{2222}\,,
\qquad
M_{23} \,=\, \tfrac{i}{4}\big( T_{1121} - T_{1112} + T_{2122} - T_{1222} \big)\,,
\ea
with $M_{ji} = M_{ij}$.
The spectrum of $M$ does not depend on which orthonormal basis of the support is chosen: a different choice merely rotates the frame in which $M$ is expressed, $M \to R\, M\, R^{T}$ with $R \in SO(3)$, leaving $\lambda_{\min}$, and hence $\tau$, unchanged.
This is not an academic point: $\rho_{AV}$ and $\rho_{BV}$ turn out to have degenerate spectra in the collinear limit [cf.\ Eq.~\eqref{eq:upm}], where the eigenbasis of $\rho$ is not unique.

Our states qualify: since $\ket{\psi}$ is pure, the Schmidt decomposition across the $AV|B$ and $BV|A$ partitions guarantees that $\rho_{AV}$ and $\rho_{BV}$ share their non-zero spectra with the complementary single-party states $\rho_B$ and $\rho_A$, so that
\be
\mathrm{rank}\, \rho_{AV} = \mathrm{rank}\, \rho_B \le 2\,,
\qquad
\mathrm{rank}\, \rho_{BV} = \mathrm{rank}\, \rho_A \le 2
\ee
hold exactly, and Osborne's formula \eqref{eq:osborne} applies.
We therefore quantify the one-to-one entanglement of the two qubit--qutrit pairs by
\be
\tilde C_{XY} \,\equiv\, \sqrt{ \tau( \rho_{XY} ) }\,,
\qquad
XY \in \{ AV, BV \}\,.
\label{eq:CXY-def}
\ee
The square root of the $I$-tangle gives the upper bound on the $I$-concurrence in Eq.~\eqref{eq:C_XY}; $\tilde C_{XY} \geq C_{XY}$.
Nonetheless, $\tilde C_{XY}$ itself is a faithful entanglement measure in its own right: non-negative, vanishing iff the state is separable, and non-increasing under local operations and classical communication (LOCC).
We adopt $\tilde C_{XY}$ as our one-to-one entanglement quantifier for $AV$ and $BV$.

At zeroth order, tracing the vector boson out of $\ket{\psi^{(0)}}$ leaves the two fermions in a rank-two state supported on the opposite-helicity subspace, with a single off-diagonal element generated by the pointer-state overlap,
\be
\rho_{AB}^{(0)} \,=\, \frac{1}{ \langle c^2 \rangle\, P }
\begin{pmatrix} c_L^2\, P & c_L c_R\, Q \\ c_L c_R\, Q & c_R^2\, P \end{pmatrix}
\ee
in the $\{ \ket{-+}, \ket{+-} \}$ basis.
Wootters' formula \eqref{eq:wootters} then gives
\be
C^{(0)}_{AB} \,=\, \hat s\; \frac{Q}{P}
\,=\, \hat s\;
\frac{ k_V^2 \sin^2\theta }{ 2 E_V^2 - k_V^2 ( 1 + \cos^2\theta ) }\,.
\label{eq:CAB-LO}
\ee
Comparing with Eq.~\eqref{eq:conc-LO} reveals a neat complementarity,
\be
\left(  C^{(0)}_{AB} \right)^{\!2}
\,+\,
\left( \tfrac{2}{\sqrt{3}} C^{(0)}_{V|AB} \right)^{\!2}
\,=\, \frac{1}{2} \left[ \left( C^{(0)}_{A|BV} \right)^{\!2}\,+\,
\left( C^{(0)}_{B|AV} \right)^{\!2} \right]\,.
\label{eq:complementarity}
\ee
The physical content of this relation is a perfect trade-off.
Its right-hand side is fixed by the couplings alone, $\tfrac12 [ ( C^{(0)}_{A|BV} )^2 + ( C^{(0)}_{B|AV} )^2 ] = \hat s^2$, while the two terms on the left share this budget in proportions set by the overlap ratio $Q/P$: as the kinematics varies, entanglement is shifted between the fermion pair and the $V|AB$ partition, one growing precisely at the expense of the other.

The origin of Eq.~\eqref{eq:complementarity} is the monogamy of entanglement.
At zeroth order the qutrit is effectively a qubit, since $\rho_V$ has rank two, and the factor $2/\sqrt{3}$ converts the qutrit-normalised concurrence into the corresponding qubit tangle, $\big( \tfrac{2}{\sqrt{3}}\, C_{V|AB} \big)^2 = 2 \big( 1 - \Tr \rho_V^2 \big)$.
For a pure three-qubit state, the Coffman--Kundu--Wootters relations \cite{Coffman:1999jd},
$C_{X|YZ}^2 = C_{XY}^2 + C_{XZ}^2 + \tau_{XYZ}$, with the residual three-tangle $\tau_{XYZ}$ common to all three partitions, combine into
\be
\frac{1}{2} \left[ C_{A|BV}^2 + C_{B|AV}^2 \right]
- \left( \tfrac{2}{\sqrt{3}}\, C_{V|AB} \right)^{\!2}
\,=\, C_{AB}^2 - \frac{1}{2} \left[ C_{AV}^2 + C_{BV}^2 \right].
\label{eq:ckw}
\ee
Equation~\eqref{eq:complementarity} is therefore equivalent to the statement that $C^{(0)}_{AV}$ and $C^{(0)}_{BV}$ vanish at this order, which we verify next.

Tracing out either fermion instead leaves a mixture of two \emph{product} states,
\be
\rho^{(0)}_{AV} \,\propto\,
c_L^2\, | {-} \ketbra {-} | \otimes | \Psi_+ \ketbra \Psi_+ |
\,+\,
c_R^2\, | {+} \ketbra {+} | \otimes | \Psi_- \ketbra \Psi_- |\,,
\ee
and likewise for $\rho^{(0)}_{BV}$, because the traced fermion perfectly tags the two branches.
Such a mixture is separable, so
\be
\tilde C^{(0)}_{AV} \,=\, \tilde C^{(0)}_{BV} \,=\, 0\,.
\ee

In the threshold region, the state \eqref{eq:threshold-state} is a product across the $V | AB$ partition: the boson decouples while the fermion pair is left in a pure two-qubit state, so
\be
\tilde C_{AV}\,,\; \tilde C_{BV} \;\to\; 0\,,
\qquad
C_{AB} \;\to\;
\frac{ c_A^2 \left( 1 + \beta^2 \sin^2\theta \right) - c_V^2 }{ c_A^2 \left( 1 + \beta^2 \sin^2\theta \right) + c_V^2 }\,,
\qquad ( m_{AB} \to 2 m_f )
\label{eq:CAB-thr}
\ee
the latter coinciding with the one-to-other value \eqref{eq:conc-thr}, as it must for a pure two-qubit state.

In the collinear regions the situation is reversed.
Tracing the boson out of Eq.~\eqref{eq:collinear-state} projects onto its three orthogonal polarisations, leaving $\rho_{AB}$ as a mixture of $\ket{-+}$, $\ket{+-}$ and the Bell-like combination $( c_R \ket{--} - c_L \ket{++} )/\sqrt{\langle c^2 \rangle}$ with weights $( c_L^2,\, c_R^2,\, \kappa^2 \langle c^2 \rangle )/[ ( 1 + \kappa^2 ) \langle c^2 \rangle ]$; Wootters' formula gives
\be
C_{AB} \,=\, \hat s\;
\max\!\left( 0,\; \frac{ \kappa^2 - 1 }{ \kappa^2 + 1 } \right).
\qquad ( \cos\theta \to \pm 1 )
\label{eq:CAB-collinear}
\ee
The concurrence $C_{AB}$ thus vanishes identically for all $\kappa \leq 1$ and revives beyond, growing towards $\hat s$ as $\kappa \to \infty$, in agreement with the $\sin\theta \to 0$ limit of Eq.~\eqref{eq:CAB-thr}.

For the $AV$ and $BV$ pairs, Osborne's formula \eqref{eq:osborne} again yields closed-form results.
Since $\hat c$ enters only quadratically, we quote them at leading order, neglecting $O( \hat c^2 )$ corrections:
\be
\tilde C_{AV} \,=\, \tilde C_{BV} \,=\,
\frac{ \min\!\left( 2 \kappa,\; \sqrt{ 1 + 2 \kappa^2 } \right) }{ 1 + \kappa^2 }
\,=\,
\begin{cases}
\;\dfrac{ 2 \kappa }{ 1 + \kappa^2 }\,, & \kappa \leq \dfrac{1}{\sqrt{2}}\,,
\\[10pt]
\;\dfrac{ \sqrt{ 1 + 2 \kappa^2 } }{ 1 + \kappa^2 }\,,
& \kappa \geq \dfrac{1}{\sqrt{2}}\,,
\end{cases}
\qquad (\cos\theta \to \pm 1, \, \hat c^2 \ll 1)
\label{eq:Ct-collinear}
\ee
At this order the two collinear regions coincide; the $O( \hat c^2 )$ corrections split $\tilde C_{AV}$ and $\tilde C_{BV}$, the two being interchanged at $\cos\theta \to -\cos\theta$ in accordance with the CP relation \eqref{eq:cp-obs}.
One can see that the two branches are attained by the following pure state decompositions of $\rho_{AV}$.
Tracing $B$ out of Eq.~\eqref{eq:collinear-state} leaves an equal mixture of the two states,
\be
\rho_{AV} = \frac{1}{2} \left(\, | u_+ \ketbra u_+ | + | u_- \ketbra u_- |  \, \right),
\qquad
\ket{u_\pm} \,=\, \frac{ \ket{\mp}_A \ket{\pm}_V \,-\, \kappa\, \ket{\pm}_A \ket{0}_V }{ \sqrt{ 1 + \kappa^2 } }\,,
\qquad \braket{ u_+ | u_- } = 0\,,
\label{eq:upm}
\ee
each of which has Schmidt coefficients $\propto ( 1, \kappa )$ and hence $C_I( \ket{u_\pm} ) = 2\kappa/(1+\kappa^2)$; this decomposition gives the first branch.
Another possible decomposition is $\rho_{AV} = \frac{1}{2} \left(\, | v_+ \ketbra v_+ | + | v_- \ketbra v_- |  \, \right)$
with $\ket{v_\pm} \equiv ( \ket{u_+} \pm \ket{u_-} )/\sqrt{2}$, which gives the second branch.
Osborne's formula \eqref{eq:osborne} returns the smaller of the two: the decomposition \eqref{eq:upm} is the optimal one for $\kappa \leq 1/\sqrt{2}$, and $\{ \ket{v_\pm} \}$ beyond.
Both measures rise from zero at $\kappa \to 0$, matching the zeroth order, peak at $\kappa = 1/\sqrt{2}$, where the two branches cross, with the value $2\sqrt{2}/3 \simeq 0.94$, and fall off again towards the threshold side ($\kappa \to \infty$), where the boson decouples.

\begin{figure}[t!]
\centering
\includegraphics[width=0.32\textwidth]{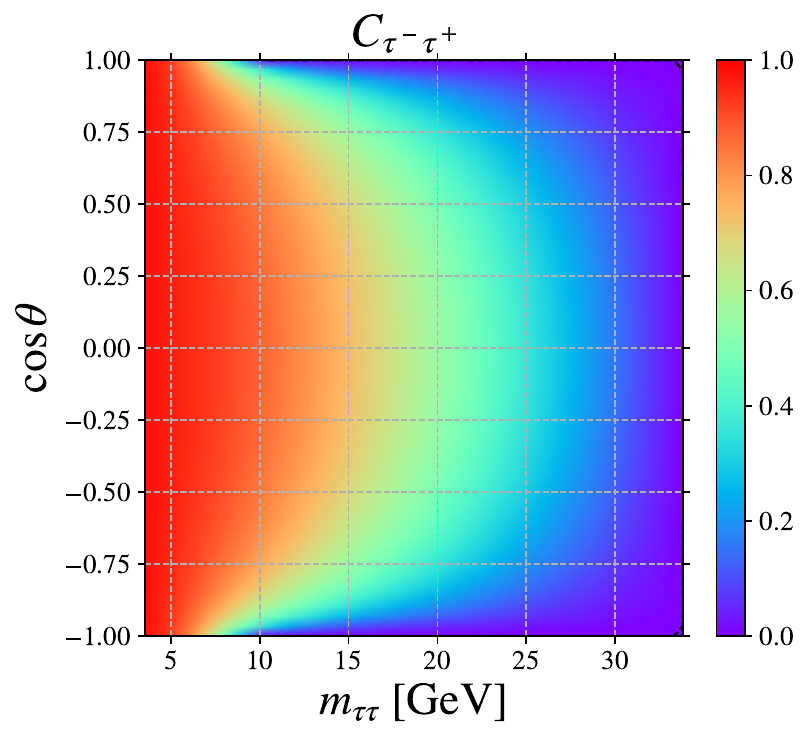}
\includegraphics[width=0.32\textwidth]{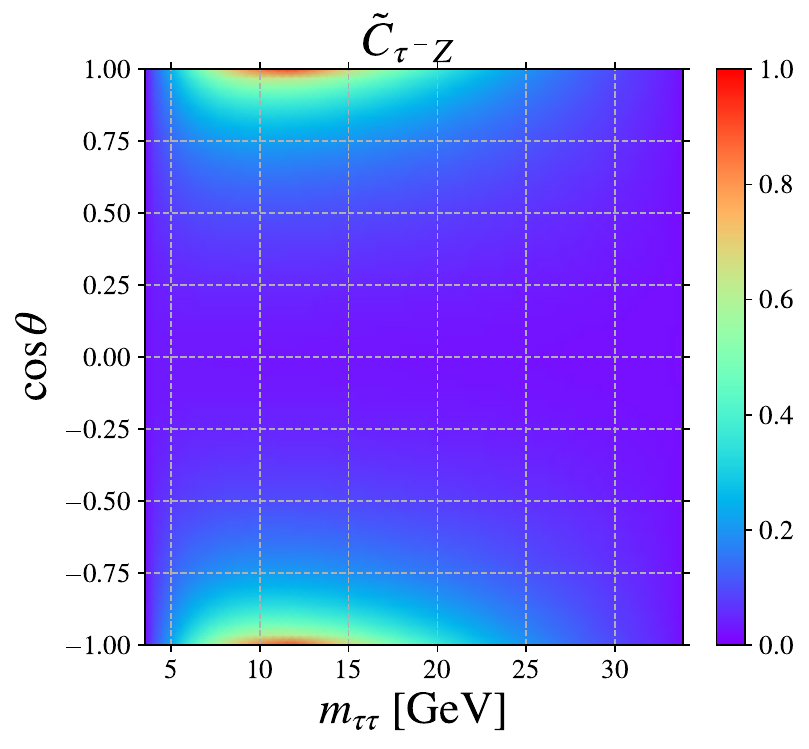}
\includegraphics[width=0.32\textwidth]{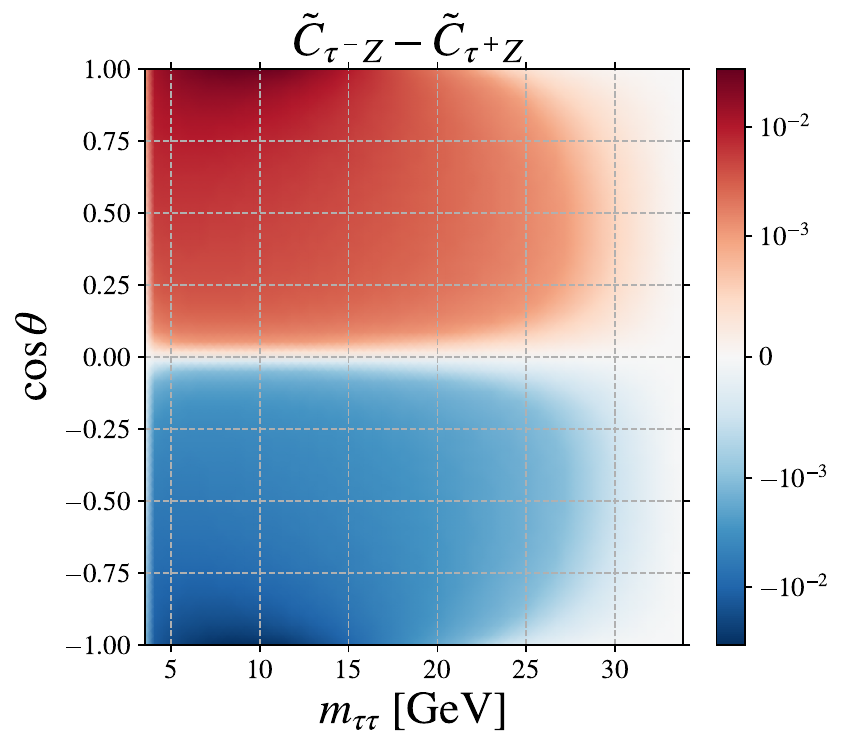}
\caption{\small One-to-one entanglement measures for $h \to \tau^- \tau^+ Z$ in the Standard Model, evaluated at tree level to all orders in $m_f$: the concurrence $C_{\tau^-\tau^+}$ (left), the measure $\tilde C_{\tau^- Z}$ (centre), and the difference $\tilde C_{\tau^- Z} - \tilde C_{\tau^+ Z}$ (right, symmetric logarithmic colour scale).}
\label{fig:one-to-one}
\end{figure}

Figure~\ref{fig:one-to-one} shows the three one-to-one measures for the Standard Model $h \to \tau^- \tau^+ Z$ process, evaluated at tree level to all orders in $m_f$.
The concurrence $C_{\tau^-\tau^+}$ (left) follows the zeroth-order form \eqref{eq:CAB-LO}: it is largest, $\simeq \hat s$, at small $m_{\tau\tau}$ (where $Q/P \to 1$), falls monotonically to zero towards the upper endpoint ($Q \to 0$), and displays the $\sin^2\theta$ suppression along the collinear edges.
An exception to the last feature arises in the threshold--collinear regions, where $C_{\tau^-\tau^+}$ revives for $\kappa > 1$, in accordance with Eq.~\eqref{eq:CAB-collinear}.
The whole map is even in $\cos\theta$.
The measure $\tilde C_{\tau^- Z}$ (centre) is small over most of the phase space, but climbs to $\simeq 0.9$ in the collinear hot spots at $\kappa \simeq 1/\sqrt{2}$ ($m_{\tau\tau} \simeq 12$~GeV), in quantitative agreement with Eq.~\eqref{eq:Ct-collinear}.
The map of $\tilde C_{\tau^+ Z}$ is its exact mirror image under $\cos\theta \to -\cos\theta$, as dictated by the CP relation \eqref{eq:cp-obs}, and the two are so close that they cannot be distinguished by eye; we therefore display their difference, $\tilde C_{\tau^- Z} - \tilde C_{\tau^+ Z}$, in the right panel.
The difference is an odd function of $\cos\theta$, again by CP, and its magnitude stays below a few $\times 10^{-2}$, confirming that the splitting of the two measures is the $O( \hat c^2 )$ effect anticipated below Eq.~\eqref{eq:Ct-collinear}.
It is largest at small $m_{\tau\tau}$ and towards the collinear edges, with $\tilde C_{\tau^- Z} > \tilde C_{\tau^+ Z}$ in the hemisphere where the $Z$ is emitted along the $\tau^-$ direction, and fades towards the upper endpoint, where the state approaches the A--B symmetric generalised GHZ form.
Comparing the three panels reveals the monogamy-like trade-off anticipated above: precisely where the $\tau^-\tau^+$ entanglement is quenched, the $\tau^\mp Z$ entanglement is largest, and vice versa.

\subsection{Genuine $2 \otimes 2 \otimes 3$ entanglement}
\label{sec:genuine223}

The tripartite $(f\bar{f}V)$ state (\ref{eq:state}) is pure, therefore,
to detect genuine $2 \otimes 2 \otimes 3$ entanglement, we can use tools introduced by Miyake in his classification \cite{Miyake:2003tee}. The Miyake classification and its relation to the 223-tangle are briefly summarised in Appendix~\ref{app:Miyake}.
Arranging the twelve amplitudes into the $3 \times 4$ matrix $A$ with rows labelled by $h_V$ and columns by the fermion configurations $( h_A h_B ) = (++, +-, -+, --)$, one forms the four $3 \times 3$ minors $m_i$ ($i = 1, \dots, 4$), obtained by deleting the $i$-th column, and the $2 \times 2 \times 3$ hyperdeterminant
\be
M \,=\, m_1 m_4 - m_2 m_3\,.
\label{eq:HDet}
\ee
Together with the local ranks $( r_A, r_B, r_V )$ of the single-party reduced states, the minors and the hyperdeterminant organise the pure $2 \otimes 2 \otimes 3$ states into the SLOCC classes of Miyake's classification, summarised in Appendix~\ref{app:Miyake}; the classes relevant below are collected in Table~\ref{tab:slocc223}.

\begin{table}[h!]
\centering
\begin{tabular}{lcl}
\hline
class & $( r_A, r_B, r_V )$ & invariants \\
\hline
C1 & $(2,2,3)$ & $M \neq 0$ \\
C2 & $(2,2,3)$ & $M = 0$, at least one $m_i \neq 0$ \\
GHZ, W & $(2,2,2)$ & all $m_i = 0$ \\
B1, B2, B3 & $(1,2,2),\, (2,1,2),\, (2,2,1)$ & all $m_i = 0$ (biseparable) \\
S & $(1,1,1)$ & all $m_i = 0$ (fully separable) \\
\hline
\end{tabular}
\caption{\small SLOCC classes of pure $2 \otimes 2 \otimes 3$ states relevant to this work, labelled by the local ranks $( r_A, r_B, r_V )$ of the single-party reduced states and by the values of the hyperdeterminant $M$ and of the $2 \times 2$ minors $m_i$.
}
\label{tab:slocc223}
\end{table}

In particular, the state is genuinely $2 \otimes 2 \otimes 3$ entangled (class C1 or C2) if and only if at least one minor is non-zero; when all minors vanish the entanglement is at most of the $2 \otimes 2 \otimes 2$ type (GHZ or W class) or biseparable.
The individual minors are basis-dependent quantities: the $|m_i|$ are insensitive to qutrit basis rotations but mix under changes of the fermion bases, and are quoted here in the helicity basis; $|M|$, by contrast, is invariant under all local unitaries.
The 223-tangle, $\tau_{223} = 3 \sqrt[3]{3}\, |M|^{1/3}$ [Eq.~\eqref{eq_app:Miyake_tangle_HDet}], is an entanglement monotone, but not a faithful detector of genuine $2 \otimes 2 \otimes 3$ entanglement: it vanishes on the C2 class, which is genuinely tripartite.
In this respect it is the exact analogue of the three-qubit 3-tangle, which vanishes on the genuinely tripartite W class; $\tau_{223} > 0$ certifies, and quantifies, the \emph{generic} (C1) type of genuine $2 \otimes 2 \otimes 3$ entanglement.

For our state, the columns of $A$ are read off from  Eqs.~\eqref{eq:psi0} and \eqref{eq:phi1} as
\be
A \,=\, \frac{1}{\mathcal{N}}\,
\Big[\, \varepsilon \Phi_+ \,\Big|\, c_R \Psi_- \,\Big|\, c_L \Psi_+ \,\Big|\, \varepsilon \Phi_- \,\Big].
\ee
Since only two columns survive at $m_f = 0$ ($\varepsilon = 0$), all four minors vanish at zeroth order: in Miyake's classification the leading state $\ket{\psi^{(0)}}$ is not genuinely $2 \otimes 2 \otimes 3$ but of the GHZ type, in accordance with the discussion of Sec.~\ref{sec:0-223}.
At $O(\varepsilon)$ the two minors built from a single mass-induced column switch on,
\be
m_1 \,=\, \frac{ \varepsilon\, c_L c_R }{ \mathcal{N}^3 } \left( \Psi_- \times \Psi_+ \right) \cdot \Phi_-\,,
\qquad
m_4 \,=\, \frac{ \varepsilon\, c_L c_R }{ \mathcal{N}^3 } \left( \Psi_- \times \Psi_+ \right) \cdot \Phi_+\,,
\label{eq:m14}
\ee
while $m_2, m_3 = O( \varepsilon^2 )$, so that $M = m_1 m_4 + O( \varepsilon^4 )$: genuine $2 \otimes 2 \otimes 3$ entanglement is a pure mass effect, switched on at $O(m_f)$.
Equation~\eqref{eq:m14} has a transparent geometric meaning: only the component of $\ket{\Phi_\pm}$ pointing \emph{out of the plane} spanned by the pointer states $\ket{\Psi_\pm}$ contributes.
This is the same component that populates the third qutrit dimension, left empty at zeroth order, and thereby renders $\lambda_{\min}( \rho_V )$ non-zero: the minors $m_1$, $m_4$ and the smallest qutrit eigenvalue shown in Fig.~\ref{fig:lmin} probe one and the same $O( \varepsilon )$ structure of the state.
Evaluating the triple products with Eqs.~\eqref{eq:pointer1} and \eqref{eq:pointer} gives the compact coupling decomposition
\be
\left( \Psi_- \times \Psi_+ \right) \cdot \Phi_\pm \,=\, \pm\, c_V\, \mathcal{V} \,+\, \cos\theta\; c_A\, \mathcal{A}\,,
\label{eq:m14-coupling}
\ee
with the kinematic functions
\ba
\mathcal{A} &=& \frac{ 8\, m_V k_V }{ E_V^2\, D_+ D_- } \left[ m_{AB}^2 k_V^2 \sin^2\theta - 2 D_+ D_- \right],
\nn \\[4pt]
\mathcal{V} &=& \frac{ 8\, m_V }{ E_V } \left[ \frac{ 2 m_V^2 }{ m_V^2 - m_{AB}^2 }
+ \frac{ m_{AB} \left( E_V ( E_V + m_{AB} )\, D \sin^2\theta + m_V^2 m_H^2 \cos^2\theta \right) }{ E_V\, D_+ D_- } \right].
\label{eq:VAfun}
\ea
Two features follow immediately.
First, under $\cos\theta \to -\cos\theta$ one finds $|m_1| \leftrightarrow |m_4|$, in accordance with the CP relation \eqref{eq:cp-obs}, while $|M|$ is even.
Second, the two terms of Eq.~\eqref{eq:m14-coupling} interfere.
If only the axial coupling were present ($c_V = 0$), both $m_1$ and $m_4$ would vanish on the equator $\cos\theta = 0$; the vector coupling displaces this common zero in opposite directions for the two minors, so that $m_1$ vanishes on the curve determined by $\cos\theta\, c_A \mathcal{A} = c_V \mathcal{V}$ and $m_4$ on its mirror image ($\cos\theta \to -\cos\theta$).
Since the kinematic ratio $\mathcal{V}/\mathcal{A}$ is of order unity, the displacement is of order $c_V/c_A$; the position of the nodal curve is therefore a direct geometric imprint of the ratio of the vector to axial-vector couplings on the entanglement pattern.
We note that, unlike the parametric hierarchies discussed in Sec.~\ref{sec:0-223}, the exact zeros along the nodal curves are not protected against radiative corrections.
Loop effects correct the minor as $m_1 \propto \varepsilon\, [ ( \Psi_- \times \Psi_+ ) \cdot \Phi_- + O( \alpha/\pi ) ]$, which displaces the zero by $\delta \cos\theta_* = O( \alpha/\pi )$.
Compared with the tree-level offset of the curves from the equator, $\cos\theta_* = O( c_V/c_A )$, this is a relative shift of $O( ( \alpha/\pi )\, c_A/c_V )$, at the level of a few per cent, without altering the qualitative picture.
Along these curves $|M|$ drops to $|m_2 m_3| = O( \varepsilon^4 )$: the state remains genuinely $2 \otimes 2 \otimes 3$ entangled ($m_4 \neq 0$, respectively $m_1 \neq 0$), but its C1-type tangle collapses, and the state passes $O( \varepsilon^2 )$-close to the degenerate C2 orbit.

At the extreme corners of the phase space, the by-now familiar pattern repeats.
Towards the threshold all minors vanish --- the qutrit freezes in $\ket{0}$, every $3 \times 3$ minor acquires two vanishing rows, and the state \eqref{eq:threshold-state} is biseparable across $V | AB$ (class B3).
In the collinear regions, Eq.~\eqref{eq:collinear-state} gives $m_2 = m_3 = 0$ exactly and
\be
|m_1| \,=\, \frac{ \kappa\, |c_L|\, c_R^2 }{ \big[ ( 1 + \kappa^2 ) \langle c^2 \rangle \big]^{3/2} }\,,
\quad
|m_4| \,=\, \frac{ \kappa\, c_L^2\, |c_R| }{ \big[ ( 1 + \kappa^2 ) \langle c^2 \rangle \big]^{3/2} }\,,
\quad
|M| \,=\, \frac{ \kappa^2\, \hat s^3 }{ 8 \left( 1 + \kappa^2 \right)^3 }\,.
\quad ( \cos\theta \to \pm 1 )
\label{eq:HDet-collinear}
\ee
The hyperdeterminant is maximised at $\kappa = 1/\sqrt{2}$ --- the same point at which $C_{V|AB}$ and $\tilde C_{AV, BV}$ peak in Sec.~\ref{sec:ent}, slightly below the population-balance point $\kappa = 1$ of Sec.~\ref{sec:0-223} --- where
\be
|M|_{\max} \,=\, \frac{ \hat s^3 }{ 54 } ,
\qquad
\tau_{223}^{\max} \,=\, \left( \tfrac{3}{2} \right)^{\!1/3} \hat s \,,
\ee
so the state is most strongly $2 \otimes 2 \otimes 3$ entangled precisely where the qutrit is most active.
For the $h \to \tau^- \tau^+ Z$ in the Standard Model, 
$\kappa = 1/\sqrt{2}$ corresponds to $m_{\tau \tau} \simeq 12$ GeV and $|M|_{\max} \simeq 0.018$ and $\tau_{223}^{\max} \simeq 1.13$.

\begin{figure}[t!]
\centering
\includegraphics[width=0.32\textwidth]{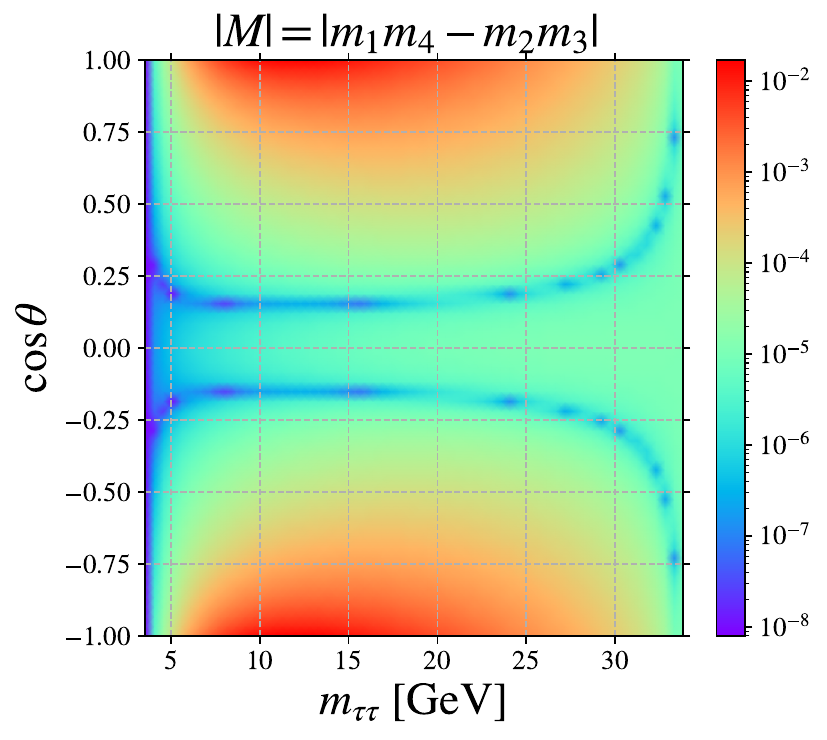}
\includegraphics[width=0.32\textwidth]{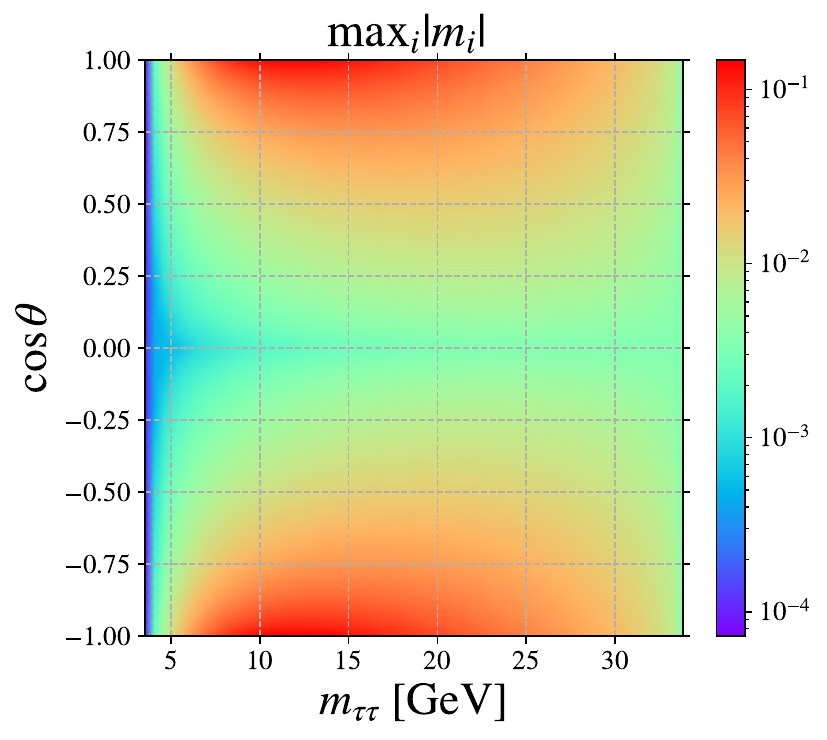}
\\[4pt]
\includegraphics[width=0.32\textwidth]{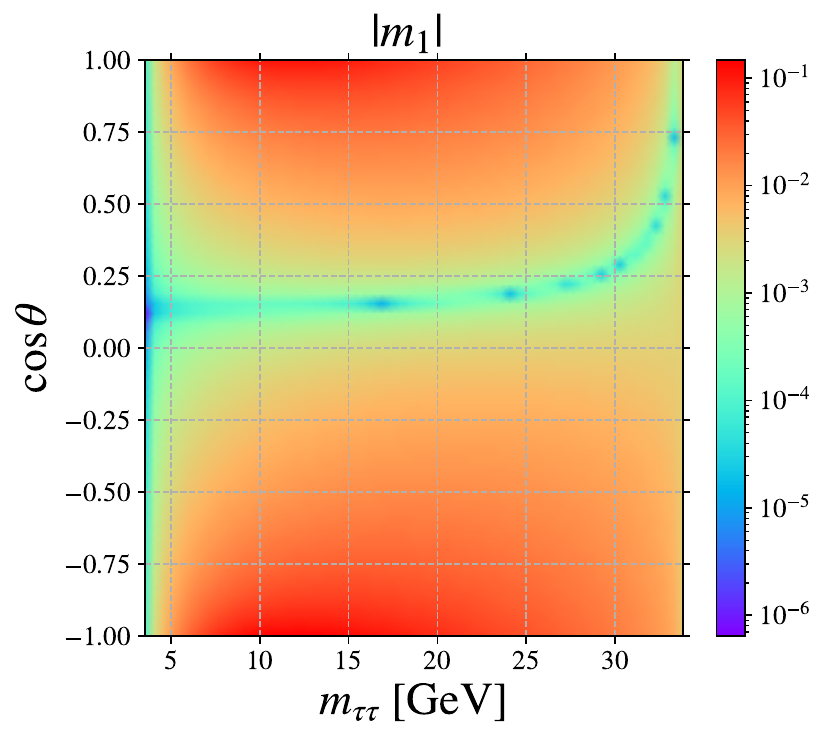}
\includegraphics[width=0.32\textwidth]{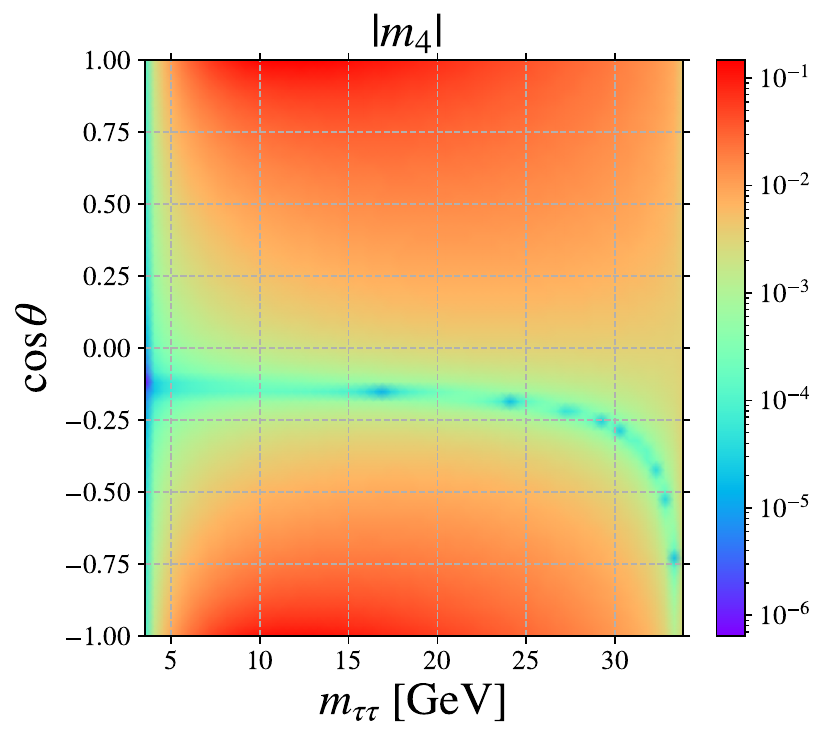}
\caption{\small
Top: the hyperdeterminant $|M| = |m_1 m_4 - m_2 m_3|$ (left) and $\max_i |m_i|$ (right); bottom: the minors $|m_1|$ (left) and $|m_4|$ (right) of the helicity-basis amplitude matrix, on the $(m_{\tau\tau},\, \cos\theta)$ plane, computed for $h \to \tau^- \tau^+ Z$ in the Standard Model at tree level to all orders in $m_f$ (logarithmic colour scales).
$\max_i |m_i| \neq 0$ detects genuine $2 \otimes 2 \otimes 3$ entanglement, while $|M| \neq 0$ selects its generic (C1) type; the blue curves in $|M|$, $|m_1|$ and $|m_4|$ are the nodal lines discussed in the text.
The remaining minors $|m_2|$ and $|m_3|$ are shown in Fig.~\ref{fig:miyake-m23}.
 }
\label{fig:miyake}
\end{figure}

Figure~\ref{fig:miyake} shows $|M|$, the genuineness detector $\max_i |m_i|$, and the two leading minors $|m_1|$ and $|m_4|$ for the Standard Model $h \to \tau^- \tau^+ Z$ process, evaluated at tree level to all orders in $m_f$ (note the logarithmic colour scales); the remaining minors $|m_2|$ and $|m_3|$ are featureless and of $O( \varepsilon^2 ) \sim 10^{-3}$ throughout, and are relegated to Appendix~\ref{app:Miyake} (Fig.~\ref{fig:miyake-m23}).
Since $O( \varepsilon^2 ) \sim 10^{-3}$ is smaller than the loop factor $\alpha/\pi \simeq 2 \times 10^{-3}$, one might worry that $m_2$ and $m_3$ are dominated by radiative corrections.
They are not: each of these minors is bilinear in the same-helicity amplitudes, which are proportional to $m_f$ at any loop order by the chirality argument of Sec.~\ref{sec:0-223}.
Radiative corrections therefore cannot generate $m_2$ and $m_3$, but only renormalise them, at the relative level of $O( \alpha/\pi )$.
The maps confirm the analytic picture.
$|m_1|$ and $|m_4|$ are $O( \varepsilon )$, reaching $\simeq 0.15$ in the collinear regions at $m_{\tau\tau} \simeq 10\!-\!15$~GeV and exhibiting the mirror nodal curves at $\cos\theta \simeq +0.15$ and $\simeq -0.15$, respectively, which bend away from the equator (upwards for $|m_1|$, downwards for $|m_4|$) towards the upper endpoint.
$|M|$ peaks at $\simeq 2 \times 10^{-2}$ in the collinear regions at $m_{\tau\tau} \simeq 12$~GeV, in agreement with Eq.~\eqref{eq:HDet-collinear}, and is suppressed in two distinct regions: towards the threshold and along the two nodal curves.
Comparing with the map of $\max_i |m_i|$ assigns each region its SLOCC class.
Over most of the plane $M \neq 0$ and the state belongs to the generic class C1 --- small but non-zero, in accordance with the small but non-vanishing $O( \varepsilon^2 )$ values of $\lambda_{\min}( \rho_Z )$ over the bulk of Fig.~\ref{fig:lmin}.
Along the nodal curves $|M| \to 0$ while $\max_i |m_i|$ remains large: the state stays genuinely $2 \otimes 2 \otimes 3$ entangled and approaches the class C2.
Towards the threshold, $|M|$ and $\max_i |m_i|$ vanish together: genuine $2 \otimes 2 \otimes 3$ entanglement is lost and the state approaches the biseparable class B3, in one-to-one correspondence with the vanishing of $\lambda_{\min}( \rho_Z )$ in Fig.~\ref{fig:lmin}.
Finally, the mild dip of $\max_i |m_i|$ along $\cos\theta = 0$ at small $m_{\tau\tau}$ reflects Eq.~\eqref{eq:m14-coupling}: at $\theta = \pi/2$ both leading minors are proportional to the small vector coupling $c_V$.
The same mechanism explains the suppression of $\lambda_{\min}( \rho_Z )$ along $\cos\theta \simeq 0$ observed in Fig.~\ref{fig:lmin}, as anticipated in Sec.~\ref{sec:0-223}: the out-of-plane qutrit component, whose square sets $\lambda_{\min}( \rho_Z )$ at $O( \varepsilon^2 )$, is $c_V$-suppressed on the equator.

\begin{table}[t!]
\centering
\begin{tabular}{lccc}
\hline
 & $O( \varepsilon^0 )$ & $O( \varepsilon^2 )$ & $O( \varepsilon^4 )$ \\
 & [$m_i = 0$] & [$M \simeq m_1 m_4$] & [$M = m_1 m_4 - m_2 m_3$] \\
\hline
bulk & GHZ & C1 & C1 \\
nodal curves & GHZ & C2 & C1 \\
threshold & B3 & B3 & B3 \\
\hline
\end{tabular}
\caption{\small SLOCC class of the $h \to \tau^- \tau^+ Z$ spin state in the three characteristic regions of the phase space (rows), as resolved at successive orders of the $m_f$ expansion (columns): at $O( \varepsilon^0 )$ all minors vanish; at $O( \varepsilon^2 )$ the hyperdeterminant is dominated by $m_1 m_4$; at $O( \varepsilon^4 )$ the $m_2 m_3$ term is included.
}
\label{tab:sloccflow}
\end{table}

The classification is thus not uniform in the $m_f$ expansion, and Table~\ref{tab:sloccflow} summarises how the SLOCC class assigned to the state evolves with the order at which the invariants are resolved.
At $O( \varepsilon^0 )$ all minors vanish: the state is of the GHZ type in the bulk and biseparable (B3) at the threshold boundary.
Including the $O( \varepsilon )$ minors, $M = m_1 m_4 + O( \varepsilon^4 )$ promotes the bulk to the generic class C1, while the nodal curves, on which $m_1 m_4 = 0$ with the other leading minor non-zero, appear as C2.
At $O( \varepsilon^4 )$ the residual $m_2 m_3$ term becomes relevant precisely there: on the nodal curves of $m_{1}$ and $m_4$ one has $M = - m_2 m_3 \neq 0$, so the state is strictly of class C1 also there; it differs, however, from an exact C2 state only by terms of $O( \varepsilon^2 )$.
The threshold region remains B3 at every order.

\section{Bell inequalities}
\label{sec:Bell}

In a tripartite system the notion of local realism splits into two inequivalent assumptions, and there are correspondingly two distinct levels of Bell nonlocality to be tested \cite{Horodecki:2025tpn}.
The stronger assumption is \emph{full local realism} (FLR), in which each of the three parties carries its own local hidden variable,
\be
P(a,b,c\,|\,A_i,B_j,V_k) \,=\, \sum_\lambda P_\lambda\;
\mathcal{A}_\lambda(a|A_i)\, \mathcal{B}_\lambda(b|B_j)\, \mathcal{V}_\lambda(c|V_k)\,.
\label{eq:FLR}
\ee
The weaker assumption is \emph{bipartite local realism} (BLR) \cite{Svetlichny:1987bqj}, in which any two of the parties may be correlated arbitrarily provided the remaining bipartition is local-real,
\be
P(a,b,c\,|\,A_i,B_j,V_k) \,=\,
p_1 \sum_\lambda P_\lambda\, \mathcal{A}_\lambda(a|A_i)\, \overline{\mathcal{A}}_\lambda(b,c|B_j,V_k)
\,+\, \big( \text{two permutations} \big)\,,
\label{eq:BLR}
\ee
with $p_i \geq 0$ and $\sum_i p_i = 1$.
Every distribution of the form \eqref{eq:FLR} is also of the form \eqref{eq:BLR}, so the two sets are nested, ${\rm FLR} \subset {\rm BLR}$, and violating a BLR bound is a strictly stronger statement than violating an FLR bound.
Moreover, every biseparable quantum state admits a distribution of the form \eqref{eq:BLR}; a violation of a BLR bound therefore certifies \emph{genuine tripartite} nonlocality, whereas a violation of an FLR bound only establishes that the correlations are not fully local-real.
In this section we probe both levels: full local realism with the tight $4 \times 4 \times 2$ observables of Wu and Zong, and bipartite local realism with the Svetlichny observable.

\subsection{Bell operators and their bounds}
\label{sec:Bell-ops}

Full local realism in a three-party system was first tested by the Mermin inequality \cite{Mermin:1990vxe}, which uses two settings per party and carries the FLR bound $2$.
It is not, however, sufficient: there exist non-FLR correlations that satisfy it \cite{laskowski}.
A strictly more powerful test is obtained by giving two of the parties four settings each, which yields the tight $4 \times 4 \times 2$ inequality of Wu and Zong \cite{WU2003262}.
Both inequalities are tight --- each is a facet of the local polytope of its own measurement scenario --- but the enlarged setting space of the $4\times 4 \times 2$ family exposes non-FLR correlations to which any two-setting inequality is blind.
We therefore work with the latter throughout.

To test FLR of the qubit--qubit--qutrit state of $H \to f \bar f V$, we employ three $4 \times 4 \times 2$ inequalities of Wu and Zong \cite{WU2003262}, which were proven to be tight in Ref.~\cite{laskowski}.
This class of inequalities was recently applied to three-body decays in which the final-state spins form a three-qubit system \cite{Horodecki:2025tpn}.
We follow the notation of Ref.~\cite{Horodecki:2025tpn}:
\ba
I_{442} &=& \big[A_1(B_1+B_2)+A_2(B_1-B_2)\big](V_1+V_2)
  + \big[A_3(B_3+B_4)+A_4(B_3-B_4)\big](V_1-V_2) ,
\nn \\
I_{424} &=& \big[A_1(V_1+V_2)+A_2(V_1-V_2)\big](B_1+B_2)
  + \big[A_3(V_3+V_4)+A_4(V_3-V_4)\big](B_1-B_2) ,
\nn \\
I_{244} &=& \big[B_1(V_1+V_2)+B_2(V_1-V_2)\big](A_1+A_2)
  + \big[B_3(V_3+V_4)+B_4(V_3-V_4)\big](A_1-A_2) .
\nn \\
\label{eq:Idef}
\ea
Here $A_i$, $B_i$ and $V_i$ denote dichotomic ($\pm 1$-valued) observables acting on the two fermions and on the vector boson, respectively.
For the qubits, these take the form $\hat a \cdot \vec\sigma$ with $\hat a$ a unit Bloch vector; observables proportional to the identity are excluded, since they cannot contribute to a Bell violation \cite{LANDAU198754}.
A dichotomic qutrit observable can be written as $V_i = \pm (2 \Pi - \mathds{1})$, with $\Pi = | v \ketbra v |$ a rank-one projector onto a normalised state $| v \rangle$ in the three-dimensional spin space of $V$.
Fixing the unphysical global phase by taking the $| + \rangle$ coefficient real and non-negative, $| v \rangle$ is parametrised by four real compact parameters,
\be
| v \rangle \,=\,
\cos\alpha \, | + \rangle
\,+\, e^{i \xi} \sin\alpha \cos\beta \, | 0 \rangle
\,+\, e^{i \eta} \sin\alpha \sin\beta \, | - \rangle ,
\label{eq:qutrit-ray}
\ee
with $\alpha, \beta \in [ 0, \pi/2 ]$ and $\xi, \eta \in [ 0, 2\pi )$.
In each operator of Eq.~\eqref{eq:Idef}, two of the parties measure four settings and the remaining one measures two; the subscripts list the numbers of settings of $(A, B, V)$ in order.

Bipartite local realism is tested by the Svetlichny observable \cite{Svetlichny:1987bqj, Horodecki:2025tpn}
\ba
I_{S} &=& A_1 B_1 V_1 + A_1 B_1 V_2 + A_1 B_2 V_1 + A_2 B_1 V_1
\nn \\
&& -\, A_2 B_2 V_2 - A_2 B_2 V_1 - A_2 B_1 V_2 - A_1 B_2 V_2 \,,
\label{eq:ISdef}
\ea
which uses two settings per party.

From these operators we define
\be
\mathcal{B}_{X} \,\equiv\, \max\, \Tr( I_{X}\, \rho)\,,
\qquad
X \,\in\, \{\, 442,\; 424,\; 244,\; S \,\}\,,
\label{eq:Bdef}
\ee
with $\rho = | \psi \ketbra \psi |$ the pure state of the qubit--qubit--qutrit system, where the maximisation is taken over all measurement settings.
Each of the $4\times4\times2$ observables obeys the local-hidden-variable (LHV) bound
$\mathcal{B}_{IJK} \leq 4$ \cite{WU2003262}, so that a value above $4$ certifies Bell nonlocality; the bound is \emph{tight} \cite{laskowski}, i.e.\ it describes a facet of the LHV polytope, so that every correlation beyond the LHV set violates it for some choice of settings.
Quantum mechanics allows values up to $8$, and general no-signalling correlations up to $16$ \cite{Horodecki:2025tpn}.
The Svetlichny observable obeys the BLR bound $\mathcal{B}_S \leq 4$, with a quantum maximum of $4\sqrt{2}$ and an algebraic maximum of $8$.
The three sets of bounds are collected in Table~\ref{tab:Bell-bounds}.
\begin{table}[h!]
\centering
\begin{tabular}{lcccc}
\hline
observable & tests & LHV/BLR & quantum & algebraic \\
\hline
Mermin \cite{Mermin:1990vxe} & FLR & $2$ & $4$ & $4$ \\
$\mathcal{B}_{442},\ \mathcal{B}_{424},\ \mathcal{B}_{244}$ \cite{WU2003262} & FLR & $4$ & $8$ & $16$ \\
$\mathcal{B}_{S}$ \cite{Svetlichny:1987bqj} & BLR & $4$ & $4\sqrt{2}$ & $8$ \\
\hline
\end{tabular}
\caption{\small Bounds on the Bell observables used in this work. All values are independent of the local Hilbert-space dimensions.}
\label{tab:Bell-bounds}
\end{table}

All of these bounds hold unchanged in the $2 \otimes 2 \otimes 3$ system.
The LHV and BLR bounds involve only the $\pm 1$ outcomes of the dichotomic observables and are therefore insensitive to the dimension of the space on which they act; the qutrit is measured with two outcomes, one eigenvalue being twofold degenerate, and never three.
The quantum bounds are likewise dimension independent: they follow from Tsirelson's inequality for the CHSH operators appearing in Eqs.~\eqref{eq:Idef} and \eqref{eq:ISdef}, together with the operator identity $(V_1+V_2)^2 + (V_1-V_2)^2 = 4\, \mathds{1}$, which requires only $V_i^2 = \mathds{1}$.
Note also that Mermin is the one case in which the quantum maximum saturates the algebraic one; the $4\times4\times2$ and Svetlichny observables both leave a genuine quantum--no-signalling gap.
The CP relation \eqref{eq:cp-obs} implies that $\mathcal{B}_{442}$ is an even function of $\cos\theta$, whereas $\mathcal{B}_{424}$ and $\mathcal{B}_{244}$ are mirror images of each other under $\cos\theta \to -\cos\theta$.
Since $I_S$ in Eq.~\eqref{eq:ISdef} is invariant under the exchange $A \leftrightarrow B$, $\mathcal{B}_S$ is likewise an even function of $\cos\theta$.
Note that, because the inequality is tight, whether a given state admits an LHV description does not depend on the assignment of the settings \cite{Horodecki:2025tpn}; the three observables of Eq.~\eqref{eq:Idef} differ only in how strongly they expose the violation for a given state, which is what we compare below.

So far, both the tightness analysis \cite{laskowski} and the optimisation of these observables \cite{Horodecki:2025tpn} have been carried out for three-qubit systems only.
One of the main results of this work is the extension of this formalism to the $2 \otimes 2 \otimes 3$ system: to the best of our knowledge, the semi-analytical optimisation of the tight $4 \times 4 \times 2$ observables over the enlarged family of qutrit measurements, presented in the following two subsections, is new.
A brute-force maximisation in Eq.~\eqref{eq:Bdef} over all measurement settings is a high-dimensional, non-convex problem; our methods carry out most of the optimisation analytically, leaving only a low-dimensional numerical search, with a reduction technique that depends on which party holds the two settings.
For $\mathcal{B}_{442}$ (Sec.~\ref{sec:442a}), all eight qubit settings are eliminated in closed form, generalising the three-qubit approach of Ref.~\cite{Horodecki:2025tpn}; for $\mathcal{B}_{424}$ and $\mathcal{B}_{244}$ (Sec.~\ref{sec:442b}), the four settings of the remaining qubit \emph{and} the four qutrit settings are eliminated by means of the qubit--qudit CHSH result of Bernal, Casas and Moreno \cite{Bernal:2024dtg}.
For the Svetlichny observable (Sec.~\ref{sec:svet}) the situation is more favourable: the optimal setting of \emph{every} party is available in closed form, so that no continuous search remains at all.
None of the three constructions relies on the dimension of the vector-boson space: the method of Sec.~\ref{sec:442a} for $\mathcal{B}_{442}$ is formulated directly for a general $2 \otimes 2 \otimes d$ system, while those of Sec.~\ref{sec:442b} for $\mathcal{B}_{424}$, $\mathcal{B}_{244}$ and of Sec.~\ref{sec:svet} for $\mathcal{B}_{S}$ extend to it straightforwardly, since the qubit--qudit CHSH result of Ref.~\cite{Bernal:2024dtg} and the trace-norm maximisation over the vector-boson settings both hold for arbitrary $d$.

\subsection{Two settings on the qutrit: $\mathcal{B}_{442}$}
\label{sec:442a}

Fix, for the moment, the two observables $V_1$ and $V_2$ of the $d$-dimensional party.
All the dependence of $\braket{ I_{442} } \equiv \Tr ( I_{442}\, \rho )$ on the state and on $V_{1,2}$ then enters through two real $3 \times 3$ correlation matrices,
\be
\left[ K_\pm \right]_{ij} \,=\, \Braket{ \sigma_i \otimes \sigma_j \otimes \left( V_1 \pm V_2 \right) }\,,
\qquad i, j = 1, 2, 3\,,
\label{eq:Kpm}
\ee
where the first (second) Pauli matrix acts on the fermion $A$ ($B$).
Writing $A_k = \hat a_k \cdot \vec\sigma$ and $B_l = \hat b_l \cdot \vec\sigma$, one has
\be
\braket{ I_{442} } \,=\,
\hat a_1 \cdot K_+ \cdot ( \hat b_1 + \hat b_2 )
+ \hat a_2 \cdot K_+ \cdot ( \hat b_1 - \hat b_2 )
+ \hat a_3 \cdot K_- \cdot ( \hat b_3 + \hat b_4 )
+ \hat a_4 \cdot K_- \cdot ( \hat b_3 - \hat b_4 )\,.
\label{eq:I442-bloch}
\ee
The maximisation over the eight unit vectors parallels the familiar treatment of the CHSH inequality and can be carried out in closed form.
Consider first the two $K_+$ terms of Eq.~\eqref{eq:I442-bloch}, which involve the four unit vectors $\hat a_1, \hat a_2, \hat b_1, \hat b_2$.
The combinations $\hat b_1 \pm \hat b_2$ are always orthogonal, $( \hat b_1 + \hat b_2 ) \cdot ( \hat b_1 - \hat b_2 ) = | \hat b_1 |^2 - | \hat b_2 |^2 = 0$, and their lengths obey $| \hat b_1 + \hat b_2 |^2 + | \hat b_1 - \hat b_2 |^2 = 4$.
The pair $( \hat b_1, \hat b_2 )$ can therefore be traded, without loss of generality, for an orthonormal pair $( \hat e_1, \hat e_2 )$ and a mixing angle $\varphi$:
\be
\hat b_1 + \hat b_2 \,=\, 2 \cos\varphi\, \hat e_1\,,
\qquad
\hat b_1 - \hat b_2 \,=\, 2 \sin\varphi\, \hat e_2\,,
\qquad
\hat e_1 \cdot \hat e_2 = 0 \,.
\label{eq:bpar}
\ee
With this parametrisation the two $K_+$ terms are written as $2 \cos\varphi \, X_1 + 2 \sin\varphi \, X_2$, where
$X_I = \hat a_I \cdot K_+ \cdot \hat e_I$, ($I = 1,2$).
The $\varphi$ angle is optimised such that $(\cos \varphi, \sin \varphi) = (X_1, X_2)/\sqrt{X_1^2 + X_2^2}$, leading to $2 \sqrt{ X_1^2 + X_2^2 }$.

It remains to maximise $X_1^2 + X_2^2 = (\hat a_1 \cdot K_+ \cdot \hat e_1)^2 + (\hat a_2 \cdot K_+ \cdot \hat e_2)^2$ over the vectors $\hat a_1, \hat a_2$ and the orthonormal pair $\hat e_1 \perp \hat e_2$.
To this end, we treat the matrix $K_+$ as a vector in $\mathbb{R}^9 = \mathbb{R}^3 \otimes \mathbb{R}^3$ and consider the Schmidt decomposition,
\be
K_+ \,=\, \sum_{i=1}^{3} \mu_i\, \hat S_i \otimes \hat S'_i\,,
\qquad
\mu_1 \geq \mu_2 \geq \mu_3 \geq 0\,,
\label{eq:schmidt}
\ee
with orthonormal triples $\{ \hat S_i \}$ and $\{ \hat S'_i \}$.
The Schmidt coefficients $\mu_i$ are the singular values of the real $3 \times 3$ matrix
$\left[ K_+ \right]_{ij}$ in Eq.~\eqref{eq:Kpm}
and $\hat S_i$, $\hat S'_i$ are its left and right singular vectors.
Choosing $\hat a_1 = \hat S_1$, $\hat a_2 = \hat S_2$, $\hat e_1 = \hat S'_1$ and $\hat e_2 = \hat S'_2$ (note that $\hat S'_1 \perp \hat S'_2$) gives $X_1 = \mu_1$ and $X_2 = \mu_2$, hence $\sqrt{ X_1^2 + X_2^2 } = \sqrt{ \mu_1^2 + \mu_2^2 }$.
By the Schur--Horn theorem, this is the maximal value that $X_1^2 + X_2^2$ can attain.
The last two terms of Eq.~\eqref{eq:I442-bloch} are maximised in the same way, with the Schmidt coefficients of $K_-$, i.e.\ the singular values of the matrix $\left[ K_- \right]_{ij}$.
The optimisation over all eight qubit settings is thus solved in closed form:
\be
\mathcal{B}_{442} \,=\, 2\, \max_{ V_1, V_2 } \left[
\sqrt{ \mu_1^2 ( K_+ ) + \mu_2^2 ( K_+ ) }
\,+\, \sqrt{ \mu_1^2 ( K_- ) + \mu_2^2 ( K_- ) } \,\right],
\label{eq:B442-max}
\ee
where $\mu_1 ( K ) \geq \mu_2 ( K ) \geq \mu_3 ( K )$ denote the singular values of $K$.

Equation~\eqref{eq:B442-max} generalises the two-qubit criterion of Ref.~\cite{Horodecki:1995nsk}, in which the maximal CHSH value of a state $\rho$ is $2 \sqrt{ \mu_1^2 + \mu_2^2 }$ in terms of the singular values of its correlation matrix; here the role of the state is played by the two hermitian, in general non-positive, effective operators $\rho_{\pm}^{AB} = \Tr_V [\, \rho\, ( \mathds{1} \otimes \mathds{1} \otimes ( V_1 \pm V_2 ) ) ]$, whose correlation matrices are $K_\pm$.
The residual maximisation in Eq.~\eqref{eq:B442-max} runs over the two observables of the $d$-dimensional party alone, with the optimal qubit settings determined analytically at each point.
Specialising to the qutrit case at hand ($d = 3$), the residual search runs over the two rank-one projectors alone, i.e.\ over the compact eight-parameter space of two copies of Eq.~\eqref{eq:qutrit-ray}.

\subsection{Two settings on a qubit: $\mathcal{B}_{424}$ and $\mathcal{B}_{244}$}
\label{sec:442b}

We describe the method for $\mathcal{B}_{424}$, where the fermion $B$ holds the two settings; $\mathcal{B}_{244}$ is obtained by exchanging the roles of the two fermions.
The structure of $I_{424}$ in Eq.~\eqref{eq:Idef} suggests grouping the terms into two CHSH-type operators acting on the qubit--qutrit pair $( A, V )$,
\be
I_{424} \,=\, I^{(1)} \left( B_1 + B_2 \right) + I^{(2)} \left( B_1 - B_2 \right),
\qquad
\begin{aligned}
I^{(1)} &= A_1 ( V_1 + V_2 ) + A_2 ( V_1 - V_2 )\,,\\
I^{(2)} &= A_3 ( V_3 + V_4 ) + A_4 ( V_3 - V_4 )\,.
\end{aligned}
\label{eq:I424-blocks}
\ee
We trade the pair $( \hat b_1, \hat b_2 )$ for an orthonormal pair $( \hat e_1, \hat e_2 )$ and a mixing angle $\varphi$, as in Eq.~\eqref{eq:bpar}.
The state and the $B$ direction associated with each $\hat e_k$ are then absorbed into two hermitian, in general non-positive, effective operators on $\Hil_A \otimes \Hil_V$,
\be
\varrho_k \,=\, \Tr_B \big[\, \rho \left( \mathds{1} \otimes \hat e_k \cdot \vec\sigma \otimes \mathds{1} \right) \big]\,,
\qquad k = 1, 2\,,
\label{eq:rhoeff}
\ee
in terms of which
$\braket{ I_{424} } = 2 \cos\varphi\, \Tr ( I^{(1)} \varrho_1 ) + 2 \sin\varphi\, \Tr ( I^{(2)} \varrho_2 )$.
The maximisations of the two traces are independent, since $I^{(1)}$ and $I^{(2)}$ contain disjoint sets of settings, and each maximum can be taken non-negative, since flipping the signs of $A_1$ and $A_2$ ($A_3$ and $A_4$) reverses the sign of $I^{(1)}$ ($I^{(2)}$).
The $\varphi$ dependence is then maximised exactly as in Sec.~\ref{sec:442a}, giving
\be
\mathcal{B}_{424} \,=\, 2 \max_{ \hat e_1 \perp \hat e_2 }
\sqrt{ \Big[ \max_{ \{ A_{1,2}, V_{1,2} \} } \Tr \big( I^{(1)} \varrho_1 \big) \Big]^2
+ \Big[ \max_{ \{ A_{3,4}, V_{3,4} \} } \Tr \big( I^{(2)} \varrho_2 \big) \Big]^2 }\,.
\label{eq:B424-chi}
\ee

Each inner maximisation in Eq.~\eqref{eq:B424-chi} is a CHSH-type problem for the qubit--qutrit pair $( A, V )$, with the state replaced by the effective operator $\varrho_k$.
For a genuine qubit--qudit state this problem was solved in closed form by Bernal, Casas and Moreno \cite{Bernal:2024dtg}.
Although the $\varrho_k$, unlike a density matrix, are in general non-positive, their construction carries over unchanged --- every step of the derivation below relies only on the hermiticity of the expanded operator, never on its positivity or unit trace --- and we spell it out in our notation.

Expanding $\varrho_k$ over the Pauli basis of $A$,

\be
\varrho_k \,=\, \frac{1}{2} \left( \mathds{1} \otimes \Omega^{(k)}_0 + \sum_{i=1}^{3} \sigma_i \otimes \Omega^{(k)}_i \right),
\qquad
\Omega^{(k)}_i \,=\, \Tr_{AB} \big[\, \rho \left( \sigma_i \otimes \hat e_k \cdot \vec\sigma \otimes \mathds{1} \right) \big]\,,
\label{eq:Omegadef}
\ee
with $\Omega^{(k)}_i$ hermitian operators on the qutrit space, and using the tracelessness of the Pauli matrices, the first block becomes

\ba
\Tr \big( I^{(1)} \varrho_1 \big)
&=& \frac{1}{2} \sum_{i=1}^3 \Tr \big[ A_1 \sigma_i \big]\, \Tr \big[ \Omega^{(1)}_i \left( V_1 + V_2 \right) \big]
\,+\, \frac{1}{2} \sum_{i=1}^3 \Tr \big[ A_2 \sigma_i \big]\, \Tr \big[ \Omega^{(1)}_i \left( V_1 - V_2 \right) \big]
\nn \\
&=&
\hat a_1 \cdot \big( \vec r_{V_1}^{\,(1)} + \vec r_{V_2}^{\,(1)} \big) + \hat a_2 \cdot \big( \vec r_{V_1}^{\,(1)} - \vec r_{V_2}^{\,(1)} \big)\,,
\label{eq:regroup}
\ea
where $\hat a_I$ $(I = 1,2)$ is the unit Bloch vector introduced in Eq.~\eqref{eq:I442-bloch} and
\ba
\vec r_{V_l}^{\,(k)} &\equiv& \left( \Tr [ \Omega^{(k)}_1 V_l ], \ \Tr [ \Omega^{(k)}_2 V_l ], \ \Tr [ \Omega^{(k)}_3 V_l ] \right).
\label{eq:rvec}
\ea
The maximum over the fermion settings is therefore reached by aligning each $\hat a_I$ with the vector it multiplies:
\be
\max_{ A_1, A_2 }\, \Tr ( I^{(1)} \varrho_1 ) \,=\, \big| \vec r_{V_1}^{\,(1)} + \vec r_{V_2}^{\,(1)} \big| \,+\, \big| \vec r_{V_1}^{\,(1)} - \vec r_{V_2}^{\,(1)} \big|\,.
\label{eq:Amax}
\ee

We then proceed exactly as in Ref.~\cite{Bernal:2024dtg}, using the identity
\be
\big(
| \vec v + \vec w | + | \vec v - \vec w | \big)^2 \,=\, 4\, \max_{R} \left[ ( R \, \vec v )_1^2 + ( R \, \vec w )_2^2 \right],
\ee
valid for any two vectors $\vec v, \vec w \in \mathbb{R}^3$, where $R \in SO(3)$ is a three-dimensional rotation and $( \vec x )_i$ stands for the $i$-th component of the vector $\vec x \in \mathbb{R}^3$.
Substituting $\vec v = \vec r_{V_1}^{\,(1)}$ and $\vec w = \vec r_{V_2}^{\,(1)}$, and using $( R\, \vec r_{V_l}^{\,(k)} )_a = \Tr [ ( R\, \vec\Omega^{(k)} )_a V_l ]$ with $( R\, \vec\Omega )_a \equiv \sum_b R_{ab} \Omega_b$, we obtain
\ba
\max_{ \{ A, V \} } \Tr ( I^{(1)} \varrho_1 )
&=& 2\, \max_{ R, V_1, V_2 } \sqrt{ \big[ R\, \vec r_{V_1}^{\,(1)} \big]_1^2 + \big[ R\, \vec r_{V_2}^{\,(1)} \big]_2^2 } \nn \\
&=& 2\, \max_{ R, V_1, V_2 } \sqrt{ \big| \Tr [ ( R\, \vec\Omega^{(1)} )_1 V_1 ] \big|^2 + \big| \Tr [ ( R\, \vec\Omega^{(1)} )_2 V_2 ] \big|^2 }\,.
\label{eq:chain}
\ea

Proceeding again in the same way as 
in Ref.~\cite{Bernal:2024dtg}, one uses the fact that, for any hermitian matrix $X$, $\max_V \Tr [ X V ] \,=\, \| X \|_1$ with $\| X \|_1 \equiv \Tr \sqrt{ X^\dagger X }$, where the maximisation runs over the dichotomic observables satisfying $V^2 = \mathds{1}$.\footnote{Here the qutrit observables are taken to range over all hermitian matrices with $V^2 = \mathds{1}$, i.e.\ the projective observables $\pm ( 2 \Pi - \mathds{1} )$ are supplemented by the trivial ones $\pm \mathds{1}$, for which the maximum is attained when $X$ has a definite sign.
The extension is harmless: the LHV bound of $4$ holds for deterministic settings as well.}
Applying the latter to each term of Eq.~\eqref{eq:chain} as well as its analogue for $I^{(2)}$ gives a concise formula, the exact analogue of the central result of Ref.~\cite{Bernal:2024dtg},
\be
\max_{ \{ A, V \} } \Tr \big( I^{(k)} \varrho_k \big) \,=\,
2 \max_{ R \in SO(3) } \sqrt{
\big\| ( R\, \vec\Omega^{(k)} )_1 \big\|_1^2 + \big\| ( R\, \vec\Omega^{(k)} )_2 \big\|_1^2 }\,.
\label{eq:bernal}
\ee

In this construction the four $A$ settings and the four qutrit observables are all determined analytically; only the rotation $R$ remains.
Combining Eqs.~\eqref{eq:B424-chi} and \eqref{eq:bernal},
\be
\mathcal{B}_{424} \,=\, 4 \max_{ \hat e_1 \perp \hat e_2 }
\sqrt{ \max_{ R } \sum_{a=1,2} \big\| ( R\, \vec\Omega^{(1)} )_a \big\|_1^2
\,+\, \max_{ R' } \sum_{a=1,2} \big\| ( R'\, \vec\Omega^{(2)} )_a \big\|_1^2 }\,,
\label{eq:B424-max}
\ee
and the residual numerical optimisation runs over the orthonormal pair $( \hat e_1, \hat e_2 )$ (three parameters) and the two rotations $R, R' \in SO(3)$ (three parameters each) --- a compact nine-parameter search space.
This construction, too, extends to any $2 \otimes 2 \otimes d$ system: the result of Ref.~\cite{Bernal:2024dtg} holds for a qudit of arbitrary dimension, and, remarkably, the dimension of the residual search space does not grow with $d$.
As noted above, $\mathcal{B}_{244}$ follows from Eq.~\eqref{eq:B424-max} upon exchanging the roles of the two fermions, and by the CP relation \eqref{eq:cp-obs} its value is the $\cos\theta \to -\cos\theta$ image of $\mathcal{B}_{424}$.

\subsection{The Svetlichny observable: $\mathcal{B}_{S}$}
\label{sec:svet}

The Svetlichny observable \eqref{eq:ISdef} is the most tractable of the four: the optimal measurement of each party can be written down in closed form, and the maximisation reduces to an alternating sequence of exact updates requiring no numerical solver.

Grouping the terms of Eq.~\eqref{eq:ISdef} by the qutrit settings gives
\be
I_S \,=\, \big( A_1 B_1 + A_1 B_2 + A_2 B_1 - A_2 B_2 \big) V_1
\,+\, \big( A_1 B_1 - A_1 B_2 - A_2 B_1 - A_2 B_2 \big) V_2 \,.
\label{eq:IS-groupV}
\ee
With $A_I = \hat a_I \cdot \vec\sigma$ and $B_I = \hat b_I \cdot \vec\sigma$ fixed, and writing $\vec s = \hat a_1 + \hat a_2$, $\vec d = \hat a_1 - \hat a_2$, the expectation value becomes
\be
\braket{ I_S } \,=\, \Tr\big( X_1 V_1 \big) \,+\, \Tr\big( X_2 V_2 \big)\,,
\label{eq:IS-traces}
\ee
where $X_1$ and $X_2$ are the hermitian qutrit operators
\ba
X_1 &=& \Tr_{AB} \Big[\, \rho \, \big( ( \vec s \cdot \vec\sigma ) \otimes ( \hat b_1 \cdot \vec\sigma ) + ( \vec d \cdot \vec\sigma ) \otimes ( \hat b_2 \cdot \vec\sigma ) \big) \Big] \,,
\nn \\[3pt]
X_2 &=& \Tr_{AB} \Big[\, \rho \, \big( ( \vec d \cdot \vec\sigma ) \otimes ( \hat b_1 \cdot \vec\sigma ) - ( \vec s \cdot \vec\sigma ) \otimes ( \hat b_2 \cdot \vec\sigma ) \big) \Big] \,.
\label{eq:IS-X}
\ea
The maximisation over the vector-boson settings is now immediate, and is the same trace-norm argument already used in Sec.~\ref{sec:442b}.
In the basis where $X_i$ is diagonal, the optimised $V_i$ is given by the diagonal matrix with the $\pm 1$ eigenvalues.
The sign of each eigenvalue is chosen to match that of the corresponding eigenvalue of $X_i$, so that every product of the two is positive.
With this choice, ${\rm Tr}(X_i V_i)$ is maximised as $\| X_i \|_1$.
Applying this to Eq.~\eqref{eq:IS-traces} gives
\be
\max_{V_1, V_2} \braket{ I_S } \,=\, \big\| X_1 \big\|_1 \,+\, \big\| X_2 \big\|_1 \,.
\label{eq:IS-qutrit}
\ee
Conversely, for fixed qutrit settings define the two real $3\times3$ correlation matrices $[ M_k ]_{ij} = \braket{ \sigma_i \otimes \sigma_j \otimes V_k }$ and $M_\pm = M_1 \pm M_2$. Grouping Eq.~\eqref{eq:ISdef} by $A$ and by $B$ in turn gives
\ba
\braket{ I_S } &=& \hat a_1 \cdot \big( M_+ \hat b_1 + M_- \hat b_2 \big) \,+\, \hat a_2 \cdot \big( M_- \hat b_1 - M_+ \hat b_2 \big)
\nn \\[3pt]
&=& \hat b_1 \cdot \big( M_+^{T} \hat a_1 + M_-^{T} \hat a_2 \big) \,+\, \hat b_2 \cdot \big( M_-^{T} \hat a_1 - M_+^{T} \hat a_2 \big) \,,
\label{eq:IS-qubits}
\ea
so that each fermion's optimal Bloch vectors are obtained by aligning them with the vectors they multiply, exactly as in the three-qubit treatment of Ref.~\cite{Horodecki:2025tpn}; the corresponding maxima are the sums of the two vector norms.

Equations~\eqref{eq:IS-qutrit} and \eqref{eq:IS-qubits} give the exact conditional optimum for each party in turn. 
Iterating them --- qutrit, then $A$, then $B$ --- yields a monotonically non-decreasing sequence which converges rapidly. 
We use a few random restarts to guard against local optima.\footnote{
We have tested the procedure on two states with known Svetlichny values.
For the GHZ state $( \ket{000} + \ket{111} )/\sqrt{2}$ it returns $\mathcal{B}_S = 4\sqrt{2}$, saturating the quantum bound, and for the biseparable state $\tfrac{1}{\sqrt{2}} ( \ket{00} + \ket{11} )_{AB} \otimes \ket{0}_V$ it returns $\mathcal{B}_S = 4$, saturating the BLR bound.
Both are reproduced to machine precision.  
} 
Unlike the constructions of Secs.~\ref{sec:442a} and \ref{sec:442b}, no continuous numerical search is necessary. 
The method makes no reference to the qutrit dimension and applies to any $2 \otimes 2 \otimes d$ system.

\subsection{Results}
\label{sec:Bell-results}

\begin{figure}[t!]
\centering
\includegraphics[width=0.32\textwidth]{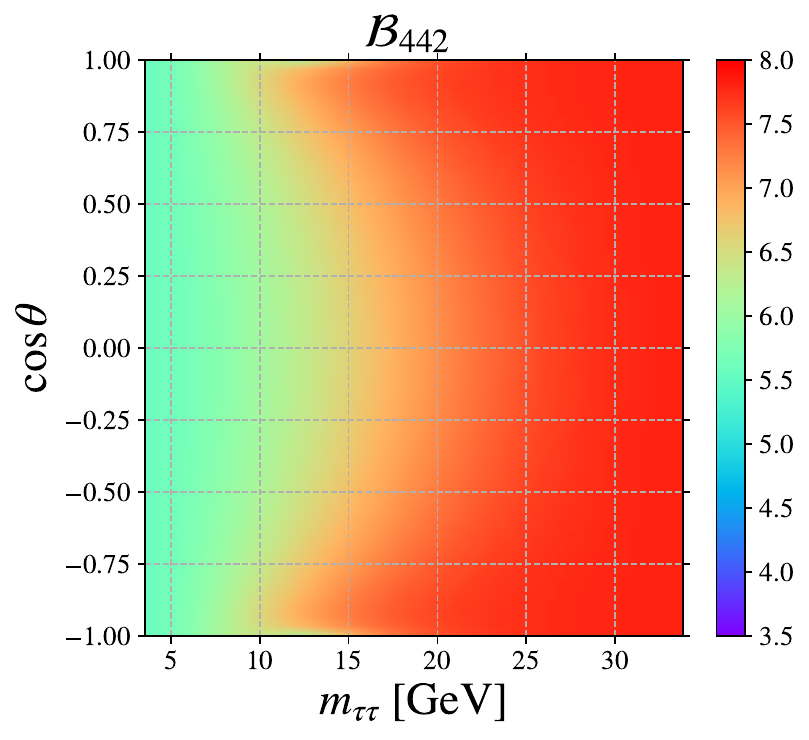}
\includegraphics[width=0.32\textwidth]{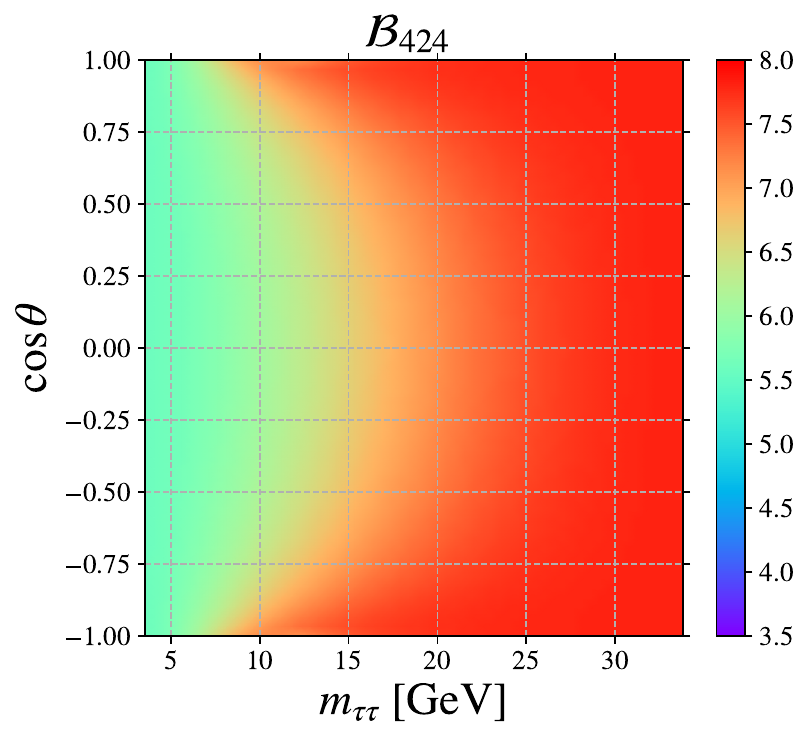}
\includegraphics[width=0.32\textwidth]{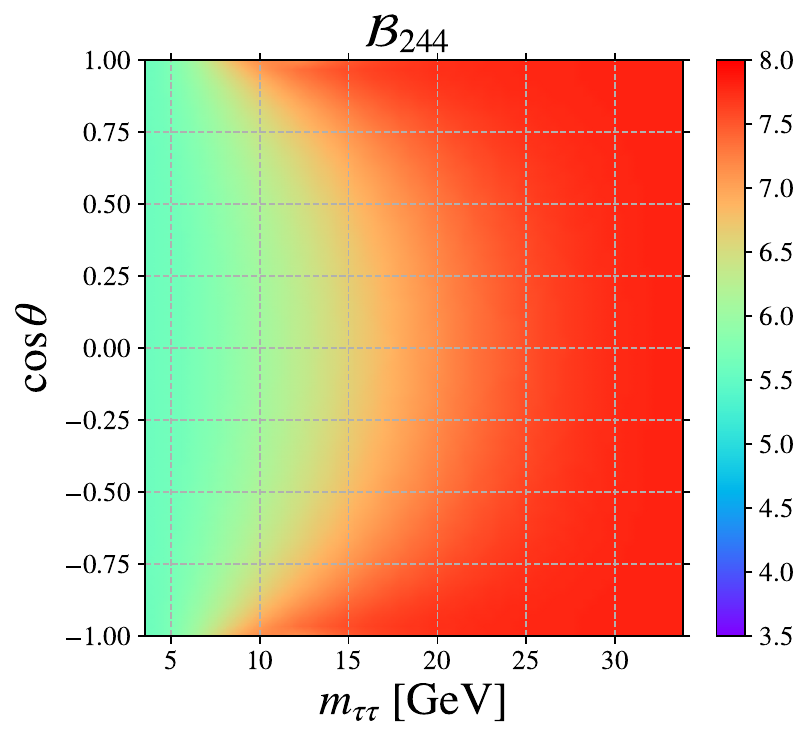}
\caption{\small
The maximised Bell observables $\mathcal{B}_{442}$ (left), $\mathcal{B}_{424}$ (centre) and $\mathcal{B}_{244}$ (right), computed with the semi-analytical methods of Secs.~\ref{sec:442a} and \ref{sec:442b}, over the phase space of $h \to \tau^- \tau^+ Z$ in the Standard Model.
The local-hidden-variable bound $\mathcal{B} \leq 4$ is violated everywhere; the quantum bound is $8$.
 }
\label{fig:Bell}
\end{figure}

\begin{figure}[t!]
\centering
\includegraphics[width=0.32\textwidth]{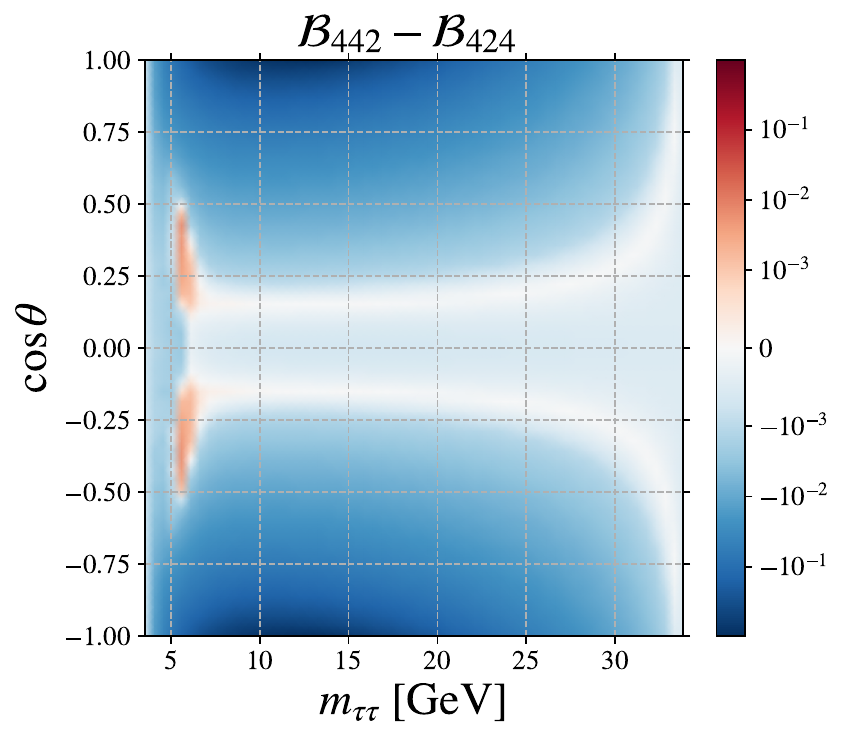}
\includegraphics[width=0.32\textwidth]{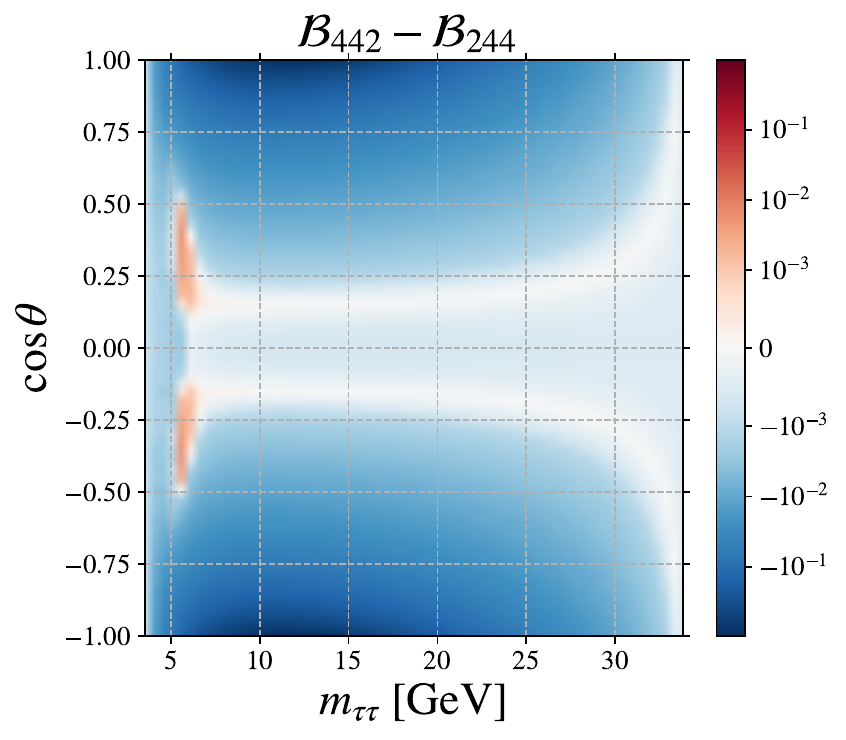}
\includegraphics[width=0.32\textwidth]{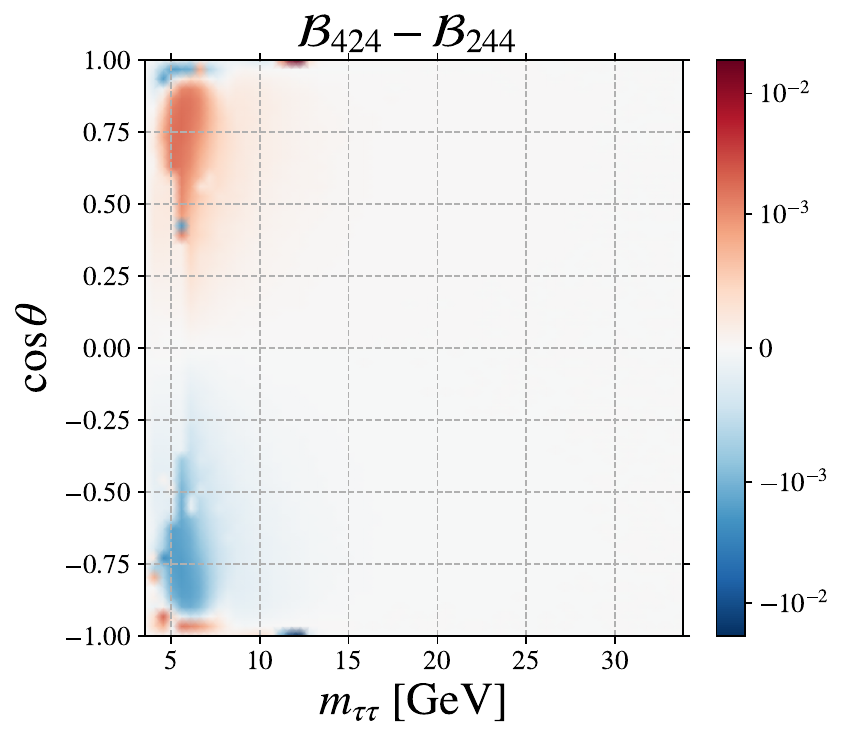}
\caption{\small
Pairwise differences of the Bell observables of Fig.~\ref{fig:Bell}: $\mathcal{B}_{442} - \mathcal{B}_{424}$ (left), $\mathcal{B}_{442} - \mathcal{B}_{244}$ (centre) and $\mathcal{B}_{424} - \mathcal{B}_{244}$ (right), on a symmetric logarithmic colour scale.
The isolated dots are numerical artefacts of the residual optimisation.
 }
\label{fig:dB}
\end{figure}

Figure~\ref{fig:Bell} shows the three tight $4 \times 4 \times 2$ observables, maximised with the semi-analytical methods of Secs.~\ref{sec:442a} and \ref{sec:442b}, across the phase space of $h \to \tau^- \tau^+ Z$ in the Standard Model; Fig.~\ref{fig:dB} shows their pairwise differences.
The headline result is that the LHV bound $4$ is violated on the \emph{entire} phase space: all three observables grow monotonically from $\simeq 5.6$ at the $\tau\tau$ threshold to $\simeq 7.9$ at the upper endpoint, within a few per cent of the quantum bound $8$.
Moreover, all three observables also grow, at fixed $m_{\tau\tau}$, from central angles towards the collinear directions, and the three maps are nearly identical to one another.
All of these features can be understood analytically from the limiting states of Sec.~\ref{sec:0-223}.

Consider first the threshold region.
There the state \eqref{eq:threshold-state} is a product across the $V | AB$ partition, so the third-party observables can contribute only trivially: in Eq.~\eqref{eq:B442-max} the correlation matrices factorise, $K_\pm = \braket{ V_1 \pm V_2 }\, K_{AB}$ with $[ K_{AB} ]_{ij} = \braket{ \sigma_i \otimes \sigma_j }$, while in Eq.~\eqref{eq:B424-max} the qutrit operators reduce to $\Omega^{(k)}_i \propto \rho_V$ with $\| \rho_V \|_1 = 1$.
In both cases the maximisation collapses to the two-qubit criterion of Ref.~\cite{Horodecki:1995nsk} for the fermion pair, and all three observables approach the common value
\be
\mathcal{B}_{442}\,,\; \mathcal{B}_{424}\,,\; \mathcal{B}_{244}
\;\to\; 4 \sqrt{ 1 + C_{AB}^2 }
\qquad ( m_{AB} \to 2 m_f )
\label{eq:B-thr}
\ee
with $C_{AB}$ the threshold concurrence \eqref{eq:CAB-thr}; numerically $4 \sqrt{ 1 + ( 0.988 )^2 } \simeq 5.6$, in agreement with the left edge of Fig.~\ref{fig:Bell}.
The inequalities are thus still violated at threshold, but the nonlocality is of a purely bipartite, CHSH type, carried by the fermion Bell pair; correspondingly, all three differences in Fig.~\ref{fig:dB} vanish towards the threshold edge.

Consider next the upper endpoint, where the pointer states become orthogonal ($Q \to 0$), and the zeroth-order state \eqref{eq:psi0} reduces to a generalised GHZ state:
\be
\ket{\psi^{(0)}} \,\to\, \ket{\psi_{\rm GHZ}(\delta)} \,\propto\, \cos\delta \ket{-+} \ket{\Psi_+^*} \,+\, \sin\delta\, \ket{+-} \ket{\Psi_-^*}
\qquad ( m_{AB} \to m_{AB}^{\rm max} )\,,
\label{eq:psi_upper}
\ee 
up to irrelevant phases, where
\be
\ket{\Psi_\pm^*} \,=\,
\pm\, \tfrac{1}{2} \left( 1 \pm \cos\theta \right) \ket{+}
\,-\, \tfrac{1}{\sqrt{2}}\, \sin\theta\, \ket{0}
\,\pm\, \tfrac{1}{2} \left( 1 \mp \cos\theta \right) \ket{-}\,,
\qquad
\braket{ \Psi_+^* | \Psi_-^* } \,=\, 0\,,
\label{eq:pointer_upper}
\ee
and $\cos\delta \equiv | c_L | / \sqrt{ \langle c^2 \rangle }$ and $\sin\delta \equiv | c_R | / \sqrt{ \langle c^2 \rangle }$, so that $\sin 2\delta = 2 | c_L c_R | / \langle c^2 \rangle = \hat s$.
The quantum bound $8$ is attained on the maximally entangled GHZ state \cite{Horodecki:2025tpn}; for the generalised GHZ family $\ket{\psi_{\rm GHZ}(\delta)}$, an explicit optimisation of Eqs.~\eqref{eq:B442-max} and \eqref{eq:B424-max} yields the closed forms
\be
\mathcal{B}_{442} \,=\, \max\!\left( 8 \sin 2\delta\,,\; 4 \sqrt{ 1 + \sin^2 2\delta } \right),
\qquad
\mathcal{B}_{424} \,=\, \mathcal{B}_{244} \,=\, \max\!\left( 8 \sin 2\delta\,,\; 4 \sqrt{ 1 + 2 \sin^2 2\delta } \right),
\label{eq:B-GHZ}
\ee
with crossovers at $\sin 2\delta = 1/\sqrt{3}$ and $1/\sqrt{2}$, respectively; the additional factor of $2$ under the square root in $\mathcal{B}_{424}$ reflects the greater freedom afforded by the four qutrit settings.
Since $\hat s \simeq 0.988$ lies well above both crossovers, all three observables approach the common, near-maximal value
\be
\mathcal{B} \;\to\; 8\, \hat s \,\simeq\, 7.9
\qquad ( m_{AB} \to m_{AB}^{\rm max} )
\label{eq:B-endpoint}
\ee
--- the red region of Fig.~\ref{fig:Bell} --- and the differences again vanish.

The behaviour in the bulk interpolates between these two limits.
The zeroth-order state depends on the kinematics only through the overlap ratio $Q/P$, so at $O( \varepsilon^0 )$ all three observables are functions of $( Q/P, \hat s )$ alone: they decrease monotonically from the generalised-GHZ value $8 \hat s$ at $Q/P \to 0$ to the bipartite value $4 \sqrt{ 1 + \hat s^2 }$ at $Q/P \to 1$, as the growing pointer-state overlap degrades the tripartite GHZ structure into a bipartite one.
Since the angle enters only through $Q/P$, with $Q \propto \sin^2\theta$, this explains the shape of the maps in Fig.~\ref{fig:Bell}.

The differences in Fig.~\ref{fig:dB} quantify how much the assignment of the two settings matters --- remarkably little, at or below the $10^{-1}$ level everywhere.
We see that $\mathcal{B}_{424} \simeq \mathcal{B}_{244} \gtrsim \mathcal{B}_{442}$ over most of the plane (blue in the first two panels), mirroring the ordering of the two expressions in Eq.~\eqref{eq:B-GHZ}, with the largest deficits along the collinear edges; small regions of the opposite ordering appear near threshold at $| \cos\theta | \sim 0.3$.
Both differences fade towards the threshold and endpoint edges, where the observables collapse onto the common values \eqref{eq:B-thr} and \eqref{eq:B-endpoint}.
Finally, since $\mathcal{B}_{244} ( \cos\theta ) = \mathcal{B}_{424} ( -\cos\theta )$ by the CP relation \eqref{eq:cp-obs}, their difference is an odd function of $\cos\theta$, as visible in the third panel, and is smaller still ($\lesssim 10^{-2}$).

\begin{figure}[t!]
\centering
\includegraphics[width=0.4\textwidth]{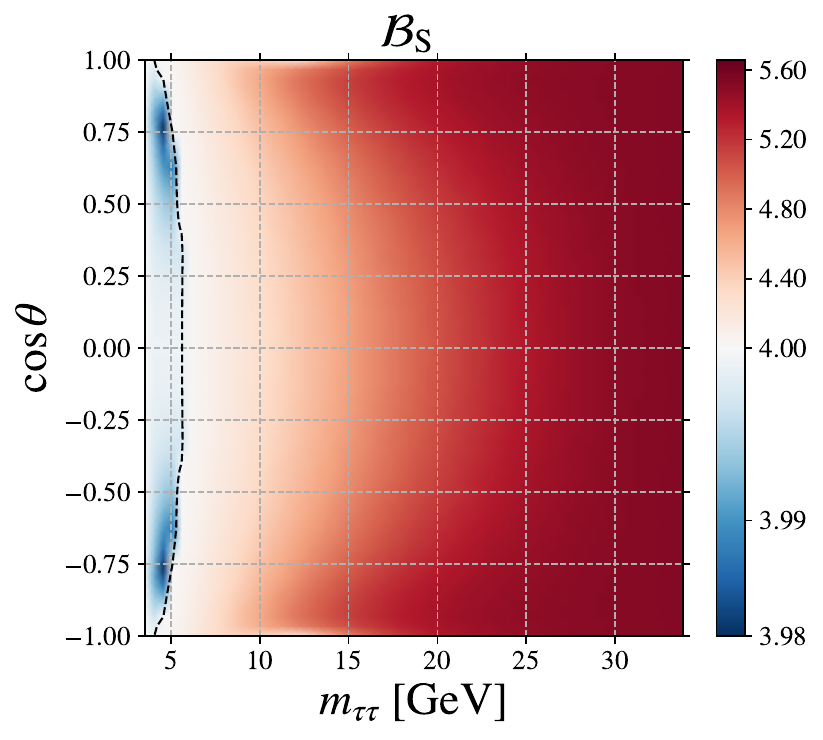}
\caption{\small
The maximised Svetlichny observable $\mathcal{B}_{S}$, computed with the method of Sec.~\ref{sec:svet}, over the phase space of $h \to \tau^- \tau^+ Z$ in the Standard Model.
The dashed curve marks the bipartite-local-real bound $\mathcal{B}_S = 4$; the quantum bound is $4\sqrt{2} \simeq 5.66$.
Note the strongly non-linear colour scale, where the blue region is visually enhanced.
 }
\label{fig:BS}
\end{figure}

We now turn to the stronger, bipartite-local-real test.
Figure~\ref{fig:BS} shows $\mathcal{B}_S$ over the same plane.
In contrast to the $4 \times 4 \times 2$ observables, which violate full local realism \emph{everywhere}, the BLR bound is violated only above $m_{\tau\tau} \simeq 6$~GeV: genuine tripartite nonlocality switches off at the threshold $m_{\tau \tau} = 2 m_\tau$.
For larger $m_{\tau\tau}$, $\mathcal{B}_S$ rises monotonically to $\simeq 5.52$ at the upper endpoint, i.e.\ $97.5\%$ of the quantum bound $4\sqrt{2}$.
As anticipated in Sec.~\ref{sec:Bell-ops}, the map is an even function of $\cos\theta$.

The absence of a violation near the threshold is not accidental.
There the state \eqref{eq:threshold-state} is a product across the $V | AB$ partition and hence biseparable, so by construction it cannot violate a BLR bound.
The optimisation reflects this transparently: in that region the optimum is attained with \emph{trivial} qutrit observables $V_1 = V_2 = \pm \mathds{1}$, for which Eq.~\eqref{eq:ISdef} collapses to $2 ( A_1 B_1 - A_2 B_2 )$ and
\be
\mathcal{B}_S \;\to\; 4\, \mu_1 \big( T_{AB} \big)
\qquad ( m_{AB} \to 2 m_f )\,,
\label{eq:BS-thr}
\ee
with $\mu_1$ the largest singular value of the fermion-pair correlation matrix $[T_{AB}]_{ij} = \braket{\sigma_i \otimes \sigma_j}$.
The vector boson thus effectively collapses in this region, and $\mathcal{B}_S$ saturates the biseparable ceiling of $4$ from below --- falling short only because the $\tau$ pair is not quite maximally correlated, $\mu_1 < 1$.


\section{Non-stabiliserness}
\label{sec:magic}

Entanglement and Bell nonlocality do not exhaust the quantum resources of a state.
By the Gottesman--Knill theorem, a quantum circuit that acts on a stabiliser state with Clifford gates and Pauli measurements can be simulated efficiently on a classical computer \cite{Gottesman:1998hu}, even though such circuits generate highly entangled states.
The resource that separates universal quantum computation from classically simulable dynamics must therefore lie beyond entanglement: it is \emph{non-stabiliserness}, commonly known as \emph{magic}.
In the associated resource theory, stabiliser states are the free states, Clifford operations are the free operations, and the magic of a state quantifies its distance from the stabiliser set --- a prerequisite for genuine quantum-computational advantage.
Among the available quantifiers, the stabiliser R\'enyi entropy (SRE) \cite{Leone:2021rzd} stands out for its computability: unlike measures defined through an extremisation over stabiliser decompositions, it is obtained directly from the Pauli expectation values of the state.
The SRE has been generalised to systems of qudits \cite{Wang:2023uog}, shown to be a genuine magic monotone for R\'enyi index $\alpha \geq 2$ \cite{Leone:2024lfr}, and recently endowed with an operational interpretation in the framework of quantum property testing \cite{Bittel:2025yhq}.

Magic has also reached particle physics.
Ref.~\cite{White:2024nuc} proposed to measure it in an elementary-particle system, showing that the spin state of top-quark pairs produced at the LHC carries an amount of magic that varies across the production phase space; following this proposal, the CMS collaboration has reported an observation of magic in $t \bar t$ events at $\sqrt{s} = 13$~TeV --- the first measurement of this quantity at the TeV scale \cite{CMS-PAS-TOP-25-001}.
These studies concern a pair of qubits.
The decay $h \to \tau^- \tau^+ Z$ offers a natural next step --- a genuinely three-party state with unequal local dimensions --- but requires an SRE defined on the asymmetric Hilbert space $\mathbb{C}^2 \otimes \mathbb{C}^2 \otimes \mathbb{C}^3$.
In this section we construct this extension and apply it to our process.

\subsection{Asymmetric magic}
\label{sec:asym-magic}

We first recall the SRE for $n$ qudits of a common prime dimension $d$ \cite{Leone:2021rzd, Wang:2023uog}.
The building blocks are the Heisenberg--Weyl (generalised Pauli) operators: on a single qudit with computational basis $\{ \ket{k} \}_{k=0}^{d-1}$, the shift and clock operators are defined by
\be
X \ket{k} \,=\, \ket{ k + 1 \;\mathrm{mod}\; d }\,,
\qquad
Z \ket{k} \,=\, \omega^k \ket{k}\,,
\qquad
\omega \,=\, e^{ 2 \pi i / d }\,,
\label{eq:XZdef}
\ee
and generate the $d^2$ unitary operators
\be
W_{ab} \,=\, X^a Z^b\,,
\qquad
a, b \in \mathbb{Z}_d\,,
\qquad
\Tr \big( W_{ab}^\dagger W_{a'b'} \big) \,=\, d\, \delta_{aa'} \delta_{bb'}\,,
\label{eq:HWdef}
\ee
which by the orthogonality relation form a complete basis of the operator space.
For $d = 2$ they reduce, up to phases, to the Pauli operators $\{ \mathds{1}, X, Z, Y \}$; for $d > 2$ they are unitary but not hermitian, so their expectation values are in general complex.
Phase conventions for $W_{ab}$ vary in the literature; they drop out of all expressions below, which involve only $| \Tr ( \rho\, W ) |$.
For $n$ qudits the Heisenberg--Weyl set consists of the $D^2$ tensor products
$\mathcal{W} = \{ W_{a_1 b_1} \otimes \cdots \otimes W_{a_n b_n} \}$, with $D = d^n$ the Hilbert-space dimension, and the orthogonality relation generalises to $\Tr ( W^\dagger W' ) = D\, \delta_{W W'}$.

For a pure state $\rho = | \psi \ketbra \psi |$, define
\be
\Xi_W \,=\, \frac{ | \braket{ \psi | W | \psi } |^2 }{ D } \,\geq\, 0 \,,
\qquad
W \in \mathcal{W}\,.
\label{eq:XiW}
\ee
Since the state expands over the basis as $\rho = \tfrac{1}{D} \sum_{W} \Tr ( \rho\, W^\dagger )\, W$, they obey the sum rule
\be
\sum_{W \in \mathcal{W}} \Xi_W
\,=\, \frac{1}{D} \sum_{W \in \mathcal{W}} | \Tr ( \rho\, W ) |^2
\,=\, \Tr \rho^2 \,=\, 1\,.
\label{eq:sumrule}
\ee
The set $\{ \Xi_W \}$ is therefore a probability distribution over the Heisenberg--Weyl operators, measuring how the weight of the state is spread across the operator basis.
The SRE of order $\alpha$ is defined as the $\alpha$-R\'enyi entropy of this distribution, offset so that stabiliser states give zero:
\be
M_\alpha ( \ket{\psi} ) \,=\, \frac{1}{1 - \alpha} \log_2 \left[\, \sum_{W \in \mathcal{W}} \Xi_W^\alpha \right] \,-\, \log_2 D\,.
\label{eq:SREdef}
\ee
The offset is natural: a pure stabiliser state obeys $| \braket{ \psi | W | \psi } | = 1$ for the $D$ elements of its stabiliser group and $0$ for all others, so $\{ \Xi_W \}$ is uniform on a support of size $D$, and $\sum_{W} \Xi_W^\alpha = D \times \frac{1}{D^\alpha}$, and therefore $M_\alpha = 0$.
Any state whose weight spreads over more of the operator basis has a larger entropy and hence $M_\alpha > 0$.
In fact, since the $W$ are unitary, $\Xi_W \leq 1/D$, which for $\alpha > 1$ gives
\be
\sum_{W \in \mathcal{W}} \Xi_W^\alpha \,\leq\, D^{ 1 - \alpha }
\qquad \Longrightarrow \qquad
M_\alpha \,\geq\, 0\,,
\label{eq:Mpos}
\ee
with equality if and only if $\Xi_W = 1/D$ on exactly $D$ operators.

The SRE is a well-behaved magic measure \cite{Leone:2021rzd, Wang:2023uog, Leone:2024lfr}:
(i) \emph{faithfulness}: $M_\alpha ( \ket\psi ) \geq 0$, with equality if and only if $\ket\psi$ is a stabiliser state;
(ii) \emph{Clifford invariance}: $M_\alpha ( C \ket\psi ) = M_\alpha ( \ket\psi )$ for any Clifford unitary $C$, since conjugation by $C$ merely permutes the Heisenberg--Weyl operators;
(iii) \emph{additivity}: $M_\alpha ( \ket\psi \otimes \ket\phi ) = M_\alpha ( \ket\psi ) + M_\alpha ( \ket\phi )$;
(iv) \emph{boundedness}: $M_\alpha ( \ket\psi ) \leq \log_2 D$, since the distribution is supported on at most $D^2$ operators;
(v) \emph{monotonicity}: for $\alpha \geq 2$, $M_\alpha$ is non-increasing under stabiliser protocols --- Clifford unitaries supplemented by computational-basis measurements with classical feed-forward --- and is thus a genuine magic monotone \cite{Leone:2024lfr}, although not a strong one (it may increase on individual measurement outcomes).
This motivates the choice $\alpha = 2$ adopted below.

Nothing in the above discussion requires the local dimensions to be equal.
Consider the asymmetric Hilbert space
\be
\Hil \,=\, \mathbb{C}^{d_1} \otimes \cdots \otimes \mathbb{C}^{d_n}\,,
\qquad
D \,=\, d_1 \cdots d_n\,,
\ee
with every $d_i$ prime, and build the composite Heisenberg--Weyl set from the local ones,
\be
\mathcal{W} \,=\, \Big\{ W^{(1)}_{a_1 b_1} \otimes \cdots \otimes W^{(n)}_{a_n b_n} \;\Big|\; a_i, b_i \in \mathbb{Z}_{d_i} \Big\}\,,
\qquad
| \mathcal{W} | \,=\, \prod_i d_i^2 \,=\, D^2\,,
\label{eq:asymW}
\ee
where $W^{(i)}_{a_i b_i}$ is defined by Eqs.~\eqref{eq:XZdef}--\eqref{eq:HWdef} with $d = d_i$.
The orthogonality relation $\Tr ( W^\dagger W' ) = D\, \delta_{W W'}$ holds factor by factor, so $\mathcal{W}$ is again a complete orthogonal operator basis, and the definitions \eqref{eq:XiW}--\eqref{eq:SREdef} carry over unchanged.
So do the properties, whose proofs never invoke the equality of the local dimensions.
The sum rule \eqref{eq:sumrule} and the bound \eqref{eq:Mpos} use only completeness and unitarity.
Faithfulness also extends straightforwardly: the equality condition in Eq.~\eqref{eq:Mpos} forces $| \braket{ \psi | W | \psi } | = 1$ on exactly $D$ operators, which close (up to phases) into an abelian subgroup $S$ of order $D$ of the composite Heisenberg--Weyl group, and $\ket\psi$ is the stabiliser state defined by $S$.\footnote{Here, the primality of the $d_i$ is necessary.
Otherwise, the common eigenstates of order-$D$ abelian subgroups need not coincide with the standard stabiliser states --- the free, classically simulable states of the Clifford resource theory --- and $M_\alpha = 0$ would no longer certify a stabiliser state \cite{Wang:2023uog}.}
Clifford invariance holds with the Clifford group defined as the normaliser of the composite Heisenberg--Weyl group: Clifford unitaries permute $\mathcal{W}$ up to phases, $W \to C\, W C^\dagger \in \mathcal{W}$; additivity is proven in Ref.~\cite{Wang:2023uog} for a tensor product of two systems of \emph{different} dimensions, which is precisely the asymmetric statement.
The monotonicity (v) has been established for systems of qubits \cite{Leone:2024lfr}; we are not aware of a published extension to mixed local dimensions, but our use of the SRE below relies only on the elementary properties (i)--(iv), which hold in full generality.

For the system at hand, $\Hil = \mathbb{C}^2_A \otimes \mathbb{C}^2_B \otimes \mathbb{C}^3_V$, the basis \eqref{eq:asymW} comprises $| \mathcal{W} | = 4 \cdot 4 \cdot 9 = 144 = D^2$ operators with $D = 12$; their explicit matrix representations are collected in Appendix~\ref{app:HW}.
Throughout this study we adopt $\alpha = 2$ --- the same choice made in Refs.~\cite{White:2024nuc, CMS-PAS-TOP-25-001} --- for which Eq.~\eqref{eq:SREdef} reduces to
\be
M_2 ( \ket{\psi} ) \,=\, - \log_2 \left[\, D \sum_{W \in \mathcal{W}} \Xi_W^2 \right]\,.
\label{eq:M2def}
\ee

The maximal value of $M_2$ is a more subtle question than the bound in (iv) suggests.
The loose bound $\log_2 D$ can be tightened to
\be
M_2 ( \ket{\psi} ) \,\leq\, \log_2 \frac{ D + 1 }{ 2 }\,,
\label{eq:SICbound}
\ee
with equality if and only if $\ket\psi$ is a fiducial state of a Weyl--Heisenberg covariant symmetric informationally complete (WH-SIC) set \cite{Leone:2021rzd, Wang:2023uog}.
Whether the bound can actually be attained therefore depends on the existence of such states for the Heisenberg--Weyl group in question, a famously open problem: for multi-qubit groups they are known \emph{not} to exist except for three qubits, so the true maximum lies strictly below Eq.~\eqref{eq:SICbound} and must be determined case by case \cite{Liu:2025frx, Ohta:2025utz}.
For two qubits, strong numerical evidence puts the maximal second-order SRE at $\log_2 ( 16/7 ) \simeq 1.19$, attained on the $480$ fiducial states of the mutually unbiased bases generated by the Weyl--Heisenberg orbits \cite{Liu:2025frx}; for three qubits --- the one exception, where the bound \eqref{eq:SICbound} is saturated --- closed-form maximal magic states have been constructed from the Barnes--Wall lattice $BW_{16}$ and classified by their entanglement structure \cite{Ohta:2025utz}.
For the asymmetric Heisenberg--Weyl group \eqref{eq:asymW} of the $2 \otimes 2 \otimes 3$ system, the existence of WH-SIC states is to our knowledge unknown.
Nevertheless, Eq.~\eqref{eq:SICbound} still provides the reference scale $M_2 \leq \log_2 ( 13/2 ) \simeq 2.70$ against which the values found below should be compared.

\subsection{Non-local asymmetric magic}
\label{sec:nlm}

The SRE defined in the previous subsection is not invariant under local unitary transformations.
Applying a local unitary to one party of a stabiliser state generally generates nonvanishing magic. 
For a multi-party state, the physically motivated quantity is therefore the magic that \emph{no} choice of local bases can remove --- the magic stored in the correlations among the parties.
This is captured by the \emph{non-local non-stabiliserness} recently introduced in \cite{Qian:2025oit}: the SRE minimised over local unitaries.
The concept has quickly found its way into high-energy physics: the magic and non-local magic of gluon and graviton helicity amplitudes have been analysed in Refs.~\cite{Gargalionis:2025iqs, Gargalionis:2026onv}, and the non-local magic generated in low-energy two-particle scattering, including nucleon--nucleon systems, in Ref.~\cite{Robin:2025ymq}.
For a two-qubit pure state $\ket{ \psi_{2q} }$, the minimisation can be carried out exactly, with the elegant result that the non-local magic is a function of the entanglement spectrum alone \cite{Qian:2025oit, Busoni:2026lvp},
\be
\mathcal{M}_{\rm nl} ( \ket{\psi_{2q}} ) \,=\, - \log_2 \left( 1 - 4 e + 16 e^2 \right),
\qquad
e \,=\, \det \rho_A\,,
\label{eq:NN2qubit}
\ee
which vanishes both for product states ($e = 0$) and for maximally entangled stabiliser states ($e = 1/4$), and peaks at $\log_2 ( 4/3 )$ for $e = 1/8$: a state must be \emph{both} entangled and magical to carry non-local magic.

The extension to our asymmetric system is immediate.
We define the non-local magic of the $2 \otimes 2 \otimes 3$ state as
\be
\mathcal{M}_{\rm nl} ( \ket\psi ) \,=\, \min_{ U_A,\, U_B,\, U_V }
M_2 \big( ( U_A \otimes U_B \otimes U_V ) \ket\psi \big)\,,
\qquad
U_{A}, U_{B} \in SU(2)\,,
\quad
U_V \in SU(3)\,,
\label{eq:NNdef}
\ee
with $M_2$ the asymmetric SRE of Eq.~\eqref{eq:M2def}.
The minimisation domain is a compact group manifold of $3 + 3 + 8 = 14$ parameters, over which we minimise numerically with repeated local searches from random starting points; since the identity belongs to the group, $\mathcal{M}_{\rm nl} ( \ket\psi ) \leq M_2 ( \ket\psi )$ in whatever basis the state is given.

The properties of $\mathcal{M}_{\rm nl}$ follow directly from those of the SRE established in Sec.~\ref{sec:asym-magic}.
(i) \emph{Non-negativity and boundedness}: $0 \leq \mathcal{M}_{\rm nl} \leq M_2$, the lower bound following from the faithfulness of $M_2$.
(ii) \emph{Local-unitary invariance}: the minimisation absorbs any local basis change, so $\mathcal{M}_{\rm nl}$ depends only on the local-unitary orbit of the state.
(iii) \emph{Faithfulness on orbits}: $\mathcal{M}_{\rm nl} ( \ket\psi ) = 0$ if and only if $\ket\psi$ is local-unitarily equivalent to a stabiliser state of the composite Heisenberg--Weyl group.
In particular, $\mathcal{M}_{\rm nl}$ vanishes for every fully product state (any single-party pure state can be rotated onto a stabiliser state) and for every stabiliser state, no matter how entangled; a nonzero value therefore certifies the joint presence of entanglement and non-stabiliserness.
It does not, however, certify \emph{genuinely tripartite} correlations: by the additivity of $M_2$, a biseparable state $\ket{\psi_{AB}} \otimes \ket{\chi_V}$ has $\mathcal{M}_{\rm nl} ( \ket{\psi_{AB}} \otimes \ket{\chi_V} ) = \mathcal{M}_{\rm nl} ( \ket{\psi_{AB}} )$, the two-qubit non-local magic \eqref{eq:NN2qubit} carried entirely by the fermion pair.
(iv) \emph{Symmetries}: $\mathcal{M}_{\rm nl}$ is symmetric under the exchange of the two qubits, and, since complex conjugation in the computational basis maps the Heisenberg--Weyl set onto itself up to phases, it obeys the same CP relation \eqref{eq:cp-obs} as the other observables of this paper.
(v) \emph{Sub-additivity}: additivity of $M_2$ implies $\mathcal{M}_{\rm nl} ( \ket\psi \otimes \ket\phi ) \leq \mathcal{M}_{\rm nl} ( \ket\psi ) + \mathcal{M}_{\rm nl} ( \ket\phi )$, magic in the correlations being potentially removable by local unitaries acting on larger parties.
Closed-form results of the type \eqref{eq:NN2qubit} have recently been extended to bipartite systems of two qutrits and two ququints \cite{Busoni:2026lvp}, but none is available for the tripartite $2 \otimes 2 \otimes 3$ system --- the local-unitary orbits are no longer labelled by the entanglement spectrum of a single cut --- and we evaluate Eq.~\eqref{eq:NNdef} numerically.

\subsection{Results}
\label{sec:magic-results}

\begin{figure}[t!]
\centering
\includegraphics[width=0.32\textwidth]{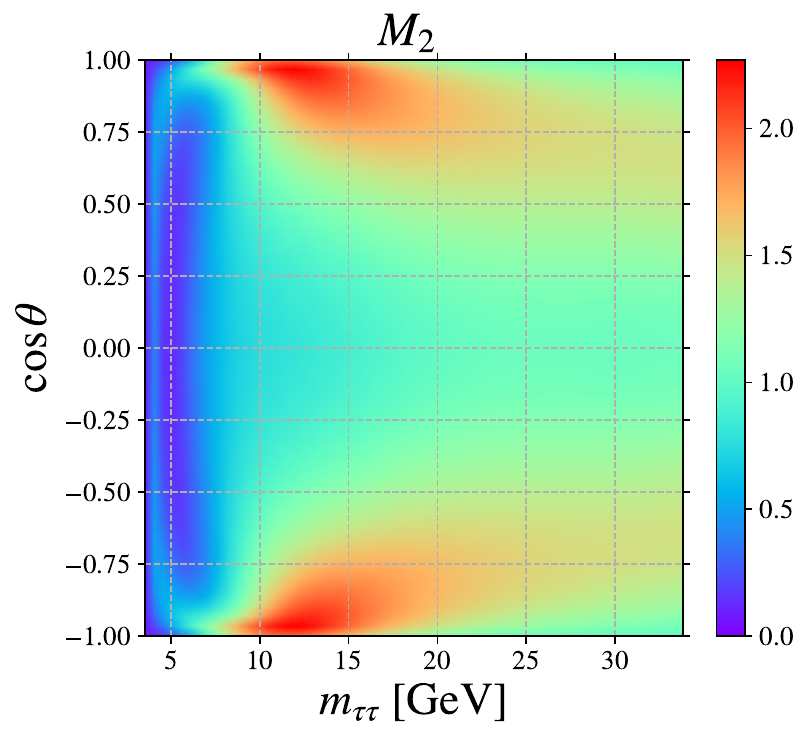}
\includegraphics[width=0.32\textwidth]{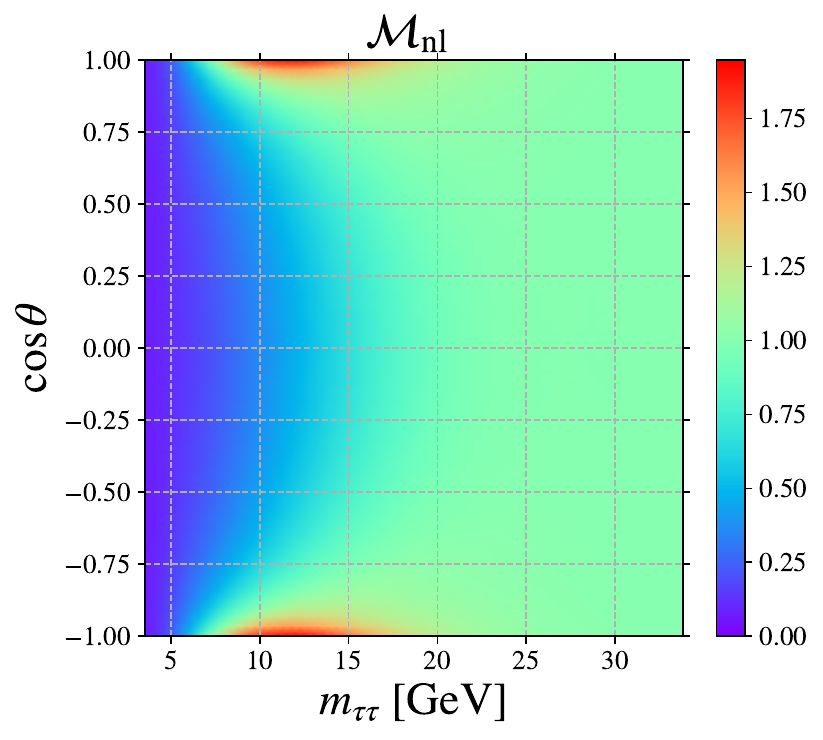}
\caption{\small
Non-stabiliserness of the $h \to \tau^- \tau^+ Z$ state in the Standard Model.
\emph{Left}: the asymmetric stabiliser R\'enyi entropy $M_2 ( \ket\psi )$ of Eq.~\eqref{eq:M2def}, evaluated in the helicity basis.
\emph{Right}: the non-local magic $\mathcal{M}_{\rm nl} ( \ket\psi )$ of Eq.~\eqref{eq:NNdef}.
Both are bounded by $\log_2 ( 13/2 ) \simeq 2.70$, cf.\ Eq.~\eqref{eq:SICbound}.
 }
\label{fig:M2}
\end{figure}

Figure~\ref{fig:M2} shows the helicity-basis SRE $M_2$ (left) and the non-local magic $\mathcal{M}_{\rm nl}$ (right) across the phase space of $h \to \tau^- \tau^+ Z$ in the Standard Model.
Both quantities vanish towards the $\tau\tau$ threshold and grow with $m_{\tau\tau}$; the non-local magic settles onto a plateau $\mathcal{M}_{\rm nl} \simeq 1.0$ over the upper half of $m_{\tau\tau}$, nearly independently of $\cos\theta$, and both maps exhibit hot spots in the collinear regions around $m_{\tau\tau} \simeq 12$~GeV, where $\mathcal{M}_{\rm nl}$ reaches $\simeq 1.9$.
All values stay well below the reference scale $\log_2 ( 13/2 ) \simeq 2.70$ of Eq.~\eqref{eq:SICbound}.
As for the observables of the previous sections, this structure can be understood analytically from the limiting states of Sec.~\ref{sec:0-223}.

At the threshold, the state \eqref{eq:threshold-state} is a product of a two-qubit state and the frozen qutrit $\ket{0}$; the latter is a computational-basis, hence stabiliser, state, so by additivity all magic resides in the fermion pair.
The non-local magic then follows from the two-qubit formula \eqref{eq:NN2qubit} with $e = C_{AB}^2 / 4$, where $C_{AB}$ is the threshold concurrence \eqref{eq:CAB-thr}:
\be
\mathcal{M}_{\rm nl} \;\to\; - \log_2 \left( 1 - C_{AB}^2 + C_{AB}^4 \right)
\qquad ( m_{AB} \to 2 m_f )\,.
\label{eq:NN-thr}
\ee
Numerically $\mathcal{M}_{\rm nl} \simeq 0.03$: the threshold state is almost a Bell pair ($C_{AB} \simeq \hat s$ close to $1$), and Eq.~\eqref{eq:NN-thr} vanishes at $C_{AB} = 1$ exactly --- maximal entanglement is stabiliser entanglement.
The helicity-basis $M_2$ admits a closed form as well; at strict threshold ($\beta = 0$),
\be
M_2 \;\to\; - \log_2 \left[\,
\Big( \tfrac{ 1 + \hat s }{2} \Big)^{\!4}
+ \frac{7}{8}\, \hat c^4 \left( 1 - \tfrac{1}{2} \sin^2 2\theta \right)
+ \Big( \tfrac{ 1 - \hat s }{2} \Big)^{\!4} \left( \cos^4 2\theta + \sin^2 2\theta \right)
\right],
\label{eq:M2-thr}
\ee
which is $O( \hat c^4 )$-small and nearly angle-independent, in agreement with the dark left edge of both panels.

At the upper endpoint ($m_{AB} \to m_{AB}^{\rm max}$) the state reduces to the generalised GHZ state $\ket{ \psi_{\rm GHZ} ( \delta ) }$ of Eq.~\eqref{eq:psi_upper} with mixing angle $\sin 2\delta = \hat s$.
As noted in Sec.~\ref{sec:0-223}, this state is effectively a $2 \otimes 2 \otimes 2$ state \emph{embedded} in the $2 \otimes 2 \otimes 3$ Hilbert space: the third party occupies only the two-dimensional subspace of the qutrit spanned by the orthogonal pointer states $\ket{ \Psi_\pm^* }$.
For the SRE, which is defined with respect to the Heisenberg--Weyl basis \eqref{eq:asymW} of the \emph{full} $2 \otimes 2 \otimes 3$ space, this embedding is not innocuous, and has a striking consequence.
To evaluate the non-local magic, recall that the minimisation in Eq.~\eqref{eq:NNdef} sweeps all local-bases, so the $M_2$ of \emph{any} basis provides an upper bound on $\mathcal{M}_{\rm nl}$.
A convenient trial basis is the canonical one, obtained from $\ket{ \psi_{\rm GHZ} ( \delta ) }$ by rotating the pointer states onto computational-basis states, $\cos\delta \ket{ 0 0 }\ket{0} + \sin\delta \ket{ 1 1 }\ket{1}$, where the Heisenberg--Weyl sum can be carried out explicitly:
\be
M_2^{\rm canonical} \,=\, - \log_2 \left( 1 - \sin^2 2\delta + \tfrac{1}{2} \sin^4 2\delta \right).
\label{eq:M2-GHZ23}
\ee
We have verified numerically that no other local-unitary position yields a smaller value, i.e.\ the bound is saturated and Eq.~\eqref{eq:M2-GHZ23} equals $\mathcal{M}_{\rm nl}$.
For a genuine three-\emph{qubit} system the same calculation yields $- \log_2 ( 1 - \sin^2 2\delta + \sin^4 2\delta )$, which vanishes for the maximally entangled GHZ state ($\delta = \pi/4$), a stabiliser state.
In the $2 \otimes 2 \otimes 3$ system the coefficient of the quartic term is halved, and the maximally entangled GHZ state carries exactly \emph{one bit} of non-local magic.
With $\sin 2\delta = \hat s$, Eq.~\eqref{eq:M2-GHZ23} gives $\mathcal{M}_{\rm nl} \simeq 1.0$ --- the plateau dominating the right panel of Fig.~\ref{fig:M2}.

In the bulk, the local-unitary invariance of $\mathcal{M}_{\rm nl}$ implies that, at $O( \varepsilon^0 )$, it can depend on the kinematics only through the overlap ratio $Q/P$, exactly as the Bell observables of Sec.~\ref{sec:Bell-results}: it decreases monotonically from the embedded-GHZ value \eqref{eq:M2-GHZ23} at $Q/P \to 0$ to the two-qubit value \eqref{eq:NN-thr} at $Q/P \to 1$, which explains both the overall gradient of the right panel and its weak dependence on $\cos\theta$.
The helicity-basis $M_2$, by contrast, is \emph{not} a local-unitary invariant: it depends on the explicit orientation of the pointer states $\ket{\Psi_\pm}$ in the helicity basis, which varies with both $m_{\tau\tau}$ and $\theta$.
The left panel accordingly shows a richer angular structure, the excess $M_2 - \mathcal{M}_{\rm nl}$ being locally generated, and locally removable, magic.

Finally, consider the collinear regions.
For the collinear states \eqref{eq:collinear-state} the Heisenberg--Weyl sum can again be carried out in closed form,
\be
M_2 ( \kappa ) \,=\, - \log_2 \left[
\frac{ ( 1 - \hat s^2 ) \left( 1 + 6 \kappa^4 + \kappa^8 \right)
+ \hat s^4 \left( \tfrac{1}{2} + 3 \kappa^4 + \kappa^8 \right) }{ ( 1 + \kappa^2 )^4 }
\right],
\label{eq:M2-collinear}
\ee
and we find numerically that along this family the helicity basis is already optimal, $M_2 ( \kappa ) = \mathcal{M}_{\rm nl} ( \kappa )$: the collinear states carry no locally removable magic.
Equation~\eqref{eq:M2-collinear} interpolates between the embedded-GHZ value \eqref{eq:M2-GHZ23} at $\kappa = 0$ and the two-qubit value \eqref{eq:NN-thr} (with $C_{AB} = \hat s$) at $\kappa \to \infty$, and is maximised at $\kappa^2 = 1/2$ in the $\hat c \to 0$ limit, where
\be
\mathcal{M}_{\rm nl}^{\max} \,=\, \log_2 \frac{27}{7} \,\simeq\, 1.95\,.
\label{eq:NN-max}
\ee
This is the origin of the hot spots: $\kappa^2 = 1/2$ corresponds to $m_{\tau\tau} \simeq 12$~GeV, precisely where the red regions of Fig.~\ref{fig:M2} sit, and it is the \emph{same} balance point at which the hyperdeterminant peaks in Sec.~\ref{sec:genuine223} --- maximal non-local magic and maximal genuine tripartite entanglement occur together, where all three qutrit polarisations are populated most evenly.

\section{Experimental prospects}
\label{sec:rate}

\begin{figure}[t!]
\centering
\includegraphics[width=0.4\textwidth]{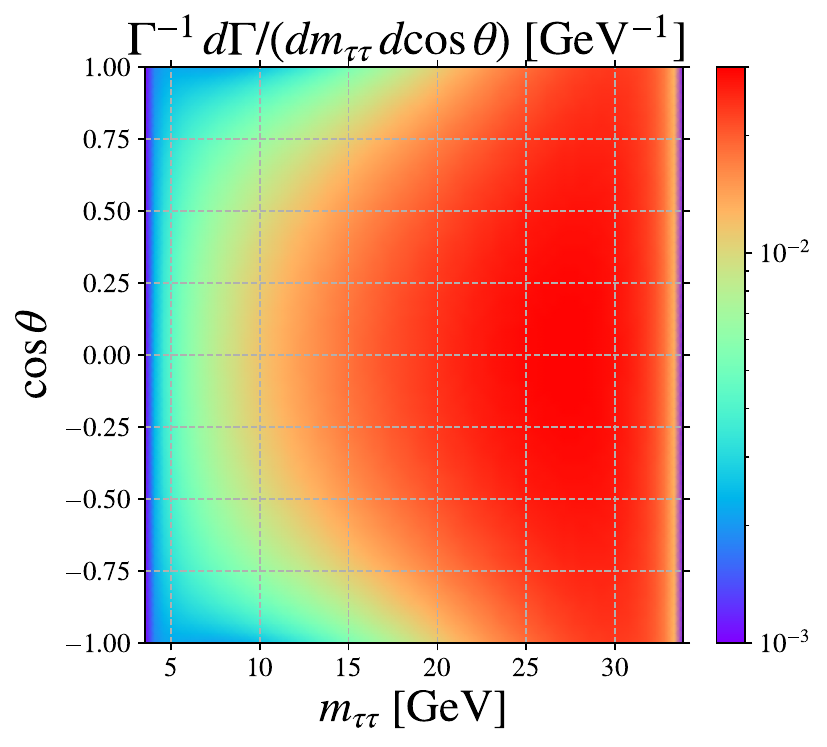}
\caption{\small
The normalised differential decay rate of $h \to \tau^- \tau^+ Z$ in the Standard Model over the $( m_{\tau\tau}, \cos\theta )$ plane (logarithmic colour scale).
The distribution is even in $\cos\theta$, in accordance with the CP relation \eqref{eq:cp-obs}, and peaks in the near-endpoint region $m_{\tau\tau} \simeq 25$--$30$~GeV.
 }
\label{fig:dGamma}
\end{figure}

The preceding sections mapped the entanglement, Bell nonlocality and non-stabiliserness of the $h \to \tau^- \tau^+ Z$ state across its phase space; we now ask where in the $( m_{\tau\tau}, \cos\theta )$ plane the decays actually occur, and what event samples can be expected.
Figure~\ref{fig:dGamma} shows the normalised differential decay rate 
$\frac{1}{ \Gamma }\frac{d\Gamma}{d m_{\tau\tau}\, d\!\cos\theta }$, computed at tree level in the Standard Model.
The rate is even in $\cos\theta$, as required by the CP relation of Sec.~\ref{sec:cp}; it is suppressed towards the di-tau threshold and in the last GeV before the upper endpoint, where the phase space closes, and peaks broadly at $m_{\tau\tau} \simeq 25$--$30$~GeV and central angles.
The statistics therefore concentrate in the near-endpoint region --- precisely where the state approaches the generalised GHZ state \eqref{eq:psi_upper}, the Bell observables their near-maximal value, and the non-local magic its plateau.
Moreover, the collinear regions around $m_{\tau\tau} \simeq 10$--$12$~GeV, which host the most distinctive $2 \otimes 2 \otimes 3$ structures --- the maxima of $C_{V|AB}$, of the $223$-tangle and of the non-local magic --- are rate-suppressed by a factor of a few relative to the peak.

A rough estimate of the available event yields is as follows.
At $\sqrt{s} = 13.6$~TeV the inclusive Higgs production cross section is $\sigma ( pp \to h ) \simeq 55$~pb \cite{LHCHiggsCrossSectionWorkingGroup:2016ypw}, and the branching ratio of the decay considered here is ${\rm BR} ( h \to \tau^- \tau^+ Z ) \approx {\rm BR} ( h \to Z Z^* ) \times {\rm BR} ( Z^* \to \tau\tau ) \sim 10^{-3}$ \cite{LHCHiggsCrossSectionWorkingGroup:2016ypw, ParticleDataGroup:2022pth}.
With the $\sim 300\ {\rm fb}^{-1}$ per experiment collected by the end of Run~3, this corresponds to $\sim 1.5 \times 10^4$ decays, growing to $\sim 1.5 \times 10^5$ at the HL-LHC ($3\ {\rm ab}^{-1}$) \cite{ZurbanoFernandez:2020cco}.
Demanding a clean tag of the on-shell boson through $Z \to e^+ e^-, \mu^+ \mu^-$ costs a further factor $\simeq 15$ \cite{ParticleDataGroup:2022pth}, leaving $O( 10^3 )$ and $O( 10^4 )$ events, respectively, before reconstruction efficiencies.
We stress that this is an order-of-magnitude yield estimate only: accessing the observables studied in this paper requires reconstructing the $\tau^\pm$ spin correlations and the $Z$ polarisation from the decay products, and a dedicated sensitivity study --- including backgrounds and the finite $\tau$ spin-analysing power --- is beyond the scope of this work.
The numbers nevertheless indicate that a first exploration of the $2 \otimes 2 \otimes 3$ correlations may become conceivable at the HL-LHC.

An $e^+ e^-$ Higgs factory \cite{FCC:2018evy, CEPCStudyGroup:2018ghi, ILC:2013jhg, LinearColliderVision:2025hlt} would collect a much smaller but far cleaner sample.
At $\sqrt{s} \simeq 240$--$250$~GeV the Higgsstrahlung cross section is $\sigma ( e^+ e^- \to Z h ) \simeq 0.2$~pb, so an integrated luminosity of $5$--$10\ {\rm ab}^{-1}$, as foreseen at the circular colliders, delivers $1$--$2 \times 10^6$ Higgs bosons and hence only $\sim 10^3$ $h \to \tau^- \tau^+ Z$ decays.
Each event, however, carries much more information than at a hadron collider: the recoil against the Higgsstrahlung $Z$ tags the event independently of the decay, all decay modes of the on-shell $Z$ become usable, backgrounds are small, and the kinematic constraints from the known initial state allow a full reconstruction of the $\tau$ momenta, from which the spin correlations can be extracted \cite{Altakach:2022ywa, Jeans:2026eys}.
A qualitatively better measurement per event may therefore compensate for the smaller yield.

\section{Conclusion}
\label{sec:concl}

We have presented a comprehensive analysis of the quantum correlations carried by the qubit--qubit--qutrit spin state produced in the three-body Higgs decay $h \to \tau^- \tau^+ Z$.
The decay of an on-shell scalar delivers a pure state.
Organising the tree-level amplitude as an expansion in the small fermion mass exposes a simple underlying structure.
At leading order the state is a GHZ-like superposition of two branches, in which the fermion helicities are perfectly correlated with two vector-boson pointer states, and the overlap of the pointer states is controlled by the kinematics.
The state simplifies further at the boundaries of the phase space: at the di-tau threshold, in the collinear limits and at the upper endpoint of the di-tau mass spectrum.
In these regions every observable considered in this paper is given by a compact closed formula.
These formulas reproduce all the main features of the exact numerical maps.

The entanglement content of the state is rich at every level.
The one-to-other concurrences of the two fermions, $C_{A|BV}$ and $C_{B|AV}$, are nearly maximal throughout, being fixed at leading order by the chiral couplings alone, while $C_{V|AB}$ traces the kinematic competition between the two pointer states and, in the collinear regions, becomes nearly maximal.
The one-to-one measures exhibit a monogamy-like trade-off: the fermion--boson entanglement peaks precisely where the fermion-pair entanglement $C_{AB}$ is quenched, and vice versa.
The genuinely tripartite correlations, quantified by the Miyake classification, the hyperdeterminant and the $223$-tangle, peak in the collinear regions where the three polarisations of the vector boson are populated most evenly.
There the state attains its maximal genuine $2 \otimes 2 \otimes 3$ character, a feature invisible to any bipartite diagnostic.

Two of our results go beyond the existing quantum information literature.
First, we derived, for the first time, semi-analytical expressions for the tight $4 \times 4 \times 2$ Bell inequalities of the $2 \otimes 2 \otimes d$ system, eliminating all qubit settings in closed form and reducing the qudit optimisation to a low-dimensional compact search.
Using this method, we found that for the $2 \otimes 2 \otimes 3$ states resulting from the considered decay the local-hidden-variable bound is violated over the entire phase space, from a purely bipartite CHSH-type violation at the threshold to a near-maximal tripartite violation at the endpoint, where $\mathcal{B}_{442}$, $\mathcal{B}_{424}$ and $\mathcal{B}_{244}$ all reach within a few per cent of the quantum bound.
Second, we extended the stabiliser R\'enyi entropy and the non-local magic to Hilbert spaces with unequal local dimensions, verifying that their defining properties survive the generalisation.
The asymmetric setting turns out to have a character of its own.
A generalised GHZ state (which is effectively three-qubit) embedded in the $2 \otimes 2 \otimes 3$ system carries almost exactly one bit of magic that no local unitaries can remove; this magic is generated by the dimensional mismatch itself.
Moreover, the non-local magic of the $h \to \tau^- \tau^+ Z$ spin state peaks at $\mathcal{M}_{\rm nl} = \log_2 \frac{27}{7} \simeq 1.95$ in the same collinear regions that maximise the genuine tripartite entanglement.
We stress that maximal entanglement does not imply large magic: the Bell and GHZ states of qubit systems are stabiliser states and carry none.
In our process, sizeable non-local magic emerges from the combination of the chiral couplings, the kinematics, and the mismatch between the effectively two-dimensional spin state of the $Z$ boson and its three-dimensional spin space.

The measurement prospects appear challenging, but not out of reach.
The differential decay rate concentrates precisely in the near-endpoint region where the state is closest to a generalised GHZ state, the Bell violation is near maximal and the non-local magic sits on its plateau.
With an order of $10^3$ leptonically tagged events expected at the LHC and $10^4$ at the HL-LHC, a first exploration of these correlations may become conceivable.
An $e^+ e^-$ Higgs factory would offer a complementary, higher-quality measurement on a smaller event sample, thanks to its clean environment and full kinematic reconstruction.

The methods developed here carry beyond the process at hand.
The small-fermion-mass expansion applies to any $H \to f \bar f V$ decay with generic couplings and masses, the semi-analytical Bell optimisation to any $2 \otimes 2 \otimes d$ system, and the asymmetric stabiliser R\'enyi entropy to arbitrary collections of unequal-dimensional parties.
Several questions remain open, among them the maximal magic states of the $2 \otimes 2 \otimes 3$ Heisenberg--Weyl group, the monotonicity of the stabiliser R\'enyi entropy for unequal local dimensions, and the feasibility of observing the presented quantum correlations in a realistic experimental setting.
We regard the present work as a step towards a systematic quantum-information programme for multipartite final states at colliders, in which the pure states produced in Higgs decays play a privileged role.

\section*{Acknowledgements}
MB, SSA and PH acknowledge support from the National Science Centre in Poland under the research grant Maestro (2021/42/A/ST2/00356). MB also acknowledges the support of Germany's Excellence Strategy -- Cluster of Excellence Matter and Light for Quantum Computing (ML4Q), EXC 2004/1 (390534769). ME was supported by the National Science Centre in Poland under the research grant Sonata BIS (2023/50/E/ST2/00472).

\appendix

\section{Vector boson pointer states}
\label{app:pointer}
This appendix completes the description of the sub-leading pointer states $\ket{\Phi_\pm}$ introduced in Eq.~\eqref{eq:pointer}. 
Contracting $\ket{\Phi_\pm}$ with the zeroth-order pointer states $\ket{\Psi_\pm}$ yields the cross
overlaps $\braket{\Psi_s|\Phi_{s'}}$ that may enter the next-to-leading-order expressions of quantum observables. 
They take the compact form
\ba
\langle\Psi_{s}|\Phi_{s'}\rangle &=& -\,s\,\frac{2\sqrt2\,k_V\sin\theta}{E_V\,D_+D_-}
\Bigg\{\;
\frac{c_V\,m_V}{m_V^2-m_{AB}^2}\Big[m_{AB}D\,(m_V^2-m_{AB}^2) - s s' \,k_V\cos\theta\,\mathcal J\Big]
\nn \\
&& ~~~
-
\frac{c_A}{m_V}\Big[s\,\mathcal G - s'\,k_Vm_{AB}^2m_V^2\cos\theta\Big]
\; \Bigg\}\,,
\label{eq:PsiPhi}
\ea
with $s,s' \in \{ \pm1\}$,
$\mathcal{J} \equiv 2D_+D_- + m_{AB}^2\left(m_V^2-m_{AB}^2\right)$ and
$\mathcal{G} \equiv m_{AB}m_V^2 D - \left(2E_V+m_{AB}\right)D_+D_-$.

The norms and mutual overlap of $\ket{\Phi_\pm}$ themselves are quadratic in the couplings,
\ba
\braket{ \Phi_\pm | \Phi_\pm } \,=\, \mathcal S\, c_V^2 \,+\, \mathcal T\, c_A^2 \,\pm\,\mathcal U\, c_V c_A , \qquad
\braket{ \Phi_+|\Phi_- } \,=\, \mathcal S'\, c_V^2+\mathcal T'\, c_A^2\,,
\label{eq:PhiNorm}
\ea
where the coefficients are given by
\ba
\mathcal S &=&
4\sin^2\theta\left[
\frac{4 m_V^4}{(m_V^2 - m_{AB}^2)^2}
+ \frac{4 m_{AB} m_V^2 (E_V + m_{AB})\, D}{(m_V^2 - m_{AB}^2)\, D_+ D_-}
+ \frac{m_{AB}^2 (m_H^2 + 2 k_V^2)\, D^2}{(D_+ D_-)^2}
\right]
\nn \\[4pt]
&& +\; 4 m_V^2 \cos^2\theta
\left( \frac{2 E_V}{m_V^2 - m_{AB}^2} + \frac{m_{AB} m_H^2}{D_+ D_-} \right)^{\!2} ,
\\[10pt]
\mathcal T &=&
\frac{4 m_{AB}^4 k_V^2 (m_H^2 + 2 k_V^2)\, \sin^2\theta \cos^2\theta}{(D_+ D_-)^2}
+ \frac{4 k_V^2}{m_V^2}
\left( 2 + \frac{m_{AB}^2 D \sin^2\theta}{D_+ D_-} \right)^{\!2} ,
\\[10pt]
\mathcal U &=&
\frac{8 m_{AB}^2 k_V\, \sin^2\theta \cos\theta}{D_+ D_-}
\left[
\frac{2 m_V^2 (E_V + m_{AB})}{m_V^2 - m_{AB}^2}
+ \frac{m_{AB} (m_H^2 + 2 k_V^2)\, D}{D_+ D_-}
\right]
\nn \\[4pt]
&& -\; 8 k_V \cos\theta
\left( 2 + \frac{m_{AB}^2 D \sin^2\theta}{D_+ D_-} \right)
\left( \frac{2 E_V}{m_V^2 - m_{AB}^2} + \frac{m_{AB} m_H^2}{D_+ D_-} \right) .
\ea
and
\ba
\mathcal S' &=&
-\,4\sin^2\theta\left[
\frac{4 m_V^4}{(m_V^2 - m_{AB}^2)^2}
+ \frac{4 m_{AB} m_V^2 (E_V + m_{AB})\, D}{(m_V^2 - m_{AB}^2)\, D_+ D_-}
+ \frac{m_{AB}^2 m_H^2\, D^2}{(D_+ D_-)^2}
\right]
\nn \\[4pt]
&& -\; 4 m_V^2 \cos^2\theta
\left( \frac{2 E_V}{m_V^2 - m_{AB}^2} + \frac{m_{AB} m_H^2}{D_+ D_-} \right)^{\!2} ,
\\[10pt]
\mathcal T' &=&
\frac{4 m_{AB}^4 k_V^2 m_H^2\, \sin^2\theta \cos^2\theta}{(D_+ D_-)^2}
+ \frac{4 k_V^2}{m_V^2}
\left( 2 + \frac{m_{AB}^2 D \sin^2\theta}{D_+ D_-} \right)^{\!2} .
\ea

\section{$I$-concurrence}
\label{app:I-conc}

The $I$-concurrence \cite{Rungta:2001zcj} is a natural generalisation of Wootters' two-qubit concurrence \cite{Wootters:1997id}, which is defined for a two-qubit pure state $\ket{\omega}$ as
\be
C_W( \ket{\omega} ) \, \equiv \, \sqrt{ \braket{ \omega | \tilde \sigma | \omega } },
\ee
where
$\tilde \sigma \equiv (\sigma_y \otimes \sigma_y) \sigma^* (\sigma_y \otimes \sigma_y)$ is the double spin-flipped matrix of $\sigma \,\equiv\, | \omega \ketbra \omega |$, and $\sigma^*$ denotes complex conjugation in the computational basis.
A natural intuition leading to generalising this definition to bipartite states of arbitrary dimension is that the single-qubit spin flip $\sigma_y \tau^* \sigma_y$ can be rewritten without any reference to $\sigma_y$ as
\be
\sigma_y \tau^* \sigma_y \,=\, \Tr \tau \cdot \mathds{1} - \tau
\,=\,\mathds{1} - \tau\,,
\ee
where $\tau$ is a normalised single-qubit state.

This form carries over to any dimension. An alternative, more precise derivation, based on a set of natural axioms \cite{Rungta:2001zcj}, leads to the same object --- the \emph{universal inverter} on a single system --- defined for a state $\rho$ of arbitrary dimension as
\be
S(\rho) \,\equiv\,
\Tr \rho \cdot \mathds{1} - \rho\,.
\ee
For a normalised bipartite state $\rho$ acting on $\Hil =\Hil_X \otimes \Hil_Y$,
the double spin-flipped matrix is then
\be
\tilde \rho \,=\, (S_X \otimes S_Y)(\rho)
\,=\,
\mathds{1}_{XY} \,-\, \rho_X \otimes \mathds{1}_Y \,-\,
\mathds{1}_X \otimes \rho_Y
\,+\, \rho\,,
\label{eq:spin-flip}
\ee
where $\rho_X$ and $\rho_Y$ are the reduced density matrices of subsystems $X$ and $Y$.
For the two-qubit case this $\tilde \rho$ reduces to
$(\sigma_y \otimes \sigma_y) \rho^* (\sigma_y \otimes \sigma_y)$.
Equation~\eqref{eq:spin-flip} presupposes that $\rho$ is a state, i.e.\ hermitian and of unit trace.
For an arbitrary operator $O$ on $\Hil_X \otimes \Hil_Y$ the same construction reads
\be
\tilde O \,\equiv\, \Tr\big( O^\dagger \big)\, \mathds{1}_{XY}
\,-\, \big( \Tr_Y O^\dagger \big) \otimes \mathds{1}_Y
\,-\, \mathds{1}_X \otimes \big( \Tr_X O^\dagger \big)
\,+\, O^\dagger \,,
\label{eq:inverter-general}
\ee
and reduces to Eq.~\eqref{eq:spin-flip} when $O = \rho$.
The trace and adjoint terms, immaterial for a state, are essential in Sec.~\ref{sec:one-to-one}, where the inverter is applied to operators that are neither hermitian nor of unit trace.
Wootters' two-qubit concurrence is then naturally extended to the $I$-concurrence with the following definition
\be
C_I( \ket{ \psi } ) \,\equiv\, \sqrt{ \bra{ \psi } \tilde \rho \ket{ \psi } },
\qquad
\tilde \rho \,=\, (S_X \otimes S_Y)( |\psi \ketbra \psi | )\,.
\ee
A direct calculation shows that the $I$-concurrence can equivalently be expressed through the reduced state alone,
\be
C_I( \ket{ \psi } ) \,=\, \sqrt{2 S_L(\rho_X)}\,,
\qquad
\rho_X \,=\, \Tr_Y | \psi \ketbra \psi |\,,
\ee
where $S_L(\rho) \equiv 1 - \Tr \rho^2$ is the linear entropy.

The $I$-concurrence is known to be a faithful entanglement measure --- non-negative, and positive if and only if the state is non-separable --- and non-increasing under LOCC \cite{Demkowicz-Dobrzanski:2006wyi}.
For a bipartite system of dimension $d_X \times d_Y$ it obeys the upper bound $C_I \leq \sqrt{2(r-1)/r}$, where $r = \min(d_X, d_Y)$ is the maximal Schmidt rank.
Normalising by this bound, we define the normalised $I$-concurrence as
\be
C( \ket{ \psi } ) \,=\, \sqrt{ \tfrac{r}{r-1} (1 - \Tr \rho^2_X) }\,,
\qquad
r = \min(d_X, d_Y)\,.
\ee

\section{Miyake classification and 223-tangle}
\label{app:Miyake}

In \cite{Miyake:2003tee} Miyake used multidimensional determinants (hyperdeterminants) and their singularities in the classification of entanglement of pure multipartite states.
For a $2\times2\times3$ state
\be
\ket{\psi} = \sum_{i,j=0}^{1} \sum_{k=0}^{2} a_{ijk} \ket{ijk},
\label{eq_app:Miyake_state}
\ee
the hyperdeterminant of the matrix $A(\psi)=[a_{ijk}]$ is defined as
\be
\mathrm{HDet}(\psi) = m_1(\psi) m_4(\psi) - m_2(\psi) m_3(\psi),
\label{app:Miyake_HDet_def}
\ee
where $m_i(\psi)$, $i=1,\dots,4$ is the $3\times3$ minor of the matrix
\be
\begin{pmatrix}
    a_{000} & a_{010} & a_{100} & a_{110}\\
    a_{001} & a_{011} & a_{101} & a_{111}\\
    a_{002} & a_{012} & a_{102} & a_{112}
\end{pmatrix}
\ee
obtained by removing the $i$-th column.

Now, values of $\textrm{HDet}(\psi)$ and minors $m_i(\psi)$ ($i=1,\dots,4$) together with the local ranks of the state $\ket{\psi}$ defined as
\be
(r_A,r_B,r_V) = (\textrm{rank}(\Tr_{BV} | \psi \ketbra \psi |), 
\textrm{rank}(\Tr_{AV} | \psi \ketbra \psi |),
\textrm{rank}(\Tr_{AB} | \psi \ketbra \psi |)),
\label{eq_app:Miyake_local_rank_def}
\ee
enable us to classify multipartite entanglement of the state (\ref{eq_app:Miyake_state}) 
in 8 classes \cite{Miyake:2003tee}:

\begin{itemize}
    \item[--] Genuine tripartite entanglement, full rank states (genuine 223 entanglement)
    \begin{itemize}
        \item[(i)] C1 class, local ranks $(r_A,r_B,r_V)=(2,2,3)$, $\textrm{HDet}(\psi)\not=0$,\\
        representative: $\ket{000}+\ket{101}+\ket{011}+\ket{112}$.
        \item[(ii)] C2 class, local ranks $(r_A,r_B,r_V)=(2,2,3)$, $\textrm{HDet}(\psi)=0$, at least one minor $m_i(\psi)\not=0$,\\
        representative: $\ket{000}+\ket{011}+\ket{112}$.
    \end{itemize}
    \item[--] Genuine tripartite entanglement, states with reduced local ranks (222 entanglement)
    \begin{itemize}
        \item[(iii)] GHZ class, local ranks $(r_A,r_B,r_V)=(2,2,2)$, $m_i=0$ for all $i=1,\dots,4$,\\
        representative: $\ket{000}+\ket{111}$.
        \item[(iv)] W class, local ranks $(r_A,r_B,r_V)=(2,2,2)$, $m_i=0$ for all $i=1,\dots,4$,\\
        representative: $\ket{010}+\ket{100}+\ket{001}$.
    \end{itemize}
    \item[--] Biseparable states
    \begin{itemize}
        \item[(v)] B1 class, local ranks $(r_A,r_B,r_V)=(1,2,2)$, 
        $m_i=0$ for all $i=1,\dots,4$,\\
        representative: $\ket{001}+\ket{010}$.
        \item[(vi)] B2 class, local ranks $(r_A,r_B,r_V)=(2,1,2)$, 
        $m_i=0$ for all $i=1,\dots,4$,\\
        representative: $\ket{001}+\ket{100}$.
        \item[(vii)] B3 class, local ranks $(r_A,r_B,r_V)=(2,2,1)$, 
        $m_i=0$ for all $i=1,\dots,4$,\\
        representative: $\ket{010}+\ket{100}$.
    \end{itemize}
    \item[--] Completely separable states
    \begin{itemize}
        \item[(viii)] S class, local ranks $(r_A,r_B,r_V)=(1,1,1)$, 
        $m_i=0$ for all $i=1,\dots,4$,\\
        representative: $\ket{000}$. 
    \end{itemize}
\end{itemize}

All of the above classes are invariant under stochastic local operations and classical communication (SLOCC); i.e. under the action of GL(2)$\times$GL(2)$\times$GL(3) group.

In discussion of multipartite entanglement of pure $2\times2\times3$ states one also uses 
so called 223-tangle, an entanglement monotone introduced in \cite{Verstraete:2003dtr} and defined as
\begin{multline}
\tau_{223}(\psi)= \sqrt[3]{\frac{27}{4}} \Big|
\sum_{i_1,\dots,i_6=0}^1 \sum_{j_1,\dots,j_6=0}^1 \sum_{k_1,\dots,k_6=0}^2
a_{i_1 j_1 k_1} a_{i_2 j_2 k_2} a_{i_3 j_3 k_3} a_{i_4 j_4 k_4} a_{i_5 j_5 k_5} a_{i_6 j_6 k_6}\\
\epsilon_{i_1 i_4} \epsilon_{i_2 i_5} \epsilon_{i_3 i_6} 
\epsilon_{j_1 j_4} \epsilon_{j_2 j_5} \epsilon_{j_3 j_6} 
\epsilon_{k_1 k_2 k_3} \epsilon_{k_4 k_5 k_6}
\Big|^{1/3},
\label{eq_app:Miyake_tangle_def}
\end{multline}
where $\epsilon_{ij}$ and $\epsilon_{ijk}$ are completely antisymmetric Levi-Civita tensors
in 2 and 3 dimensions, respectively. However, as one can verify by a direct calculation, 
the 223-tangle is closely related to the hyperdeterminant:
\be
\tau_{223}(\psi) = 3 \sqrt[3]{3} \big|
\textrm{HDet}(\psi)
\big|^{1/3}.
\label{eq_app:Miyake_tangle_HDet}
\ee

\begin{figure}[t!]
\centering
\includegraphics[width=0.32\textwidth]{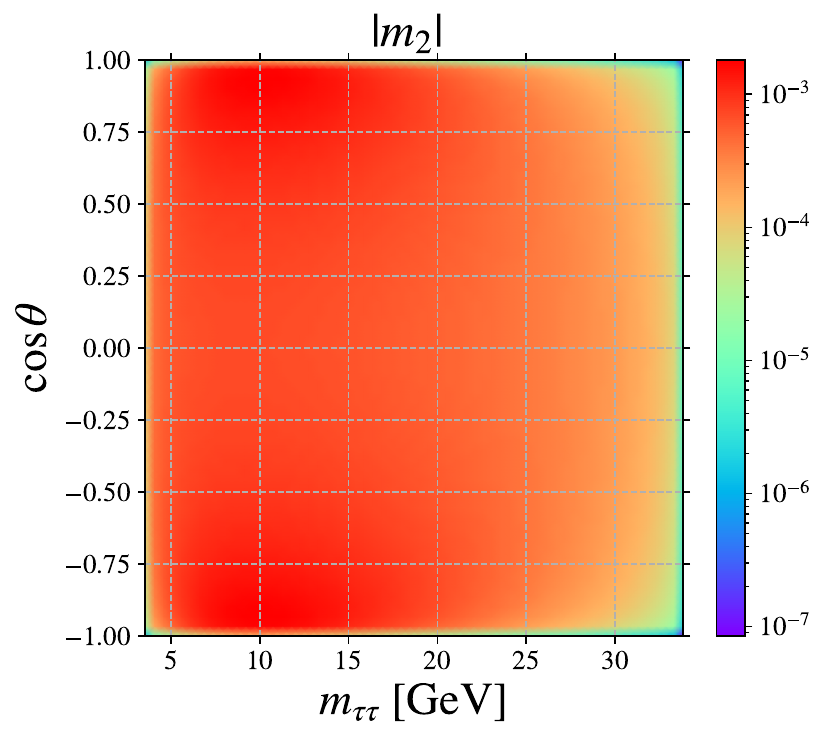}
\includegraphics[width=0.32\textwidth]{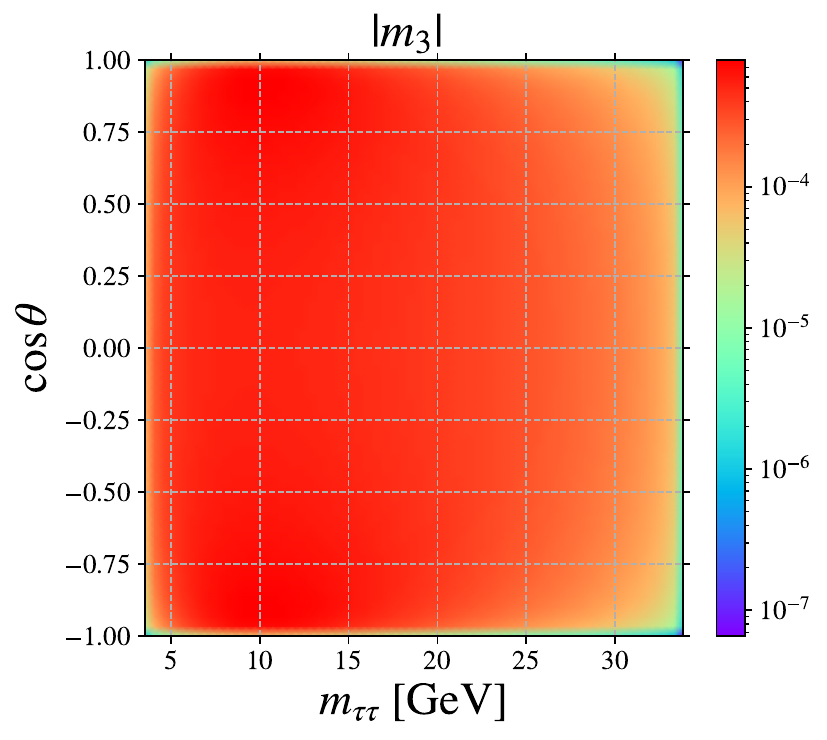}
\caption{\small
The Miyake minors $|m_2|$ (left) and $|m_3|$ (right) of the helicity-basis amplitude matrix, on the $(m_{\tau\tau},\, \cos\theta)$ plane, computed for $h \to \tau^- \tau^+ Z$ in the Standard Model at tree level to all orders in $m_f$ (logarithmic colour scales).
Both are of $O(\varepsilon^2)$, cf.\ Eq.~\eqref{eq:m14}.
 }
\label{fig:miyake-m23}
\end{figure}

For completeness, Fig.~\ref{fig:miyake-m23} shows the two minors $|m_2|$ and $|m_3|$ of the $h \to \tau^- \tau^+ Z$ spin state, omitted from Fig.~\ref{fig:miyake}: both are of $O(\varepsilon^2)$ [cf.\ Eq.~\eqref{eq:m14}] and essentially featureless across the phase space.

\section{Heisenberg--Weyl operators for the $2 \otimes 2 \otimes 3$ system}
\label{app:HW}

This appendix lists the explicit matrix representations of the Heisenberg--Weyl operators used in the stabiliser R\'enyi entropy of Sec.~\ref{sec:asym-magic}, in the conventions of Eqs.~\eqref{eq:XZdef}--\eqref{eq:HWdef}: $W_{ab} = X^a Z^b$ in the computational basis $\{ \ket{k} \}$, with no additional phase factors.
The composite basis \eqref{eq:asymW} is formed by the tensor products $W^{(A)}_{a_1 b_1} \otimes W^{(B)}_{a_2 b_2} \otimes W^{(V)}_{a_3 b_3}$, in this order.

For each qubit ($d = 2$, $\omega = -1$) the four operators are
\be
W_{00} = \begin{pmatrix} 1 & 0 \\ 0 & 1 \end{pmatrix} , \quad
W_{01} = \begin{pmatrix} 1 & 0 \\ 0 & -1 \end{pmatrix} , \quad
W_{10} = \begin{pmatrix} 0 & 1 \\ 1 & 0 \end{pmatrix} , \quad
W_{11} = \begin{pmatrix} 0 & -1 \\ 1 & 0 \end{pmatrix} ,
\label{eq:HWqubit}
\ee
i.e.\ $\{ \mathds{1}, \sigma_z, \sigma_x, - i \sigma_y \}$: the Pauli matrices up to a phase, which is immaterial in Eq.~\eqref{eq:XiW}.

For the qutrit ($d = 3$, $\omega = e^{ 2 \pi i / 3 }$) the shift and clock generators read
\be
X \,=\, \begin{pmatrix} 0 & 0 & 1 \\ 1 & 0 & 0 \\ 0 & 1 & 0 \end{pmatrix} ,
\qquad
Z \,=\, \begin{pmatrix} 1 & 0 & 0 \\ 0 & \omega & 0 \\ 0 & 0 & \omega^2 \end{pmatrix} ,
\label{eq:XZqutrit}
\ee
and the nine operators $W_{ab} = X^a Z^b$, with matrix elements $[ W_{ab} ]_{jk} = \omega^{ b k }\, \delta_{ j,\, k + a \,\mathrm{mod}\, 3 }$, are
\ba
W_{00} &=& \begin{pmatrix} 1 & 0 & 0 \\ 0 & 1 & 0 \\ 0 & 0 & 1 \end{pmatrix} , \quad
W_{01} \,=\, \begin{pmatrix} 1 & 0 & 0 \\ 0 & \omega & 0 \\ 0 & 0 & \omega^2 \end{pmatrix} , \quad
W_{02} \,=\, \begin{pmatrix} 1 & 0 & 0 \\ 0 & \omega^2 & 0 \\ 0 & 0 & \omega \end{pmatrix} ,
\nn \\
W_{10} &=& \begin{pmatrix} 0 & 0 & 1 \\ 1 & 0 & 0 \\ 0 & 1 & 0 \end{pmatrix} , \quad
W_{11} \,=\, \begin{pmatrix} 0 & 0 & \omega^2 \\ 1 & 0 & 0 \\ 0 & \omega & 0 \end{pmatrix} , \quad
W_{12} \,=\, \begin{pmatrix} 0 & 0 & \omega \\ 1 & 0 & 0 \\ 0 & \omega^2 & 0 \end{pmatrix} ,
\nn \\
W_{20} &=& \begin{pmatrix} 0 & 1 & 0 \\ 0 & 0 & 1 \\ 1 & 0 & 0 \end{pmatrix} , \quad
W_{21} \,=\, \begin{pmatrix} 0 & \omega & 0 \\ 0 & 0 & \omega^2 \\ 1 & 0 & 0 \end{pmatrix} , \quad
W_{22} \,=\, \begin{pmatrix} 0 & \omega^2 & 0 \\ 0 & 0 & \omega \\ 1 & 0 & 0 \end{pmatrix} .
\label{eq:HWqutrit}
\ea
One checks directly that $\Tr ( W_{ab}^\dagger W_{a'b'} ) = d\, \delta_{aa'} \delta_{bb'}$ within each factor, so that the $4 \cdot 4 \cdot 9 = 144$ composite operators satisfy $\Tr ( W^\dagger W' ) = 12\, \delta_{W W'}$ and form a complete orthogonal basis of the operator space, as used in Eq.~\eqref{eq:sumrule}.

\bibliography{reference}

@article{LANDAU198754,
title = {On the violation of Bell's inequality in quantum theory},
journal = {Physics Letters A},
volume = {120},
number = {2},
pages = {54-56},
year = {1987},
issn = {0375-9601},
doi = {https://doi.org/10.1016/0375-9601(87)90075-2},
url = {https://www.sciencedirect.com/science/article/pii/0375960187900752},
author = {Lawrence J. Landau}
}

@article{Rungta:2001zcj,
    author = "Rungta, Pranaw and Bu{\v{z}}ek, V. and Caves, Carlton M. and Hillery, M. and Milburn, G. J.",
    title = "{Universal state inversion and concurrence in arbitrary dimensions}",
    eprint = "quant-ph/0102040",
    archivePrefix = "arXiv",
    doi = "10.1103/PhysRevA.64.042315",
    journal = "Phys. Rev. A",
    volume = "64",
    number = "4",
    pages = "042315",
    year = "2001"
}

@article{Demkowicz-Dobrzanski:2006wyi,
    author = "Demkowicz-Dobrzanski, Rafal and Buchleitner, Andreas and Kus, Marek and Mintert, Florian",
    title = "{Evaluable multipartite entanglement measures: are multipartite concurrences entanglement monotones?}",
    eprint = "quant-ph/0607084",
    archivePrefix = "arXiv",
    doi = "10.1103/PhysRevA.74.052303",
    journal = "Phys. Rev. A",
    volume = "74",
    pages = "052303",
    year = "2006"
}

@article{Osborne:2002vcf,
    author = "Osborne, Tobias J.",
    title = "{Entanglement for rank-2 mixed states}",
    eprint = "quant-ph/0203087",
    archivePrefix = "arXiv",
    doi = "10.1103/PhysRevA.72.022309",
    journal = "Phys. Rev. A",
    volume = "72",
    pages = "022309",
    year = "2005"
}

@article{Wootters:1997id,
    author = "Wootters, William K.",
    title = "{Entanglement of formation of an arbitrary state of two qubits}",
    eprint = "quant-ph/9709029",
    archivePrefix = "arXiv",
    doi = "10.1103/PhysRevLett.80.2245",
    journal = "Phys. Rev. Lett.",
    volume = "80",
    pages = "2245--2248",
    year = "1998"
}

@article{Fabbrichesi:2022ovb,
    author = "Fabbrichesi, Marco and Floreanini, Roberto and Gabrielli, Emidio",
    title = "{Constraining new physics in entangled two-qubit systems: top-quark, tau-lepton and photon pairs}",
    eprint = "2208.11723",
    archivePrefix = "arXiv",
    primaryClass = "hep-ph",
    doi = "10.1140/epjc/s10052-023-11307-2",
    journal = "Eur. Phys. J. C",
    volume = "83",
    number = "2",
    pages = "162",
    year = "2023"
}

@article{Qian:2025oit,
    author = "Qian, Dongheng and Wang, Jing",
    title = "{Quantum nonlocal nonstabilizerness}",
    eprint = "2502.06393",
    archivePrefix = "arXiv",
    primaryClass = "quant-ph",
    doi = "10.1103/PhysRevA.111.052443",
    journal = "Phys. Rev. A",
    volume = "111",
    number = "5",
    pages = "052443",
    year = "2025"
}

@article{Leone:2021rzd,
    author = "Leone, Lorenzo and Oliviero, Salvatore F. E. and Hamma, Alioscia",
    title = "{Stabilizer R{\'e}nyi Entropy}",
    eprint = "2106.12587",
    archivePrefix = "arXiv",
    primaryClass = "quant-ph",
    doi = "10.1103/PhysRevLett.128.050402",
    journal = "Phys. Rev. Lett.",
    volume = "128",
    number = "5",
    pages = "050402",
    year = "2022"
}

@article{Wang:2023uog,
    author = "Wang, Yiran and Li, Yongming",
    title = "{Stabilizer R{\'e}nyi entropy on qudits}",
    doi = "10.1007/s11128-023-04186-9",
    journal = "Quant. Inf. Proc.",
    volume = "22",
    number = "12",
    pages = "444",
    year = "2023"
}

@article{Verstraete:2003dtr,
    author = "Verstraete, Frank and Dehaene, Jeroen and Moor, Bart De",
    title = "{Normal forms and entanglement measures for multipartite quantum states}",
    eprint = "quant-ph/0105090",
    archivePrefix = "arXiv",
    doi = "10.1103/PhysRevA.68.012103",
    journal = "Phys. Rev. A",
    volume = "68",
    number = "1",
    pages = "012103",
    year = "2003"
}

@article{Miyake:2003tee,
    author = "Miyake, Akimasa",
    title = "{Classification of multipartite entangled states by multidimensional determinants}",
    eprint = "quant-ph/0206111",
    archivePrefix = "arXiv",
    doi = "10.1103/PhysRevA.67.012108",
    journal = "Phys. Rev. A",
    volume = "67",
    number = "1",
    pages = "012108",
    year = "2003"
}

@article{Altakach:2022ywa,
    author = "Altakach, Mohammad Mahdi and Lamba, Priyanka and Maltoni, Fabio and Mawatari, Kentarou and Sakurai, Kazuki",
    title = "{Quantum information and CP measurement in $H \to \tau^+ \tau^-$ at future lepton colliders}",
    eprint = "2211.10513",
    archivePrefix = "arXiv",
    primaryClass = "hep-ph",
    doi = "10.1103/PhysRevD.107.093002",
    journal = "Phys. Rev. D",
    volume = "107",
    number = "9",
    pages = "093002",
    year = "2023"
}

@article{Lee:2026wlk,
    author = "Lee, Lawrence and Lawless, John and Riggall, Caroline",
    title = "{Higgs Boson Spookiness: Probing Quantum Nonlocality with Spacetime-Resolved $H\rightarrow \tau^+\tau^-$ Decays}",
    eprint = "2603.28868",
    archivePrefix = "arXiv",
    primaryClass = "hep-ph",
    month = "3",
    year = "2026"
}

@article{Ma:2023yvd,
    author = "Ma, Kai and Li, Tong",
    title = "{Testing Bell inequality through ${\boldsymbol h{\bf\rightarrow}\boldsymbol\tau\boldsymbol\tau }$ at CEPC*}",
    eprint = "2309.08103",
    archivePrefix = "arXiv",
    primaryClass = "hep-ph",
    doi = "10.1088/1674-1137/ad62d8",
    journal = "Chin. Phys. C",
    volume = "48",
    number = "10",
    pages = "103105",
    year = "2024"
}

@article{Ehataht:2023zzt,
    author = {Ehat{\"a}ht, Karl and Fabbrichesi, Marco and Marzola, Luca and Veelken, Christian},
    title = "{Probing entanglement and testing Bell inequality violation with $e^+ e^- \to \tau^+ \tau^-$ at Belle II}",
    eprint = "2311.17555",
    archivePrefix = "arXiv",
    primaryClass = "hep-ph",
    doi = "10.1103/PhysRevD.109.032005",
    journal = "Phys. Rev. D",
    volume = "109",
    number = "3",
    pages = "032005",
    year = "2024"
}

@article{Han:2025ewp,
    author = "Han, Tao and Low, Matthew and Su, Youle",
    title = "{Entanglement and Bell nonlocality in {\ensuremath{\tau}}$^{+}${\ensuremath{\tau}}$^{-}$ at the BEPC}",
    eprint = "2501.04801",
    archivePrefix = "arXiv",
    primaryClass = "hep-ph",
    reportNumber = "PITT-PACC-2412",
    doi = "10.1007/JHEP10(2025)217",
    journal = "JHEP",
    volume = "10",
    pages = "217",
    year = "2025"
}

@article{Jeans:2026eys,
    author = "Jeans, Daniel",
    title = "{Experimental aspects of the quantum tomography of tau lepton pairs at a Higgs factory collider}",
    eprint = "2603.12609",
    archivePrefix = "arXiv",
    primaryClass = "hep-ex",
    doi = "10.1140/epjc/s10052-026-15804-y",
    journal = "Eur. Phys. J. C",
    volume = "86",
    number = "5",
    pages = "549",
    year = "2026"
}

@article{Zhang:2025mmm,
    author = "Zhang, Yulei and Zhou, Bai-Hong and Liu, Qi-Bin and Wu, Tong Arthur and Li, Shu and Han, Tao and Hsu, Shih-Chieh and Low, Matthew",
    title = "{Entanglement and Bell nonlocality in {\ensuremath{\tau}}$^{+}${\ensuremath{\tau}}$^{-}$ at the LHC using machine learning for neutrino reconstruction}",
    eprint = "2504.01496",
    archivePrefix = "arXiv",
    primaryClass = "hep-ph",
    doi = "10.1007/JHEP04(2026)190",
    journal = "JHEP",
    volume = "04",
    pages = "190",
    year = "2026"
}

@article{Guhne:2008qic,
    author = {G{\"u}hne, Otfried and T{\'o}th, G{\'e}za},
    title = "{Entanglement detection}",
    eprint = "0811.2803",
    archivePrefix = "arXiv",
    primaryClass = "quant-ph",
    doi = "10.1016/j.physrep.2009.02.004",
    journal = "Phys. Rept.",
    volume = "474",
    pages = "1--75",
    year = "2009"
}

@article{Afik:2022kwm,
    author = "Afik, Yoav and de Nova, Juan Ram{\'o}n Mu{\~n}oz",
    title = "{Quantum information with top quarks in QCD}",
    eprint = "2203.05582",
    archivePrefix = "arXiv",
    primaryClass = "quant-ph",
    doi = "10.22331/q-2022-09-29-820",
    journal = "Quantum",
    volume = "6",
    pages = "820",
    year = "2022"
}

@article{Cheng:2023qmz,
    author = "Cheng, Kun and Han, Tao and Low, Matthew",
    title = "{Optimizing fictitious states for Bell inequality violation in bipartite qubit systems with applications to the $t\bar{t}$ system}",
    eprint = "2311.09166",
    archivePrefix = "arXiv",
    primaryClass = "hep-ph",
    reportNumber = "PITT-PACC-2321",
    doi = "10.1103/PhysRevD.109.116005",
    journal = "Phys. Rev. D",
    volume = "109",
    number = "11",
    pages = "116005",
    year = "2024"
}

@article{Aguilar-Saavedra:2022wam,
    author = "Aguilar-Saavedra, J. A. and Bernal, A. and Casas, J. A. and Moreno, J. M.",
    title = "{Testing entanglement and Bell inequalities in H\textrightarrow{}ZZ}",
    eprint = "2209.13441",
    archivePrefix = "arXiv",
    primaryClass = "hep-ph",
    doi = "10.1103/PhysRevD.107.016012",
    journal = "Phys. Rev. D",
    volume = "107",
    number = "1",
    pages = "016012",
    year = "2023"
}

@article{Fabbrichesi:2023cev,
    author = "Fabbrichesi, Marco and Floreanini, Roberto and Gabrielli, Emidio and Marzola, Luca",
    title = "{Bell inequalities and quantum entanglement in weak gauge boson production at the LHC and future colliders}",
    eprint = "2302.00683",
    archivePrefix = "arXiv",
    primaryClass = "hep-ph",
    doi = "10.1140/epjc/s10052-023-11935-8",
    journal = "Eur. Phys. J. C",
    volume = "83",
    number = "9",
    pages = "823",
    year = "2023"
}

@article{Aoude:2023hxv,
    author = "Aoude, Rafael and Madge, Eric and Maltoni, Fabio and Mantani, Luca",
    title = "{Probing new physics through entanglement in diboson production}",
    eprint = "2307.09675",
    archivePrefix = "arXiv",
    primaryClass = "hep-ph",
    reportNumber = "IRMP-CP3-23-37",
    doi = "10.1007/JHEP12(2023)017",
    journal = "JHEP",
    volume = "12",
    pages = "017",
    year = "2023"
}

@article{Ashby-Pickering:2022umy,
    author = "Ashby-Pickering, Rachel and Barr, Alan J. and Wierzchucka, Agnieszka",
    title = "{Quantum state tomography, entanglement detection and Bell violation prospects in weak decays of massive particles}",
    eprint = "2209.13990",
    archivePrefix = "arXiv",
    primaryClass = "quant-ph",
    doi = "10.1007/JHEP05(2023)020",
    journal = "JHEP",
    volume = "05",
    pages = "020",
    year = "2023"
}

@article{Grossi:2024jae,
    author = "Grossi, Michele and Pelliccioli, Giovanni and Vicini, Alessandro",
    title = "{From angular coefficients to quantum observables: a phenomenological appraisal in di-boson systems}",
    eprint = "2409.16731",
    archivePrefix = "arXiv",
    primaryClass = "hep-ph",
    reportNumber = "COMETA-2024-24, MPP-2024-183, TIF-UNIMI-2024-15",
    doi = "10.1007/JHEP12(2024)120",
    journal = "JHEP",
    volume = "12",
    pages = "120",
    year = "2024"
}

@article{Aoude:2026eeg,
    author = "Aoude, Rafael and Camacho, Jos{\'e} Manuel and Durupt, Valentin and Garc{\'\i}a-Mir, Guillermo and Maltoni, Fabio and Moreno Ll{\'a}cer, Mar{\'\i}a and Satrioni, Leonardo and Sakurai, Kazuki and Vos, Marcel",
    title = "{Experimental prospects for quantum decoherence measurements at colliders}",
    eprint = "2604.16268",
    archivePrefix = "arXiv",
    primaryClass = "hep-ph",
    month = "4",
    year = "2026"
}

@article{Cheng:2026zfb,
    author = "Cheng, Kun and Han, Tao and Singh, Harman and Su, Youle",
    title = "{Decoherence and More Coherence in the Radiative Decay of the $Z$ Boson}",
    eprint = "2607.12015",
    archivePrefix = "arXiv",
    primaryClass = "hep-ph",
    month = "7",
    year = "2026"
}

@article{Gu:2025ijz,
    author = "Gu, Jiayin and Lin, Shi-Jia and Shao, Ding Yu and Wang, Lian-Tao and Yang, Si-Xiang",
    title = "{Decoherence in high energy collisions as renormalization group flow}",
    eprint = "2510.13951",
    archivePrefix = "arXiv",
    primaryClass = "hep-ph",
    month = "10",
    year = "2025"
}

@article{Aoude:2025ovu,
    author = "Aoude, Rafael and Barr, Alan J. and Maltoni, Fabio and Satrioni, Leonardo",
    title = "{Decoherence effects in entangled fermion pairs at colliders}",
    eprint = "2504.07030",
    archivePrefix = "arXiv",
    primaryClass = "quant-ph",
    doi = "10.1103/yjgh-f2gw",
    journal = "Phys. Rev. D",
    volume = "113",
    number = "7",
    pages = "076007",
    year = "2026"
}

@article{DelGratta:2025qyp,
    author = "Del Gratta, Morgan and Fabbri, Federica and Lamba, Priyanka and Maltoni, Fabio and Pagani, Davide",
    title = "{Quantum properties of H {\textrightarrow} VV$^{*}$: precise predictions in the SM and sensitivity to new physics}",
    eprint = "2504.03841",
    archivePrefix = "arXiv",
    primaryClass = "hep-ph",
    doi = "10.1007/JHEP09(2025)013",
    journal = "JHEP",
    volume = "09",
    pages = "013",
    year = "2025"
}

@article{Afik:2025ejh,
    author = "Afik, Yoav and others",
    title = "{Quantum Information meets High-Energy Physics: Input to the update of the European Strategy for Particle Physics}",
    eprint = "2504.00086",
    archivePrefix = "arXiv",
    primaryClass = "hep-ph",
    month = "3",
    year = "2025"
}

@article{Morales:2023gow,
    author = "Morales, R. A.",
    title = "{Exploring Bell inequalities and quantum entanglement in vector boson scattering}",
    eprint = "2306.17247",
    archivePrefix = "arXiv",
    primaryClass = "hep-ph",
    doi = "10.1140/epjp/s13360-023-04784-7",
    journal = "Eur. Phys. J. Plus",
    volume = "138",
    number = "12",
    pages = "1157",
    year = "2023"
}

@article{Bi:2023uop,
    author = "Bi, Qi and Cao, Qing-Hong and Cheng, Kun and Zhang, Hao",
    title = "{New observables for testing Bell inequalities in W boson pair production}",
    eprint = "2307.14895",
    archivePrefix = "arXiv",
    primaryClass = "hep-ph",
    doi = "10.1103/PhysRevD.109.036022",
    journal = "Phys. Rev. D",
    volume = "109",
    number = "3",
    pages = "036022",
    year = "2024"
}

@article{Barr:2024djo,
    author = "Barr, Alan J. and Fabbrichesi, Marco and Floreanini, Roberto and Gabrielli, Emidio and Marzola, Luca",
    title = "{Quantum entanglement and Bell inequality violation at colliders}",
    eprint = "2402.07972",
    archivePrefix = "arXiv",
    primaryClass = "hep-ph",
    doi = "10.1016/j.ppnp.2024.104134",
    journal = "Prog. Part. Nucl. Phys.",
    volume = "139",
    pages = "104134",
    year = "2024"
}

@article{Barr:2021zcp,
    author = "Barr, Alan J.",
    title = "{Testing Bell inequalities in Higgs boson decays}",
    eprint = "2106.01377",
    archivePrefix = "arXiv",
    primaryClass = "hep-ph",
    doi = "10.1016/j.physletb.2021.136866",
    journal = "Phys. Lett. B",
    volume = "825",
    pages = "136866",
    year = "2022"
}

@article{Fabbrichesi:2023jep,
    author = "Fabbrichesi, M. and Floreanini, R. and Gabrielli, E. and Marzola, L.",
    title = "{Stringent bounds on HWW and HZZ anomalous couplings with quantum tomography at the LHC}",
    eprint = "2304.02403",
    archivePrefix = "arXiv",
    primaryClass = "hep-ph",
    doi = "10.1007/JHEP09(2023)195",
    journal = "JHEP",
    volume = "09",
    pages = "195",
    year = "2023"
}

@article{Fabbri:2023ncz,
    author = "Fabbri, Federica and Howarth, James and Maurin, Theo",
    title = "{Isolating semi-leptonic $H\rightarrow WW^{*}$decays for Bell inequality tests}",
    eprint = "2307.13783",
    archivePrefix = "arXiv",
    primaryClass = "hep-ph",
    doi = "10.1140/epjc/s10052-023-12371-4",
    journal = "Eur. Phys. J. C",
    volume = "84",
    number = "1",
    pages = "20",
    year = "2024"
}

@article{Bernal:2023ruk,
    author = "Bernal, Alexander and Caban, Pawe\l{} and Rembieli\'nski, Jakub",
    title = "{Entanglement and Bell inequalities violation in $H\rightarrow ZZ$ with anomalous coupling}",
    eprint = "2307.13496",
    archivePrefix = "arXiv",
    primaryClass = "hep-ph",
    doi = "10.1140/epjc/s10052-023-12216-0",
    journal = "Eur. Phys. J. C",
    volume = "83",
    number = "11",
    pages = "1050",
    year = "2023"
}

@article{Aguilar-Saavedra:2024whi,
    author = "Aguilar-Saavedra, J. A.",
    title = "{Tripartite entanglement in H\textrightarrow{}ZZ,WW decays}",
    eprint = "2403.13942",
    archivePrefix = "arXiv",
    primaryClass = "hep-ph",
    reportNumber = "IFT-UAM/CSIC-24-45",
    doi = "10.1103/PhysRevD.109.113004",
    journal = "Phys. Rev. D",
    volume = "109",
    number = "11",
    pages = "113004",
    year = "2024"
}

@article{Subba:2024mnl,
    author = "Subba, Amir and Rahaman, Rafiqul",
    title = "{On bipartite and tripartite entanglement at present and future particle colliders}",
    eprint = "2404.03292",
    archivePrefix = "arXiv",
    primaryClass = "hep-ph",
    month = "4",
    year = "2024"
}

@article{Bernal:2024xhm,
    author = "Bernal, Alexander and Caban, Pawe\l{} and Rembieli\'nski, Jakub",
    title = "{Entanglement and Bell inequality violation in vector diboson systems produced in decays of spin-0 particles}",
    eprint = "2405.16525",
    archivePrefix = "arXiv",
    primaryClass = "hep-ph",
    month = "5",
    year = "2024"
}

@article{Sullivan:2024wzl,
    author = "Sullivan, Matthew",
    title = "{Constraining New Physics with $h\rightarrow VV$ Tomography}",
    eprint = "2410.10980",
    archivePrefix = "arXiv",
    primaryClass = "hep-ph",
    month = "10",
    year = "2024"
}

@article{Aguilar-Saavedra:2024jkj,
    author = "Aguilar-Saavedra, J. A.",
    title = "{$H \to ZZ$ as a double-slit experiment}",
    eprint = "2411.13464",
    archivePrefix = "arXiv",
    primaryClass = "hep-ph",
    month = "11",
    year = "2024"
}

@article{Dong:2023xiw,
    author = "Dong, Zhongtian and Gon\c{c}alves, Dorival and Kong, Kyoungchul and Navarro, Alberto",
    title = "{Entanglement and Bell inequalities with boosted $t\overline{t}$}",
    eprint = "2305.07075",
    archivePrefix = "arXiv",
    primaryClass = "hep-ph",
    doi = "10.1103/PhysRevD.109.115023",
    journal = "Phys. Rev. D",
    volume = "109",
    number = "11",
    pages = "115023",
    year = "2024"
}

@article{Han:2023fci,
    author = "Han, Tao and Low, Matthew and Wu, Tong Arthur",
    title = "{Quantum Entanglement and Bell Inequality Violation in Semi-Leptonic Top Decays}",
    eprint = "2310.17696",
    archivePrefix = "arXiv",
    primaryClass = "hep-ph",
    reportNumber = "PITT-PACC-2316",
    month = "10",
    year = "2023"
}

@article{ParticleDataGroup:2022pth,
    author = "Workman, R. L. and others",
    collaboration = "Particle Data Group",
    title = "{Review of Particle Physics}",
    doi = "10.1093/ptep/ptac097",
    journal = "PTEP",
    volume = "2022",
    pages = "083C01",
    year = "2022"
}

@article{ATLAS:2023fsd,
    author = "Aad, Georges and others",
    collaboration = "ATLAS",
    title = "{Observation of quantum entanglement with top quarks at the ATLAS detector}",
    eprint = "2311.07288",
    archivePrefix = "arXiv",
    primaryClass = "hep-ex",
    reportNumber = "CERN-EP-2023-230",
    doi = "10.1038/s41586-024-07824-z",
    journal = "Nature",
    volume = "633",
    number = "8030",
    pages = "542--547",
    year = "2024"
}

@article{Maltoni:2024tul,
    author = "Maltoni, Fabio and Severi, Claudio and Tentori, Simone and Vryonidou, Eleni",
    title = "{Quantum detection of new physics in top-quark pair production at the LHC}",
    eprint = "2401.08751",
    archivePrefix = "arXiv",
    primaryClass = "hep-ph",
    doi = "10.1007/JHEP03(2024)099",
    journal = "JHEP",
    volume = "03",
    pages = "099",
    year = "2024"
}

@article{Aguilar-Saavedra:2023hss,
    author = "Aguilar-Saavedra, J. A.",
    title = "{Postdecay quantum entanglement in top pair production}",
    eprint = "2307.06991",
    archivePrefix = "arXiv",
    primaryClass = "hep-ph",
    reportNumber = "IFT-UAM/CSIC-23-92",
    doi = "10.1103/PhysRevD.108.076025",
    journal = "Phys. Rev. D",
    volume = "108",
    number = "7",
    pages = "076025",
    year = "2023"
}

@article{Cheng:2024btk,
    author = "Cheng, Kun and Han, Tao and Low, Matthew",
    title = "{Optimizing entanglement and Bell inequality violation in top antitop events}",
    eprint = "2407.01672",
    archivePrefix = "arXiv",
    primaryClass = "hep-ph",
    reportNumber = "PITT-PACC-2401",
    doi = "10.1103/PhysRevD.111.033004",
    journal = "Phys. Rev. D",
    volume = "111",
    number = "3",
    pages = "033004",
    year = "2025"
}

@article{Afik:2020onf,
    author = "Afik, Yoav and de Nova, Juan Ram\'on Mu\~noz",
    title = "{Entanglement and quantum tomography with top quarks at the LHC}",
    eprint = "2003.02280",
    archivePrefix = "arXiv",
    primaryClass = "quant-ph",
    doi = "10.1140/epjp/s13360-021-01902-1",
    journal = "Eur. Phys. J. Plus",
    volume = "136",
    number = "9",
    pages = "907",
    year = "2021"
}

@article{Fabbrichesi:2021npl,
    author = "Fabbrichesi, M. and Floreanini, R. and Panizzo, G.",
    title = "{Testing Bell Inequalities at the LHC with Top-Quark Pairs}",
    eprint = "2102.11883",
    archivePrefix = "arXiv",
    primaryClass = "hep-ph",
    doi = "10.1103/PhysRevLett.127.161801",
    journal = "Phys. Rev. Lett.",
    volume = "127",
    number = "16",
    pages = "161801",
    year = "2021"
}

@article{Severi:2021cnj,
    author = "Severi, Claudio and Boschi, Cristian Degli Esposti and Maltoni, Fabio and Sioli, Maximiliano",
    title = "{Quantum tops at the LHC: from entanglement to Bell inequalities}",
    eprint = "2110.10112",
    archivePrefix = "arXiv",
    primaryClass = "hep-ph",
    doi = "10.1140/epjc/s10052-022-10245-9",
    journal = "Eur. Phys. J. C",
    volume = "82",
    number = "4",
    pages = "285",
    year = "2022"
}

@article{Aguilar-Saavedra:2022uye,
    author = "Aguilar-Saavedra, J. A. and Casas, J. A.",
    title = "{Improved tests of entanglement and Bell inequalities with LHC tops}",
    eprint = "2205.00542",
    archivePrefix = "arXiv",
    primaryClass = "hep-ph",
    reportNumber = "IFT-UAM/CSIC-22-45",
    doi = "10.1140/epjc/s10052-022-10630-4",
    journal = "Eur. Phys. J. C",
    volume = "82",
    number = "8",
    pages = "666",
    year = "2022"
}

@article{Afik:2022dgh,
    author = "Afik, Yoav and de Nova, Juan Ram\'on Mu\~noz",
    title = "{Quantum Discord and Steering in Top Quarks at the LHC}",
    eprint = "2209.03969",
    archivePrefix = "arXiv",
    primaryClass = "quant-ph",
    doi = "10.1103/PhysRevLett.130.221801",
    journal = "Phys. Rev. Lett.",
    volume = "130",
    number = "22",
    pages = "221801",
    year = "2023"
}

@article{Severi:2022qjy,
    author = "Severi, Claudio and Vryonidou, Eleni",
    title = "{Quantum entanglement and top spin correlations in SMEFT at higher orders}",
    eprint = "2210.09330",
    archivePrefix = "arXiv",
    primaryClass = "hep-ph",
    doi = "10.1007/JHEP01(2023)148",
    journal = "JHEP",
    volume = "01",
    pages = "148",
    year = "2023"
}

@article{Aoude:2022imd,
    author = "Aoude, Rafael and Madge, Eric and Maltoni, Fabio and Mantani, Luca",
    title = "{Quantum SMEFT tomography: Top quark pair production at the LHC}",
    eprint = "2203.05619",
    archivePrefix = "arXiv",
    primaryClass = "hep-ph",
    reportNumber = "CP3-22-14",
    doi = "10.1103/PhysRevD.106.055007",
    journal = "Phys. Rev. D",
    volume = "106",
    number = "5",
    pages = "055007",
    year = "2022"
}

@article{Dong:2024xsg,
    author = "Dong, Zhongtian and Gon\c{c}alves, Dorival and Kong, Kyoungchul and Larkoski, Andrew J. and Navarro, Alberto",
    title = "{Hadronic top quark polarimetry with ParticleNet}",
    eprint = "2407.01663",
    archivePrefix = "arXiv",
    primaryClass = "hep-ph",
    doi = "10.1016/j.physletb.2025.139314",
    journal = "Phys. Lett. B",
    volume = "862",
    pages = "139314",
    year = "2025"
}

@article{Maltoni:2024csn,
    author = "Maltoni, Fabio and Severi, Claudio and Tentori, Simone and Vryonidou, Eleni",
    title = "{Quantum tops at circular lepton colliders}",
    eprint = "2404.08049",
    archivePrefix = "arXiv",
    primaryClass = "hep-ph",
    doi = "10.1007/JHEP09(2024)001",
    journal = "JHEP",
    volume = "09",
    pages = "001",
    year = "2024"
}

@article{White:2024nuc,
    author = "White, Chris D. and White, Martin J.",
    title = "{Magic states of top quarks}",
    eprint = "2406.07321",
    archivePrefix = "arXiv",
    primaryClass = "hep-ph",
    reportNumber = "ADP-24-10/T1249",
    doi = "10.1103/PhysRevD.110.116016",
    journal = "Phys. Rev. D",
    volume = "110",
    number = "11",
    pages = "116016",
    year = "2024"
}

@article{Han:2024ugl,
    author = "Han, Tao and Low, Matthew and McGinnis, Navin and Su, Shufang",
    title = "{Measuring Quantum Discord at the LHC}",
    eprint = "2412.21158",
    archivePrefix = "arXiv",
    primaryClass = "hep-ph",
    reportNumber = "PITT-PACC-2316",
    month = "12",
    year = "2024"
}

@article{Nason:2025hix,
    author = "Nason, Paolo and Re, Emanuele and Rottoli, Luca",
    title = "{Spin Correlations in $t{\bar t}$ Production and Decay at the LHC in QCD Perturbation Theory}",
    eprint = "2505.00096",
    archivePrefix = "arXiv",
    primaryClass = "hep-ph",
    reportNumber = "LAPTH-017/25",
    month = "4",
    year = "2025"
}

@article{Sakurai:2023nsc,
    author = "Sakurai, Kazuki and Spannowsky, Michael",
    title = "{Three-Body Entanglement in Particle Decays}",
    eprint = "2310.01477",
    archivePrefix = "arXiv",
    primaryClass = "quant-ph",
    reportNumber = "IPPP/23/54",
    doi = "10.1103/PhysRevLett.132.151602",
    journal = "Phys. Rev. Lett.",
    volume = "132",
    number = "15",
    pages = "151602",
    year = "2024"
}

@article{Liu:2025frx,
    author = "Liu, Qiaofeng and Low, Ian and Yin, Zhewei",
    title = "{Maximal magic for two-qubit states}",
    eprint = "2502.17550",
    archivePrefix = "arXiv",
    primaryClass = "quant-ph",
    doi = "10.1088/2058-9565/ae3028",
    journal = "Quantum Sci. Technol.",
    volume = "11",
    number = "1",
    pages = "015035",
    year = "2026"
}

@article{Ohta:2025utz,
    author = "Ohta, Misaki and Sakurai, Kazuki",
    title = "{Extremal Magic States from Symmetric Lattices}",
    eprint = "2506.11725",
    archivePrefix = "arXiv",
    primaryClass = "quant-ph",
    month = "6",
    year = "2025"
}

@techreport{CMS-PAS-TOP-25-001,
      collaboration = "CMS",
      title         = "{Observation of magic states of top quark pairs produced
                       in proton-proton collisions at $\sqrt{s}=13~\mathrm{TeV}$}",
      institution   = "CERN",
      number        = "CMS-PAS-TOP-25-001",
      reportNumber  = "CMS-PAS-TOP-25-001",
      address       = "Geneva",
      year          = "2025",
      url           = "https://cds.cern.ch/record/2926751"
}

@article{Bittel:2025yhq,
    author = "Bittel, Lennart and Leone, Lorenzo",
    title = "{Operational interpretation of the Stabilizer Entropy}",
    eprint = "2507.22883",
    archivePrefix = "arXiv",
    primaryClass = "quant-ph",
    doi = "10.22331/q-2026-04-15-2069",
    journal = "Quantum",
    volume = "10",
    pages = "2069",
    year = "2026"
}

@article{Horodecki:2025tpn,
    author = "Horodecki, Pawe{\l} and Sakurai, Kazuki and Shaleena, Abhyoudai S. and Spannowsky, Michael",
    title = "{Three-body non-locality in particle decays}",
    eprint = "2502.19470",
    archivePrefix = "arXiv",
    primaryClass = "quant-ph",
    reportNumber = "IPPP/25/12",
    doi = "10.1007/JHEP10(2025)160",
    journal = "JHEP",
    volume = "10",
    pages = "160",
    year = "2025"
}

@article{laskowski,
  title = {Tight Multipartite Bell's Inequalities Involving Many Measurement Settings},
  author = {Laskowski, Wies\l{}aw and Paterek, Tomasz and \ifmmode \dot{Z}\else \.{Z}\fi{}ukowski, Marek and Brukner, \ifmmode \check{C}\else \v{C}\fi{}aslav},
  journal = {Phys. Rev. Lett.},
  volume = {93},
  issue = {20},
  pages = {200401},
  numpages = {4},
  year = {2004},
  month = {Nov},
  publisher = {American Physical Society},
  doi = {10.1103/PhysRevLett.93.200401},
  url = {https://link.aps.org/doi/10.1103/PhysRevLett.93.200401}
}

@article{WU2003262,
title = {A new Bell inequality for three spin-half particle system},
journal = {Physics Letters A},
volume = {307},
number = {5},
pages = {262-264},
year = {2003},
issn = {0375-9601},
doi = {https://doi.org/10.1016/S0375-9601(02)01672-9},
url = {https://www.sciencedirect.com/science/article/pii/S0375960102016729},
author = {Xiao-Hua Wu and Hong-Shi Zong}
}

@article{Bernal:2024dtg,
    author = "Bernal, Alexander and Casas, J. Alberto and Moreno, Jes{\'u}s M.",
    title = "{Maximal Clauser-Horne-Shimony-Holt violation for qubit-qudit states}",
    eprint = "2404.02092",
    archivePrefix = "arXiv",
    primaryClass = "quant-ph",
    doi = "10.1103/g3mw-6rl1",
    journal = "Phys. Rev. A",
    volume = "112",
    number = "4",
    pages = "042404",
    year = "2025"
}

@article{Aguilar-Saavedra:2022mpg,
    author = "Aguilar-Saavedra, J. A.",
    title = "{Laboratory-frame tests of quantum entanglement in H\textrightarrow{}WW}",
    eprint = "2209.14033",
    archivePrefix = "arXiv",
    primaryClass = "hep-ph",
    reportNumber = "IFT-UAM/CSIC-22-119",
    doi = "10.1103/PhysRevD.107.076016",
    journal = "Phys. Rev. D",
    volume = "107",
    number = "7",
    pages = "076016",
    year = "2023"
}

@article{Goncalves:2025mvl,
    author = "Gon{\c{c}}alves, Dorival and Kaladharan, Ajay and Krauss, Frank and Navarro, Alberto",
    title = "{Quantum entanglement is quantum: ZZ production at the LHC}",
    eprint = "2505.12125",
    archivePrefix = "arXiv",
    primaryClass = "hep-ph",
    doi = "10.1007/JHEP12(2025)122",
    journal = "JHEP",
    volume = "12",
    pages = "122",
    year = "2025"
}

@article{Aguilar-Saavedra:2025byk,
    author = "Aguilar-Saavedra, J. A.",
    title = "{Quantum tomography beyond the leading order}",
    eprint = "2505.11870",
    archivePrefix = "arXiv",
    primaryClass = "hep-ph",
    reportNumber = "IFT-UAM/CSIC-25-49",
    doi = "10.1140/epjc/s10052-025-14710-z",
    journal = "Eur. Phys. J. C",
    volume = "85",
    number = "9",
    pages = "969",
    year = "2025"
}

@article{Ruzi:2024cbt,
    author = "Ruzi, Alim and Wu, Youpeng and Ding, Ran and Qian, Sitian and Levin, Andrew Micheal and Li, Qiang",
    title = "{Testing Bell inequalities and probing quantum entanglement at a muon collider}",
    eprint = "2408.05429",
    archivePrefix = "arXiv",
    primaryClass = "hep-ph",
    doi = "10.1007/JHEP10(2024)211",
    journal = "JHEP",
    volume = "10",
    pages = "211",
    year = "2024"
}

@article{Wu:2024ovc,
    author = "Wu, Youpeng and Jiang, Ruobing and Ruzi, Alim and Ban, Yong and Yan, Xueqing and Li, Qiang",
    title = "{Testing Bell inequalities and probing quantum entanglement at CEPC}",
    eprint = "2410.17025",
    archivePrefix = "arXiv",
    primaryClass = "hep-ph",
    doi = "10.1103/PhysRevD.111.036008",
    journal = "Phys. Rev. D",
    volume = "111",
    number = "3",
    pages = "036008",
    year = "2025"
}

@article{Ding:2025mzj,
    author = "Ding, Ran and Ruzi, Alim and Qian, Sitian and Levin, Andrew and Wu, Youpeng and Li, Qiang",
    title = "{Quantum Entanglement between gauge boson pairs at a Muon Collider}",
    eprint = "2504.09832",
    archivePrefix = "arXiv",
    primaryClass = "hep-ph",
    month = "4",
    year = "2025"
}

@article{Cheng:2025cuv,
    author = "Cheng, Kun and Yan, Bin",
    title = "{Bell Inequality Violation of Light Quarks in Back-to-Back Dihadron Pair Production at Lepton Colliders}",
    eprint = "2501.03321",
    archivePrefix = "arXiv",
    primaryClass = "hep-ph",
    reportNumber = "PITT-PACC-2414",
    month = "1",
    year = "2025"
}

@article{Altomonte:2024upf,
    author = "Altomonte, Clelia and Barr, Alan J. and Eckstein, Micha\l{} and Horodecki, Pawe\l{} and Sakurai, Kazuki",
    title = "{Prospects for quantum process tomography at high energies}",
    eprint = "2412.01892",
    archivePrefix = "arXiv",
    primaryClass = "hep-ph",
    month = "12",
    year = "2024"
}

@article{Ruzi:2025jql,
    author = "Ruzi, Alim and Wu, Youpeng and Ding, Ran and Li, Qiang",
    title = "{Searching Quantum Entanglement in $p\ p\to Z\ Z$ process}",
    eprint = "2506.16077",
    archivePrefix = "arXiv",
    primaryClass = "hep-ph",
    month = "6",
    year = "2025"
}

@article{Goncalves:2025xer,
    author = "Gon{\c{c}}alves, Dorival and Kaladharan, Ajay and Navarro, Alberto",
    title = "{Higher-order corrections to quantum observables in h {\textrightarrow} WW$^{*}$}",
    eprint = "2506.19951",
    archivePrefix = "arXiv",
    primaryClass = "hep-ph",
    doi = "10.1007/JHEP11(2025)158",
    journal = "JHEP",
    volume = "11",
    pages = "158",
    year = "2025"
}

@article{Goncalves:2026njf,
    author = "Gon{\c{c}}alves, Dorival and Kaladharan, Ajay and Navarro, Alberto",
    title = "{Quantum Tomography and Entanglement in Semi-Leptonic $h\to VV^*$ Decays at Higher Orders}",
    eprint = "2604.16218",
    archivePrefix = "arXiv",
    primaryClass = "hep-ph",
    month = "4",
    year = "2026"
}

@article{De:2025dpo,
    author = "De, Songshaptak and Dey, Atri and Samui, Tousik",
    title = "{Unfolding quantum entanglement from $h\to ZZ^*\to jj\ell\ell$ at a muon collider}",
    eprint = "2512.09121",
    archivePrefix = "arXiv",
    primaryClass = "hep-ph",
    reportNumber = "IMSc/2025/07",
    month = "12",
    year = "2025"
}

@article{CMS:2024pts,
    author = "Hayrapetyan, Aram and others",
    collaboration = "CMS",
    title = "{Observation of quantum entanglement in top quark pair production in proton{\textendash}proton collisions at $\sqrt{s} = 13$ TeV}",
    eprint = "2406.03976",
    archivePrefix = "arXiv",
    primaryClass = "hep-ex",
    reportNumber = "CMS-TOP-23-001, CERN-EP-2024-137",
    doi = "10.1088/1361-6633/ad7e4d",
    journal = "Rept. Prog. Phys.",
    volume = "87",
    number = "11",
    pages = "117801",
    year = "2024"
}

@article{CMS:2024zkc,
    author = "Hayrapetyan, Aram and others",
    collaboration = "CMS",
    title = "{Measurements of polarization and spin correlation and observation of entanglement in top quark pairs using lepton+jets events from proton-proton collisions at s=13{\,}{\,}TeV}",
    eprint = "2409.11067",
    archivePrefix = "arXiv",
    primaryClass = "hep-ex",
    reportNumber = "CMS-TOP-23-007, CERN-EP-2024-231",
    doi = "10.1103/PhysRevD.110.112016",
    journal = "Phys. Rev. D",
    volume = "110",
    number = "11",
    pages = "112016",
    year = "2024"
}

@article{ATLAS:2026hye,
    author = "Aad, Georges and others",
    collaboration = "ATLAS",
    title = "{Measurements of $Z$-boson pair entanglement in decays of Higgs bosons at the ATLAS experiment}",
    eprint = "2603.26463",
    archivePrefix = "arXiv",
    primaryClass = "hep-ex",
    reportNumber = "CERN-EP-2026-061",
    month = "3",
    year = "2026"
}

@article{Fang:2026ddi,
    author = "Fang, Yi-Jing and Bhoonah, Amit and Cheng, Kun and Han, Tao and Liu, Yandong and Zhang, Hao",
    title = "{Spin Correlation and Quantum Entanglement of Fermion Pairs in Transversely Polarized $e^-e^+$ Collisions}",
    eprint = "2604.11887",
    archivePrefix = "arXiv",
    primaryClass = "hep-ph",
    reportNumber = "PITT-PACC-2604",
    month = "4",
    year = "2026"
}

@article{Guo:2026yhz,
    author = "Guo, Yu-Chen and Han, Tao and Low, Matthew and Su, Youle",
    title = "{Quantum Tomography of Fermion Pairs in $e^+e^-$ Collisions: Longitudinal Beam Polarization Effects}",
    eprint = "2602.02719",
    archivePrefix = "arXiv",
    primaryClass = "hep-ph",
    month = "2",
    year = "2026"
}

@article{Choi:2026omc,
    author = "Choi, Seong Youl and Kang, Dong Woo and Lee, Jae Sik and Park, Chan Beom",
    title = "{Quantum entanglement and Bell nonlocality in top-quark pair production at a photon linear collider}",
    eprint = "2603.12830",
    archivePrefix = "arXiv",
    primaryClass = "hep-ph",
    reportNumber = "IUEP-HEP-26-01",
    month = "3",
    year = "2026"
}

@article{Durupt:2025wuk,
    author = "Durupt, Valentin and Maltoni, Fabio and Mattelaer, Olivier",
    title = "{Automated computation of spin-density matrices and quantum observables for collider physics}",
    eprint = "2510.17730",
    archivePrefix = "arXiv",
    primaryClass = "hep-ph",
    reportNumber = "IRMP-CP3-25-34",
    month = "10",
    year = "2025"
}

@article{Lin:2025eci,
    author = "Lin, Shi-Jia and Liu, Ming-Jun and Shao, Ding Yu and Wei, Shu-Yi",
    title = "{Spin correlations and Bell nonlocality in $ \Lambda \overline{\Lambda} $ pair production from e$^{+}$e$^{-}$ collisions with a thrust cut}",
    eprint = "2507.15387",
    archivePrefix = "arXiv",
    primaryClass = "hep-ph",
    doi = "10.1007/JHEP11(2025)082",
    journal = "JHEP",
    volume = "11",
    pages = "082",
    year = "2025"
}

@article{Aguilar-Saavedra:2025njw,
    author = "Aguilar-Saavedra, J. A.",
    title = "{Momentum entanglement at colliders: the $H \to WW,ZZ$ case}",
    eprint = "2512.02104",
    archivePrefix = "arXiv",
    primaryClass = "hep-ph",
    reportNumber = "IFT-UAM/CSIC-25-154",
    month = "12",
    year = "2025"
}

@article{Pelliccioli:2026ltl,
    author = "Pelliccioli, Giovanni and Re, Emanuele",
    title = "{SMEFT effects on spin correlations and entanglement at NLO QCD in di-boson production at hadron colliders}",
    eprint = "2601.09540",
    archivePrefix = "arXiv",
    primaryClass = "hep-ph",
    reportNumber = "COMETA-2026-01, LAPTH-003/26",
    month = "1",
    year = "2026"
}

@article{Altakach:2026fpl,
    author = "Altakach, Mohammad Mahdi and Lamba, Priyanka and Maltoni, Fabio and Sakurai, Kazuki",
    title = "{Quantum properties of heavy-fermion pairs at a lepton collider with polarised beams}",
    eprint = "2601.09558",
    archivePrefix = "arXiv",
    primaryClass = "hep-ph",
    month = "1",
    year = "2026"
}

@article{Gabrielli:2026tnl,
    author = "Gabrielli, Emidio and Marzola, Luca",
    title = "{Quantum entanglement and bell nonlocality at future lepton colliders}",
    eprint = "2602.03960",
    archivePrefix = "arXiv",
    primaryClass = "hep-ph",
    doi = "10.1140/epjp/s13360-026-07551-6",
    journal = "Eur. Phys. J. Plus",
    volume = "141",
    number = "3",
    pages = "331",
    year = "2026"
}

@article{Low:2025aqq,
    author = "Low, Matthew",
    title = "{Addressing local realism through Bell tests at colliders}",
    eprint = "2508.10979",
    archivePrefix = "arXiv",
    primaryClass = "hep-ph",
    doi = "10.1103/15c3-mg5l",
    journal = "Phys. Rev. D",
    volume = "112",
    number = "9",
    pages = "096008",
    year = "2025"
}

@article{Afik:2025grr,
    author = "Afik, Yoav and Kats, Yevgeny and de Nova, Juan Ram{\'o}n Mu{\~n}oz and Soffer, Abner and Uzan, David",
    title = "{Entanglement and Bell nonlocality with bottom-quark pairs at hadron colliders}",
    eprint = "2406.04402",
    archivePrefix = "arXiv",
    primaryClass = "hep-ph",
    doi = "10.1103/fhkc-kfhr",
    journal = "Phys. Rev. D",
    volume = "111",
    number = "11",
    pages = "L111902",
    year = "2025"
}

@article{Morales:2024jhj,
    author = "Morales, R. A.",
    title = "{Tripartite entanglement and Bell non-locality in loop-induced Higgs boson decays}",
    eprint = "2403.18023",
    archivePrefix = "arXiv",
    primaryClass = "hep-ph",
    doi = "10.1140/epjc/s10052-024-12921-4",
    journal = "Eur. Phys. J. C",
    volume = "84",
    number = "6",
    pages = "581",
    year = "2024"
}

@article{Fabbrichesi:2025zpw,
    author = "Fabbrichesi, M. and Floreanini, R. and Gabrielli, E. and Marzola, L.",
    title = "{Quantum-entangled pions}",
    eprint = "2506.05464",
    archivePrefix = "arXiv",
    primaryClass = "hep-ph",
    doi = "10.1103/2yst-qkcv",
    journal = "Phys. Rev. D",
    volume = "112",
    number = "9",
    pages = "L091301",
    year = "2025"
}

@article{Fabbrichesi:2024wcd,
    author = "Fabbrichesi, M. and Marzola, L.",
    title = "{Quantum tomography with {\ensuremath{\tau}} leptons at the FCC-ee: Entanglement, Bell inequality violation, $\sin \theta_W$, and anomalous couplings}",
    eprint = "2405.09201",
    archivePrefix = "arXiv",
    primaryClass = "hep-ph",
    doi = "10.1103/PhysRevD.110.076004",
    journal = "Phys. Rev. D",
    volume = "110",
    number = "7",
    pages = "076004",
    year = "2024"
}

@article{Fabbrichesi:2025ywl,
    author = "Fabbrichesi, Marco and Low, Matthew and Marzola, Luca",
    title = "{Trace distance between density matrices: A nifty tool in new-physics searches}",
    eprint = "2501.03311",
    archivePrefix = "arXiv",
    primaryClass = "hep-ph",
    doi = "10.1103/kdmh-3yb4",
    journal = "Phys. Rev. D",
    volume = "112",
    number = "1",
    pages = "013003",
    year = "2025"
}

@article{Fabbrichesi:2025psr,
    author = "Fabbrichesi, Marco and Floreanini, Roberto and Marzola, Luca",
    title = "{Local vs. nonlocal entanglement in top-quark pairs at the LHC}",
    eprint = "2505.02902",
    archivePrefix = "arXiv",
    primaryClass = "hep-ph",
    doi = "10.1007/JHEP11(2025)005",
    journal = "JHEP",
    volume = "11",
    pages = "005",
    year = "2025"
}

@article{LinearColliderVision:2025hlt,
    author = "Abramowicz, H. and others",
    collaboration = "Linear Collider Vision",
    title = "{A linear collider vision for the future of particle physics}",
    eprint = "2503.19983",
    archivePrefix = "arXiv",
    primaryClass = "hep-ex",
    reportNumber = "FERMILAB-PUB-25-0216-CSAID-TD",
    doi = "10.1140/epjs/s11734-026-02153-w",
    journal = "Eur. Phys. J. ST",
    volume = "235",
    number = "6",
    pages = "1641--1796",
    year = "2026"
}

@article{ILC:2013jhg,
    editor = "Baer, Howard and others",
    collaboration = "ILC",
    title = "{The International Linear Collider Technical Design Report - Volume 2: Physics}",
    eprint = "1306.6352",
    archivePrefix = "arXiv",
    primaryClass = "hep-ph",
    reportNumber = "ILC-REPORT-2013-040, ANL-HEP-TR-13-20, BNL-100603-2013-IR, IRFU-13-59, CERN-ATS-2013-037, COCKCROFT-13-10, CLNS-13-2085, DESY-13-062, FERMILAB-TM-2554, IHEP-AC-ILC-2013-001, INFN-13-04-LNF, JAI-2013-001, JINR-E9-2013-35, JLAB-R-2013-01, KEK-REPORT-2013-1, KNU-CHEP-ILC-2013-1, LLNL-TR-635539, SLAC-R-1004, ILC-HIGRADE-REPORT-2013-003",
    month = "6",
    year = "2013"
}

@article{Horodecki:2009zz,
    author = "Horodecki, Ryszard and Horodecki, Pawel and Horodecki, Michal and Horodecki, Karol",
    title = "{Quantum entanglement}",
    eprint = "quant-ph/0702225",
    archivePrefix = "arXiv",
    doi = "10.1103/RevModPhys.81.865",
    journal = "Rev. Mod. Phys.",
    volume = "81",
    pages = "865--942",
    year = "2009"
}

@article{Coffman:1999jd,
    author = "Coffman, Valerie and Kundu, Joydip and Wootters, William K.",
    title = "{Distributed entanglement}",
    eprint = "quant-ph/9907047",
    archivePrefix = "arXiv",
    doi = "10.1103/PhysRevA.61.052306",
    journal = "Phys. Rev. A",
    volume = "61",
    pages = "052306",
    year = "2000"
}

@article{Goncalves:2026nnx,
    author = "Gon\c{c}alves, Dorival and Navarro, Alberto and Sakurai, Kazuki",
    title = "{Tripartite Entanglement in $e^+ e^- \to t \bar{t} Z$}",
    eprint = "2606.11296",
    archivePrefix = "arXiv",
    primaryClass = "hep-ph",
    month = "6",
    year = "2026"
}

@article{FCC:2018evy,
    author = "Abada, A. and others",
    collaboration = "FCC",
    title = "{FCC-ee: The Lepton Collider}: {Future Circular Collider Conceptual Design Report Volume 2}",
    reportNumber = "CERN-ACC-2018-0057",
    doi = "10.1140/epjst/e2019-900045-4",
    journal = "Eur. Phys. J. ST",
    volume = "228",
    number = "2",
    pages = "261--623",
    year = "2019"
}

@article{CEPCStudyGroup:2018ghi,
    author = "Dong, Mingyi and others",
    editor = "Guimar{\~a}es da Costa, Jo{\~a}o Barreiro and others",
    collaboration = "CEPC Study Group",
    title = "{CEPC Conceptual Design Report: Volume 2 - Physics {\&} Detector}",
    eprint = "1811.10545",
    archivePrefix = "arXiv",
    primaryClass = "hep-ex",
    reportNumber = "IHEP-CEPC-DR-2018-02, IHEP-EP-2018-01, IHEP-TH-2018-01",
    month = "11",
    year = "2018"
}

@article{LHCHiggsCrossSectionWorkingGroup:2016ypw,
    author = "de Florian, D. and others",
    collaboration = "LHC Higgs Cross Section Working Group",
    title = "{Handbook of LHC Higgs Cross Sections: 4. Deciphering the Nature of the Higgs Sector}",
    eprint = "1610.07922",
    archivePrefix = "arXiv",
    primaryClass = "hep-ph",
    reportNumber = "CERN-2017-002-M",
    doi = "10.23731/CYRM-2017-002",
    journal = "CERN Yellow Rep. Monogr.",
    volume = "2",
    pages = "1--869",
    year = "2017"
}

@article{ZurbanoFernandez:2020cco,
    author = "Zurbano Fernandez, I. and others",
    editor = {B{\'e}jar Alonso, I. and Br{\"u}ning, O. and Fessia, P. and Rossi, L. and Tavian, L. and Zerlauth, M.},
    title = "{High-Luminosity Large Hadron Collider (HL-LHC): Technical design report}",
    reportNumber = "CERN-2020-010",
    doi = "10.23731/CYRM-2020-0010",
    volume = "10/2020",
    month = "12",
    year = "2020"
}

@article{Gargalionis:2025iqs,
    author = "Gargalionis, John and Moynihan, Nathan and Trifinopoulos, Sokratis and Wallace, Ewan N. V. and White, Chris D. and White, Martin J.",
    title = "{Spin versus Magic: Lessons from Gluon and Graviton Scattering}",
    eprint = "2508.14967",
    archivePrefix = "arXiv",
    primaryClass = "hep-th",
    journal = "Phys. Rev. D",
    volume = "113",
    number = "1",
    pages = "016007",
    year = "2026"
}

@article{Gargalionis:2026onv,
    author = "Gargalionis, John and Moynihan, Nathan and Reichenberg Ashby, Michael L. and Wallace, Ewan N. V. and White, Chris D. and White, Martin J.",
    title = "{Non-local nonstabiliserness in Gluon and Graviton Scattering}",
    eprint = "2603.04148",
    archivePrefix = "arXiv",
    primaryClass = "hep-th",
    month = "3",
    year = "2026"
}

@article{Robin:2025ymq,
    author = "Robin, C. E. P. and Savage, M. J.",
    title = "{Anti-Flatness and Non-Local Magic in Two-Particle Scattering Processes}",
    eprint = "2510.23426",
    archivePrefix = "arXiv",
    primaryClass = "quant-ph",
    month = "10",
    year = "2025"
}

@article{Busoni:2026lvp,
    author = "Busoni, Giorgio and Gargalionis, John and Wallace, Ewan N. V. and White, Martin J.",
    title = "{Analytic formulae for non-local magic in bipartite systems of qutrits and ququints}",
    eprint = "2603.09155",
    archivePrefix = "arXiv",
    primaryClass = "quant-ph",
    month = "3",
    year = "2026"
}

@article{Gottesman:1998hu,
    author = "Gottesman, Daniel",
    title = "{The Heisenberg representation of quantum computers}",
    eprint = "quant-ph/9807006",
    archivePrefix = "arXiv",
    reportNumber = "LA-UR-98-2848",
    month = "7",
    year = "1998"
}

@article{Leone:2024lfr,
    author = "Leone, Lorenzo and Bittel, Lennart",
    title = "{Stabilizer entropies are monotones for magic-state resource theory}",
    eprint = "2404.11652",
    archivePrefix = "arXiv",
    primaryClass = "quant-ph",
    doi = "10.1103/PhysRevA.110.L040403",
    journal = "Phys. Rev. A",
    volume = "110",
    number = "4",
    pages = "L040403",
    year = "2024"
}

@article{Horodecki:1995nsk,
    author = "Horodecki, Ryszard and Horodecki, Pawel and Horodecki, Michal",
    title = "{Violating Bell inequality by mixed spin-1/2 states: necessary and sufficient condition}",
    doi = "10.1016/0375-9601(95)00214-N",
    journal = "Phys. Lett. A",
    volume = "200",
    number = "5",
    pages = "340--344",
    year = "1995"
}

@article{Chu:2026yxm,
    author = "Chu, Jinwei and Kharel, Savan",
    title = "{Magic without a phase: phase-independent stabilizer R{\'e}nyi entropy in gluon scattering}",
    eprint = "2607.13134",
    archivePrefix = "arXiv",
    primaryClass = "hep-th",
    month = "7",
    year = "2026"
}

@article{Lamba:2026jfm,
    author = "Lamba, Priyanka and Maltoni, Fabio and Miniati, Olimpia and Vryonidou, Eleni",
    title = "{Quantum detection of CP violation in the $t\bar{t}$ system: production}",
    eprint = "2607.25029",
    archivePrefix = "arXiv",
    primaryClass = "hep-ph",
    month = "7",
    year = "2026"
}

@article{Lamba:2026sni,
    author = "Lamba, Priyanka and Maltoni, Fabio and Miniati, Olimpia and Vryonidou, Eleni",
    title = "{Quantum detection of CP violation in the $t\bar{t}$ system: tomography}",
    eprint = "2607.25034",
    archivePrefix = "arXiv",
    primaryClass = "hep-ph",
    month = "7",
    year = "2026"
}

@article{Mermin:1990vxe,
    author = "Mermin, N. David",
    title = "{Extreme quantum entanglement in a superposition of macroscopically distinct states}",
    doi = "10.1103/PhysRevLett.65.1838",
    journal = "Phys. Rev. Lett.",
    volume = "65",
    number = "15",
    pages = "1838",
    year = "1990"
}

@article{Svetlichny:1987bqj,
    author = "Svetlichny, George",
    title = "{Distinguishing three-body from two-body nonseparability by a Bell-type inequality}",
    doi = "10.1103/physrevd.35.3066",
    journal = "Phys. Rev. D",
    volume = "35",
    number = "10",
    pages = "3066--3069",
    year = "1987"
}

\end{document}